\documentclass[11pt,journal,onecolumn]{IEEEtran}

\pdfoutput=1

\usepackage{amsfonts}
\usepackage{upgreek}
\usepackage{amssymb}
\usepackage{setspace}
\usepackage{multicol}
\usepackage{cite}
\usepackage{ctable}
\usepackage{amsmath}
\usepackage{array}
\usepackage{mdwmath}
\usepackage{mdwtab}
\usepackage{eqparbox}
\usepackage{url}
\usepackage{graphicx}
\usepackage{amsmath}
\usepackage{epsf}
\usepackage{mathrsfs}
\usepackage{cite}
\usepackage{dsfont}
\usepackage{epsfig}
\usepackage{enumerate}
\usepackage{cleveref}
\usepackage[caption=false]{subfig}

\newcommand{\subparagraph}{}
\usepackage{titlesec}
\titlespacing{\subsubsection}{\parindent}{2.25ex plus 1ex minus .2ex}{1.5ex plus .2ex}

\newtheorem{theorem}{Theorem}

\newtheorem{remark}{Remark}

\newtheorem{definition}{Definition}
\newtheorem{notation}{Notation}

\newcommand{\xd}{\mathbf{d}}

\newcommand{\xH}{\mathbf{H}}

\newcommand{\xR}{\mathbf{R}}

\newcommand{\xxC}{\mathcal{C}}
\newcommand{\xxD}{\mathcal{D}}

\newcommand{\xxI}{\mathcal{I}}
\newcommand{\xxN}{\mathcal{N}}

\newcommand{\xxS}{\mathcal{S}}

\newcommand{\bbZ}{\mathbb{Z}}

\newcommand{\xxSt}{\mathcal{S}^\textup{(t)}}
\newcommand{\xxSr}{\mathcal{S}^\textup{(r)}}

\allowdisplaybreaks

\newcommand{\Def}{\triangleq}

\newcommand{\DoFa}{\underline{\mathsf{DoF}}}
\newcommand{\DoF}{\mathsf{DoF}}
\newcommand{\lcm}{\mathrm{lcm}}
\newcommand{\ie}{i.e., }

\usepackage{wasysym}
\newcommand{\phase}[2]{$\bullet$ \underline{\emph{Phase #1 (#2)}}:}
\newcommand{\round}[2]{$\RHD$ \emph{Round #1 (#2)}:}

\DeclareMathOperator*{\argmax}{arg\,max}

\crefname{lemma}{Lemma}{Lemmas}
\Crefname{lemma}{Lemma}{Lemmas}
\crefname{notation}{Notation}{Notations}
\Crefname{notation}{Notation}{Notations}
\crefname{remark}{Remark}{Remarks}
\Crefname{remark}{Remark}{Remarks}
\crefname{theorem}{Theorem}{Theorems}
\Crefname{theorem}{Theorem}{Theorems}
\crefname{definition}{Definition}{Definitions}
\Crefname{definition}{Definition}{Definitions}
\crefname{section}{Section}{Sections}
\Crefname{section}{Section}{Sections}
\crefname{figure}{Fig.}{Figs.}
\Crefname{figure}{Figure}{Figures}
\crefname{table}{Table}{Tables}
\Crefname{table}{Table}{Tables}
\crefname{equation}{}{Eqs.}
\Crefname{equation}{Equation}{Equations}

\long\def\symbolfootnote[#1]#2{\begingroup%
\def\thefootnote{\fnsymbol{footnote}}\footnote[#1]{#2}\endgroup}

\graphicspath{{./Figs/}}

\begin{document}
\title{
Interference and X Networks with Noisy Cooperation and Feedback}
\author{\IEEEauthorblockN{Mohammad Javad Abdoli, Akbar Ghasemi, and Amir Keyvan Khandani}\\
\IEEEauthorblockA{Department of Electrical and Computer Engineering\\
University of Waterloo, Waterloo, ON, Canada N2L 3G1\\
Emails: \{mjabdoli, aghasemi, khandani\}@uwaterloo.ca}\\
} 
\maketitle
\let\thefootnote\relax\footnotetext{Financial supports provided by Natural Sciences and Engineering Research Council of Canada (NSERC) and Ontario Ministry of Research \& Innovation (ORF-RE) are gratefully acknowledged.

Part of this work will be presented in IEEE International Symposium on Information Theory (ISIT), Cambridge, MA, July 2012 \cite{Abdoli2012Fullduplex}.}

\begin{abstract}
The Gaussian $K$-user interference and $M\times K$ X channels are investigated with no instantaneous channel state information (CSI) at transmitters. First, it is assumed that the CSI is fed back to all nodes after a finite delay (delayed CSIT), and furthermore, the transmitters operate in full-duplex mode, \ie they can transmit and receive simultaneously. Achievable results are obtained on the degrees of freedom (DoF) of these channels under the above assumption. It is observed that, in contrast with no CSIT and full CSIT models, when CSIT is delayed, the achievable DoFs for both channels with full-duplex transmitter cooperation are greater than the best available achievable results on their DoF without transmitter cooperation. Our results are the first to show that  the full-duplex transmitter cooperation can potentially improve the channel DoF with delayed CSIT. Then, $K$-user interference and $K\times K$ X channels are considered with output feedback, wherein the channel output of each receiver is causally fed back to its corresponding transmitter. Our achievable results with output feedback demonstrate strict DoF improvements over those with the full-duplex delayed CSIT when $K>5$ in the $K$-user interference channel and $K>2$ in the $K\times K$ X channel. Next, the combination of delayed CSIT and output feedback, known as Shannon feedback, is studied and strictly higher DoFs compared to the output feedback model are achieved in the $K$-user interference channel when $K=5$ or $K>6$, and in the $K\times K$ X channel when $K>2$. Although being strictly greater than $1$ and increasing with size of the networks, the achievable DoFs in all the models studied in this paper approach limiting values not greater than $2$.
\end{abstract}
\newpage
\section{Introduction}
The crucial role of feedback in reliability, throughput, and complexity of transmission over communication networks has made it an indispensable ingredient of all modern communication systems. In spite of the first result by Shannon that shows the capacity of a memoryless point-to-point channel is not increased with feedback \cite{shannon1956zero}, it has been proved that feedback enlarges the capacity region of several multi-user channels. The capacity regions of the additive white Gaussian noise (AWGN) multiple-access, broadcast, and interference channels are enlarged with noiseless output feedback as shown in \cite{ozarow1984MACcapacity,ozarow1984BCCachievable,suh2010feedback}. It was shown in \cite{Bhaskaran2008GaussianBCC} that even a single output feedback link from one of receivers enlarges the capacity region of two-user AWGN broadcast channel. In fading AWGN channels, since it is commonly assumed that each receiver obtains the channel state information (CSI) instantaneously and perfectly, the channel output(s) and/or the CSI can be fed back to the transmitter(s). Without any feedback, and hence, without CSI at any transmitter (no CSIT), the capacity regions of single-input single-output (SISO) fading two-user broadcast and two-user Z-interference channels have been characterized to within constant number of bits (see \cite{tse2009fadingBC,Zhu2011ZIC} and references therein). Also, it was shown in \cite{vaze2009DoF_NoCSIT} that a large class of multiple-input single-output (MISO) multi-user channels including broadcast, interference, X, and cognitive radio channels can achieve no more that one degree of freedom (DoF) with no CSIT. As a first order approximation of the channel capacity, the DoF of a channel characterizes its sum-capacity in high signal-to-noise-ratio (SNR) regime, \ie
\begin{align}
C(\textup{SNR})= \DoF \times \log_2(\textup{SNR})+o(\log_2(\textup{SNR})),
\end{align}
where $C(\textup{SNR})$ is the sum-capacity for a given SNR and $\DoF$ is the channel sum-DoF, or simply, DoF. 

When there is CSI feedback to transmitter(s) and the channel variations are not too fast, it is commonly assumed that the CSI obtained through feedback links is valid at least over the current channel use, and hence, the transmitter(s) have access to perfect and instantaneous CSI (full CSIT). In this case, using the interference alignment technique, the $K$-user SISO interference channel (IC) and $M\times K$ SISO X channel were shown to have $K/2$ and $MK/(M+K-1)$ DoF, respectively, in \cite{cadambe2008interference,cadambe2009X}. If the channel coefficients are i.i.d.\ over time and the feedback delay is greater than a channel use period, the CSI obtained through feedback links is outdated. This makes the ``full CSIT assumption'' practically implausible, since the CSIT expires prior to the beginning of each channel use. Nevertheless, it has been established that the outdated CSIT (known as delayed CSIT) yields DoF gain in broadcast channels \cite{maddah2010DoF_BCC_Delayed_Arxiv,vaze2011DoF_BCC_Delayed,abdoli2011BCC} and interference and X channels \cite{Maleki_Jafar_Retro,Ghasemi2011Xchannel,Ghasemi2012MIMOX,Abdoli2011Allerton,Abdoli2011IC-X-Arxiv,vaze2011DoF_IC_DelayedCSIT_Arxiv,ghasemi2011interference}. The two-user MIMO interference channel with both delayed CSIT and output feedback has been recently studied in \cite{Vaze2011ShannonFeedback,Tandon2011ShannonFeedback}.

Output feedback in multi-user channels with distributed transmitters, such as IC and X channel, naturally provides some level of transmitter cooperation. As such, there are connections between communication over these channels with feedback and that with transmitter cooperation. A common cooperation setup is to enable transmitters to operate in full-duplex mode, \ie to transmit and receive simultaneously. The two-user IC with full-duplex transmitters and with full CSIT was investigated in \cite{Host2006CooperativeDiversity,Prabhakaran2011FullDuplex,yang2011interference,tandon2011dependence,cao2007achievable}. In \cite{Prabhakaran2011FullDuplex,yang2011interference,cao2007achievable} achievable schemes are proposed based on further splitting the common and/or private information of the HK scheme into two parts, namely, non-cooperative and cooperative part. The cooperative part is decoded at the other transmitter as well to be able to cooperate in delivering the information to the desired receiver. By developing an upper bound, the sum-capacity of the two-user Gaussian IC with full-duplex transmitters was obtained to within a constant number of bits in \cite{Prabhakaran2011FullDuplex}. Moreover, it was shown in \cite{Host2005Multiplexing,Cadambe2009FullDuplex} that under the full CSIT assumption, the full-duplex cooperation and/or output feedback cannot increase DoF of the $K$-user IC and $M\times K$ X channel. In other words, the full-duplex cooperation as well as output feedback can only yield ``additive'' capacity increase in the aforementioned channels when the full CSI is available at transmitters. With no CSIT also the full-duplex transmitter cooperation cannot help these channels to achieve more than one DoF, since the MISO broadcast channel DoF is equal to one with no CSIT\cite{vaze2009DoF_NoCSIT}.

In this paper, we address the problem of communication over the $K$-user SISO IC and $M\times K$ SISO X channel with no \emph{instantaneous} CSIT, and study the impact of full-duplex transmitter cooperation and/or different types of feedback on DoF of these channels. Specifically, after presenting the problem formulation in \cref{Sec:ProblemFormulation}, we start by giving some illustrative examples for the interference channel in \Cref{Sec:IC-Illustrative} and X channel in \Cref{Sec:X-Illustrative}. These examples highlight our interference alignment ideas for the channels with a few number of users. Then, we present our main results in \cref{Sec:MainResults}, and provide the proofs in subsequent sections. In particular, we consider these channels with delayed CSIT and full-duplex transmitter cooperation in \cref{Sec:ICX-DCSIT-FD}. Regarding the full-duplex CSI, we assume that the source nodes (transmitters) have only access to their incoming full-duplex CSI. We propose transmission schemes that achieve DoFs greater than the best previously reported DoFs for these channels with delayed CSIT but without transmitter cooperation \cite{Abdoli2011IC-X-Arxiv}.

In \cref{Sec:ICX-OF}, we consider the same channels with output feedback, wherein we assume that each transmitter has a causal access to the output of its paired receiver through a feedback link. This is indeed a limited output feedback (in contrast to providing each transmitter with the outputs of more than one receiver), however, the term ``limited'' will be henceforth dropped for brevity. Therefore, in the X channel, we hereafter consider only $M=K$ with a one-to-one mapping between transmitters and receivers for feedback assignment. The $3$-user IC and $2\times 2$ X channel with output feedback were previously investigated in \cite{Maleki_Jafar_Retro}, wherein $6/5$ and $4/3$ DoF were respectively achieved. While achieving the same DoFs for the $3$-user IC and $2\times 2$ X channel, our main contribution here is proposing multi-phase transmission schemes for the general $K$-user cases that achieve DoF values strictly increasing in $K$.

Next, we study the $K$-user SISO IC and $K\times K$ SISO X channel with delayed CSIT \emph{and} output feedback in \cref{Sec:ICX-SF}. Under this assumption, which is referred to as \emph{Shannon feedback}, we propose multi-phase transmission schemes capturing both the delayed CSI and output feedback to cooperatively transmit over the channel. The achieved DoFs are strictly increasing in $K$ and greater than those we achieved with output feedback for $K=5$ and $K>6$ in the $K$-user IC and for $K>2$ in the $K\times K$ X channel. The achievable results will be compared and discussed in \cref{Sec:Comparison}, and finally, the paper is concluded in \cref{Sec:Conclusions}.

\section{Problem Formulation}
\label{Sec:ProblemFormulation}
\begin{figure}
\centering
\includegraphics[scale=.9]{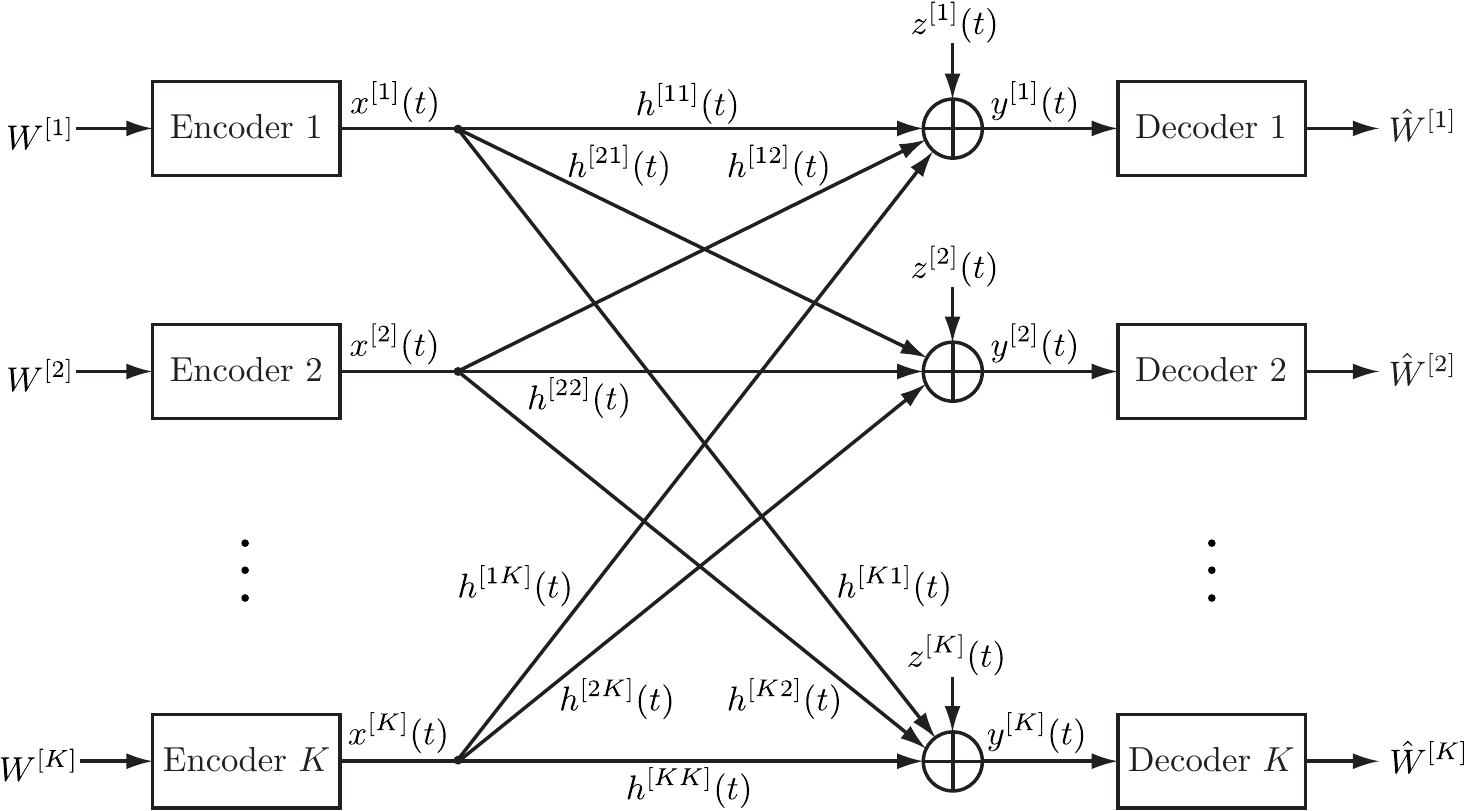}
\caption{$K$-user SISO interference channel}
\label{Fig:K-user-IC}
\end{figure}

Let us make the following definitions:
\begin{definition}[$K$-user SISO AWGN Interference Channel]
\label{Def:IC}
A set of $K$ transmitters and $K$ receivers, depicted in \cref{Fig:K-user-IC}, where transmitter $i$ (TX$_i$), $1\leq i \leq K$, wishes to communicate a message $W^{[i]}\in \{1, 2, 3, \cdots, 2^{\tau R^{[i]}}\}$ of rate $R^{[i]}$ to receiver $i$ (RX$_i$) over a block of  $\tau$ channel uses (or time slots). In time slot  $t$, $t=1,2,\cdots, \tau$, signal $x^{[i]}(t)\in \mathbb{C}$ is transmitted by TX$_i$, $1\leq i \leq K$, and signal $y^{[j]}(t)\in \mathbb{C}$ is received by RX$_j$, $1\leq j\leq K$, where
\begin{align}
\label{Eq:IC-input-output}
y^{[j]}(t)=\sum_{i=1}^Kh^{[ji]}(t)x^{[i]}(t)+z^{[j]}(t),
\end{align}
and $h^{[ji]}(t)\in \mathbb{C}$ is the channel coefficient from TX$_i$ to RX$_j$, and $z^{[j]}(t)\sim \xxC\xxN(0,1)$ is the additive white Gaussian noise at RX$_j$. The transmitted signal $x^{[i]}(t)$, $1\leq i \leq K$, is subject to power constraint $P$, \ie $\mathbb{E}[|x^{[i]}(t)|^2]\leq P$. The $K\times K$ channel matrix $\xH(t)$ in time slot $t$ is defined as $\xH(t)\Def\left(h^{[ji]}(t)\right)_{1\leq i,j \leq K}$. The channel coefficients are independent and identically distributed (i.i.d.) across all nodes as well as time slots. The channel coefficients are assumed to be drawn according to a finite-variance continuous distribution. Each receiver RX$_j$, $1\leq j \leq K$, knows all its incoming channel coefficients in time slot $t$, \ie $\{h^{[ji]}(t)\}_{i=1}^K$, perfectly and instantaneously.
\end{definition}

\begin{figure}
\centering
\includegraphics[scale=.9]{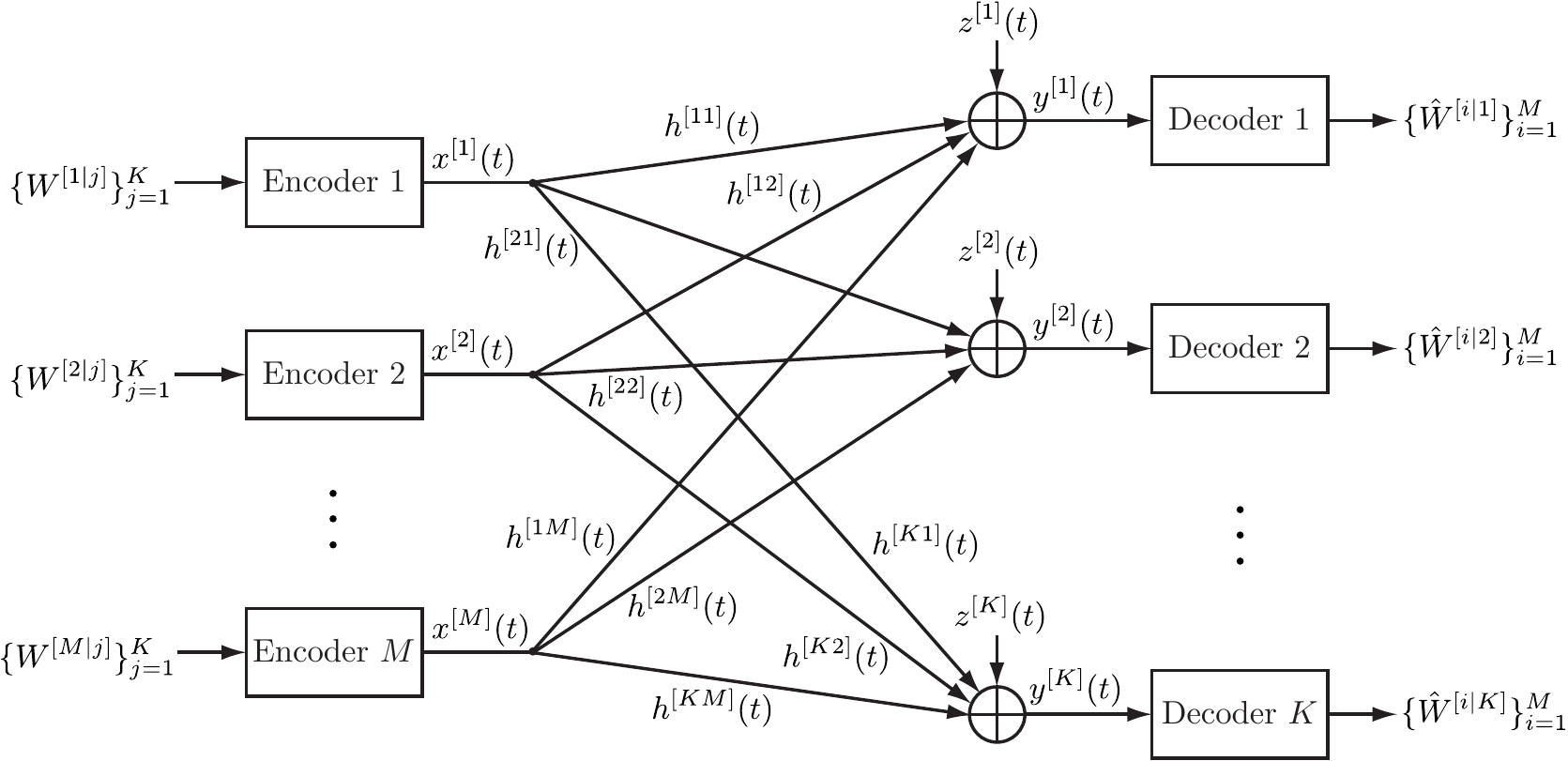}
\caption{$M\times K$ SISO X channel}
\label{Fig:M-K-X}
\end{figure}

\begin{definition}[$M\times K$ SISO AWGN X Channel]
\label{Def:XChannel}
A set of $M$ transmitters and $K$ receivers as depicted in \cref{Fig:M-K-X}, where TX$_i$, $1\leq i \leq M$, has a message $W^{[i|j]}\in \{1, 2, 3, \cdots, 2^{\tau R^{[i|j]}}\}$ of rate $R^{[i|j]}$ for each receiver RX$_j$, $1\leq j \leq K$. The input-output relationship of this channel is also given by \cref{Eq:IC-input-output} with the summation taken over the $M$ transmitters and with the same channel parameters and power constraint $P$ at each transmitter. The channel matrix $\xH(t)$ here is a $K\times M$ matrix defined as $\xH(t)\Def\left(h^{[ji]}(t)\right)_{1\leq i\leq M, 1\leq j \leq K}$. Similar to the IC, each receiver RX$_j$, $1\leq j \leq K$, knows all its incoming channel coefficients in time slot $t$, \ie $\{h^{[ji]}(t)\}_{i=1}^M$, perfectly and instantaneously.
\end{definition}

\begin{definition}[Feedback Models]
\label{Def:FeedbackModels}
We assume that each receiver knows channel coefficients of the other receivers with one time slot delay. Moreover, three different feedback models are considered in this paper, which are defined as follows:
\begin{itemize}
\item \emph{Delayed CSIT}: The channel matrix $\xH(t)$ will become available at all transmitters with one time slot delay via noiseless feedback links. 
\item \emph{Output Feedback}: Each channel output $y^{[i]}(t)$, $1\leq i \leq K$, will become available at TX$_i$ with one time slot delay via a noiseless feedback link. Therefore, for the X channel, we only consider $M=K$ under the output feedback assumption.
\item \emph{Shannon Feedback}: The transmitters have access to both delayed CSIT and output feedback as defined above. Therefore, for the X channel, we only consider $M=K$ under the Shannon feedback assumption.
\end{itemize}
\end{definition}

\begin{definition}[Full-duplux Transmitter Cooperation]
\label{Def:FullDuplex}
The transmitters are said to operate in full-duplex mode if they can transmit and receive simultaneously. In full-duplex mode, the received signal of TX$_i$ in time slot $t$ is given by
\begin{align}
&\textup{$K$-user IC}:& \tilde{y}^{[i]}(t)&=\sum_{i'=1}^K\tilde{h}^{[ii']}(t)x^{[i']}(t)+\tilde{z}^{[i]}(t),\quad 1\leq i \leq K. \\
&\textup{$M\times K$ X channel}:& \tilde{y}^{[i]}(t)&=\sum_{i'=1}^M\tilde{h}^{[ii']}(t)x^{[i']}(t)+\tilde{z}^{[i]}(t),\quad 1\leq i \leq M.
\end{align}
The noise terms and channel coefficients are assumed to be i.i.d. across all transmitters and time. No feedback link is available between the transmitters, and hence, TX$_i$ is assumed to have only its incoming full-duplex channel coefficients, \ie $\{\tilde{h}^{[ii']}(t)\}_{i'=1}^K$ in the IC and $\{\tilde{h}^{[ii']}(t)\}_{i'=1}^M$ in the X channel, perfectly and instantaneously.
\end{definition}

\begin{definition}[Block Code with Feedback]
A $(2^{\tau\xR},\tau)$ code of block length $\tau$ and rate $\xR= \left(R^{[i]}\right)_{i=1}^K$ with feedback in the $K$-user IC is defined as $K$ sets of encoding functions $\{\varphi^{[i]}_{t,\tau}\}_{t=1}^\tau$, $1\leq i \leq K$, such that
\begin{align}
x^{[i]}(t)=\varphi_{t,\tau}^{[i]}(W^{[i]},\xxI^{[i]}(t)),\quad 1\leq t \leq \tau,
\end{align}
together with $K$ decoding functions $\psi^{[i]}_\tau$, $1\leq i \leq K$, such that $\hat{W}^{[i]}_\tau=\psi^{[i]}_\tau(\left\{y^{[i]}(t)\right\}_{t=1}^\tau)$, where $\xxI^{[i]}(t)$ is the side information available at TX$_i$ \emph{before} time slot $t$, which will be defined later in this section. Similarly, A $(2^{\tau\xR},\tau)$ code of block length $\tau$ and rate $\xR= \left(R^{[i|j]}\right)_{1\leq i \leq M, 1\leq j \leq K}$ with feedback in the $M\times K$ X channel is defined as $M$ sets of encoding functions $\{\varphi^{[i]}_{t,\tau}\}_{t=1}^\tau$, $1\leq i \leq M$, such that
\begin{align}
x^{[i]}(t)=\varphi_{t,\tau}^{[i]}(\{W^{[i|j]}\}_{j=1}^K,\xxI^{[i]}(t)),\quad 1\leq t \leq \tau,
\end{align}
together with $K$ decoding functions $\psi^{[j]}_\tau$, $1\leq j \leq K$, such that $\{\hat{W}^{[i|j]}_\tau\}_{i=1}^M=\psi^{[j]}_\tau(\left\{y^{[j]}(t)\right\}_{t=1}^\tau)$.
\end{definition}

\begin{definition}[Transmitter Side Information]
\label{Def:TransmitterSideInf}
Using \cref{Def:FeedbackModels,Def:FullDuplex}, the following feedback and/or transmitter cooperation models will be investigated in this paper, each of which is equivalent to a certain transmitter side information:
\begin{enumerate}[(a)]
\item The $K$-user IC and $M\times K$ X channel with delayed CSIT and full-duplex transmitter cooperation:
\begin{align}
&\textup{$K$-user IC}:& \xxI^{[i]}(t)&\Def\left\{\tilde{y}^{[i]}(t'),\xH(t')\right\}_{t'=1}^{t-1}{\cup} \left\{\vphantom{y^{[i]}}\right.\hspace{-1mm}\tilde{h}^{[ii']}(t'):1\leq i' \leq K \hspace{-1mm}\left.\vphantom{y^{[i]}}\right\}_{t'=1}^t\hspace{-1mm},\quad 1\leq i \leq K. \\
&\textup{$M\times K$ X channel}:& \xxI^{[i]}(t)&\Def\left\{\tilde{y}^{[i]}(t'),\xH(t')\right\}_{t'=1}^{t-1}{\cup} \left\{\vphantom{y^{[i]}}\right.\hspace{-1mm}\tilde{h}^{[ii']}(t'):1\leq i' \leq M\hspace{-1mm}\left.\vphantom{y^{[i]}}\right\}_{t'=1}^t\hspace{-1mm},\quad 1\leq i \leq M.
\end{align}
\item The $K$-user IC and $K\times K$ X channel with output feedback:
\begin{align}
\xxI^{[i]}(t)\Def\left\{y^{[i]}(t')\right\}_{t'=1}^{t-1},\quad 1\leq i \leq K.
\end{align}
\item The $K$-user IC and $K\times K$ X channel with Shannon feedback:
\begin{align}
\xxI^{[i]}(t)\Def\left\{y^{[i]}(t'),\xH(t')\right\}_{t'=1}^{t-1},\quad 1\leq i \leq K.
\end{align}
\end{enumerate}
\end{definition}

\begin{definition}[Probability of Error, Achievable Rate, and Capacity Region]
Defining the probability of error of a code as the probability of decoding any of the transmitted messages incorrectly, a rate tuple $\xR$ is said to be achievable if there exists a sequence $\{(2^{\tau\xR},\tau)\}_{\tau=1}^\infty$ of codes such that their probability of error goes to zero as $\tau \to \infty$. The closure of the set of all achievable rate tuples $\xR$ is called the capacity region of the channel with power constraint $P$ and is denoted by $\xxC(P)$.
\end{definition}
\begin{definition}[DoF]
If $\xR=(R_1,R_2,\cdots,R_N)\in \xxC(P)$ is an achievable rate tuple, then $\xd\Def \lim_{P\to \infty}\frac{\xR}{\log_2 P}$ is called an achievable DoF tuple and $d_1+d_2+\cdots+d_N$ is called an achievable sum-DoF or simply achievable DoF. The closure of the set of all achievable DoF tuples is called the DoF region and denoted by $\xxD$, and the channel sum-DoF, or simply DoF, is defined as $\max_{\xd\in\xxD}d_1+d_2+\cdots+d_N$.
\end{definition}

In the following two sections, we elaborate on our transmission schemes for examples of the IC with a few number of users. Each channel will be investigated under each of the following assumptions defined in \cref{Def:TransmitterSideInf}:
\begin{enumerate}[(a)]
\item \label{Item:FD-Assumption} Full-duplex transmitter cooperation and delayed CSIT (which is also called \emph{full-duplex delayed CSIT} in this paper);
\item \label{Item:OF-Assumption} Output feedback;
\item \label{Item:SF-Assumption} Shannon feedback.
\end{enumerate}

\section{Illustrative Examples: Interference Channel}
\label{Sec:IC-Illustrative}
Note that for the two-user IC, none of the assumptions (\ref{Item:FD-Assumption})-(\ref{Item:SF-Assumption}) can help to achieve more than one DoF. This follows from the fact that DoF of this channel with full CSIT is equal to $1$, and full-duplex cooperation and/or output feedback cannot increase the channel DoF with full CSIT\cite{Cadambe2009FullDuplex}. Hence, we start by the $3$-user IC and present our transmission scheme under each of the assumptions  (\ref{Item:FD-Assumption})-(\ref{Item:SF-Assumption}). Subsequently, we consider the $4$-user IC to illustrate how our transmission techniques are generalized to the IC with more users. Let us introduce some notations which will be used \emph{only} in this section and \cref{Sec:X-Illustrative}: 
\begin{notation}
In the IC, we denote fresh information symbols of TX$_1$, TX$_2$, TX$_3$, and TX$_4$ (intended for their paired receivers) by $u$, $v$, $w$, and $s$ variables, respectively. Each of these symbols is selected from a Gaussian codeword which is intended to be decoded at its corresponding receiver. 
\end{notation}
\begin{notation}
\label{Notation:L}
The transmission schemes are multiphase. A linear combination of transmitted symbols which is received by RX$_1$ is denoted by $L_a(\cdot)$ if we are in phase $1$ of the scheme, and by $L'_a(\cdot)$ or $L'_{a,t}(\cdot)$ if we are in phase $2$, where $t$ is the time index. Similarly, $L_b(\cdot)$, $L_c(\cdot)$, and $L_d(\cdot)$ and their primed versions denote the linear combinations available at RX$_2$, RX$_3$, and RX$_4$, respectively. A linear combination which is available at a receiver but is \emph{not} desired by that receiver is coloured by a colour specified to that receiver. In particular, ``blue'', ``red'', ``green'', and ``yellow'' are assigned to RX$_1$ to RX$_4$, respectively.
\end{notation}

\subsection{$3$-user Interference Channel}
\label{Subsec:3-user-IC}
The schemes we propose for the $3$-user IC under the assumptions (\ref{Item:FD-Assumption})-(\ref{Item:SF-Assumption}) are motivated by the scheme proposed in \cite{Maleki_Jafar_Retro} for the $3$-user IC with output feedback, \ie assumption  (\ref{Item:OF-Assumption}). Indeed, the scheme proposed here for the $3$-user IC with output feedback is a modified version of the scheme proposed in \cite{Maleki_Jafar_Retro} and achieves the same DoF of $6/5$. The modification is such that our scheme can be systematically generalized to larger networks. For the full-duplex delayed CSIT and Shannon feedback, our transmission schemes also achieve $6/5$ DoF. Each scheme operates in $2$ distinct phases. Since phase $1$ is the same for all three schemes, we present phase $1$ only once, and then present phase $2$ under each assumption separately.
 
\phase{$1$}{$3$-user IC}

This phase takes $3$ time slots, during which $6$ information symbols $\{u_1,u_2\}$, $\{v_1,v_2\}$, and $\{w_1,w_2\}$ are fed to the system respectively by TX$_1$, TX$_2$, and TX$_3$ as follows:

$\rhd$ \emph{First time slot}: TX$_1$ and TX$_2$ transmit $u_1$ and $v_1$, respectively, while TX$_3$ is silent. Hence, ignoring the noise, RX$_1$ and RX$_2$ each receive one linear equation in terms of $u_1$ and $v_1$ by the end of the first time slot as follows:
\begin{align}
&\textup{RX}_1:\hspace{1cm}L_a(u_1,v_1)=h^{[11]}(1)u_1+h^{[12]}(1)v_1, \\
&\textup{RX}_2:\hspace{1cm}L_b(u_1,v_1)=h^{[21]}(1)u_1+h^{[22]}(1)v_1.
\end{align}
Therefore, if we deliver another linearly independent combination of $u_1$ and $v_1$ to RX$_1$, it will be able to decode both transmitted symbols (the desired symbol $u_1$ and the interference symbol $v_1$). Similarly, if we deliver a linearly independent combination of $u_1$ and $v_1$ to RX$_2$, it can decode both $u_1$ which is interference and $v_1$ which is a desired symbol. 

\begin{remark}
\label{Remark:Noise}
Since the noise variance is bounded, it does not affect the DoF. Hence, we ignore the noise in our analysis throughout this paper.
\end{remark}

Now, we observe that RX$_3$ has also received a linear combination of $u_1$ and $v_1$, \ie ignoring the noise,
\begin{align}
\textup{RX}_3:\hspace{1cm}L_c(u_1,v_1)=h^{[31]}(1)u_1+h^{[32]}(1)v_1.
\end{align}
Note first that this quantity does not contain any information about the information symbols of RX$_3$ ($w$ symbols). Therefore, it is \emph{not} desired by RX$_3$. However, since the channel coefficients are i.i.d. across the nodes, $L_c(u_1,v_1)$ is linearly independent of each of $L_a(u_1,v_1)$ and $L_b(u_1,v_1)$ almost surely. Therefore, if we somehow deliver $L_c(u_1,v_1)$ to both RX$_1$ and RX$_2$, each of them will be able to decode its own desired symbol (together with the interference symbol). Hence, $L_c(u_1,v_1)$ is a new ``symbol'' which is simultaneously desired by both RX$_1$ and RX$_2$ and is available at RX$_3$. 

Transmission in the second and third time slots is done similar to the first time slot, except that roles of the nodes are exchanged:

$\rhd$ \emph{Second time slot}: TX$_2$ and TX$_3$ transmit $v_2$ and $w_1$, respectively, while TX$_1$ is silent. After this time slot, the linear combination $L_a(v_2,w_1)$ will be desired by both RX$_2$ and RX$_3$.

$\rhd$ \emph{Third time slot}: TX$_3$ and TX$_1$ transmit $w_2$ and $u_2$, respectively, while TX$_2$ is silent. After this time slot, $L_b(u_2,w_2)$ will be desired by both RX$_3$ and RX$_1$.

\begin{figure}
\centering
\includegraphics[scale=.9]{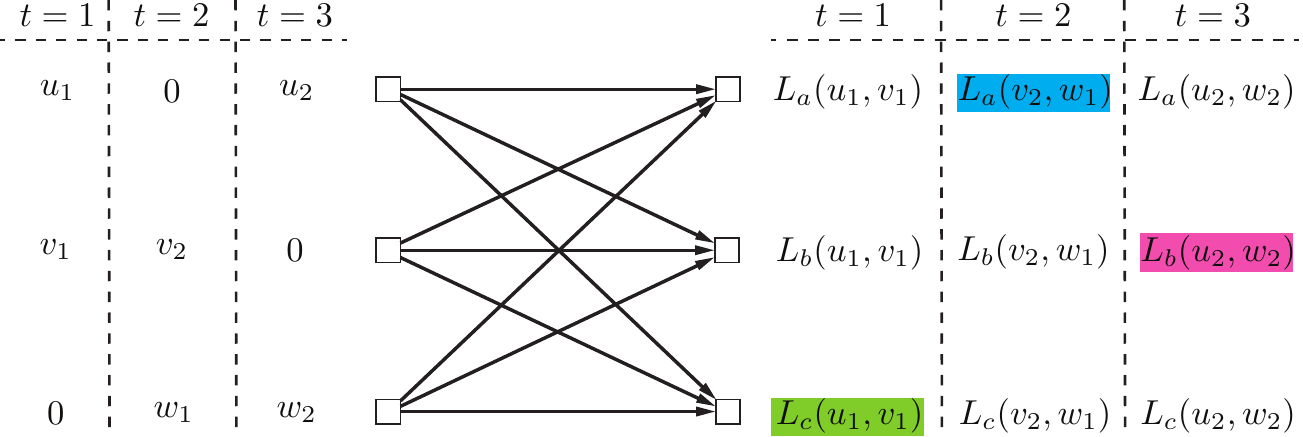}
\caption{Phase $1$ of the transmission scheme for $3$-user IC. Each coloured linear combination is the one which is (i) available at a receiver, (ii) not desired by that receiver, and (iii) desired by the other receivers.}
\label{Fig:3-IC-Scheme-Phase1}
\end{figure}

The transmission in phase $1$ is visually illustrated in \cref{Fig:3-IC-Scheme-Phase1}. Note in the figure that in each time slot, the coloured quantity denotes the quantity which is available and undesired at the corresponding receiver by the end of that time slot. It only remains to deliver these coloured symbols, \ie $L_c(u_1,v_1)$, $L_a(v_2,w_1)$, and $L_b(u_2,w_2)$ to the pairs of receivers where they are desired as discussed above. This will be accomplished in phase $2$ through cooperation between the transmitters. The type of cooperation is determined by the channel feedback/cooperation assumption, that is, the assumptions  (\ref{Item:FD-Assumption})-(\ref{Item:SF-Assumption}). However, under each assumption, phase $2$ takes $2$ time slots, and thus, the overall achieved DoF will be $6/5$. In the following, we present the phase $2$ under each assumption separately:

\phase{$2$}{Full-duplex $3$-user IC with Delayed CSIT}

Recall that in the first time slot, TX$_1$ and TX$_2$ respectively transmitted $u_1$ and $v_1$, and TX$_3$ was silent. According to full-duplex operation of the transmitters, TX$_1$ receives a noisy version of $v_1$ and TX$_2$ receives a noisy version of $u_1$ by the end of this time slot. This along with the delayed CSIT assumption enables both TX$_1$ and TX$_2$ to reconstruct a noisy version of $L_c(u_1,v_1)$, whose noise can be ignored as mentioned in \cref{Remark:Noise}. Similarly, both TX$_2$ and TX$_3$ will reconstruct $L_a(v_2,w_1)$ after the second time slot, and both TX$_3$ and TX$_1$ will reconstruct $L_b(u_2,w_2)$ after the third time slot. Therefore, this phase takes $2$ time slots as follows:

$\rhd$ \emph{Fourth time slot}: The symbols $L_c(u_1,v_1)$, $L_a(v_2,w_1)$, and $L_b(u_2,w_2)$ are transmitted by TX$_1$, TX$_2$, and TX$_3$, respectively. Then, RX$_1$ receives the following linear combination
\begin{align}
L'_{a,4}\left(L_a(v_2,w_1),L_b(u_2,w_2),L_c(u_1,v_1)\right)=h^{[11]}(4)L_c(u_1,v_1)+h^{[12]}(4)L_a(v_2,w_1)+h^{[13]}(4)L_b(u_2,w_2), \nonumber
\end{align}
and since it already has the undesired quantity $L_a(v_2,w_1)$, it can cancel it to obtain an equation solely in terms of $L_c(u_1,v_1)$ and $L_b(u_2,w_2)$. Remember that both $L_c(u_1,v_1)$ and $L_b(u_2,w_2)$ are going to be delivered to RX$_1$.

Also, RX$_2$ receives 
\begin{align}
L'_{b,4}\left(L_a(v_2,w_1),L_b(u_2,w_2),L_c(u_1,v_1)\right)=h^{[21]}(4)L_c(u_1,v_1)+h^{[22]}(4)L_a(v_2,w_1)+h^{[23]}(4)L_b(u_2,w_2), \nonumber
\end{align}
and RX$_3$ receives 
\begin{align}
L'_{c,4}\left(L_a(v_2,w_1),L_b(u_2,w_2),L_c(u_1,v_1)\right)=h^{[31]}(4)L_c(u_1,v_1)+h^{[32]}(4)L_a(v_2,w_1)+h^{[33]}(4)L_b(u_2,w_2) \nonumber
\end{align}
by the end of the fourth time slot. Similarly, RX$_2$, having the undesired quantity $L_b(u_2,w_2)$, will obtain an equation in terms of two desired quantities $L_a(v_2,w_1)$ and $L_c(u_1,v_1)$. Also, RX$_3$ will similarly obtain an equation solely in terms of $L_a(v_2,w_1)$ and $L_b(u_2,w_2)$.

$\rhd$ \emph{Fifth time slot}: This time slot is an exact repetition of the fourth time slot. Hence, since the channel coefficients are i.i.d. in time, by the end of this time slot, each receiver obtains a linearly independent equation in terms of its own two desired quantities, and thus, can decode both desired quantities. 

\begin{figure}
\centering
\subfloat[Full-duplex delayed CSIT.]{\label{Fig:3-IC-FD-Phase2}
\includegraphics[scale=.9]{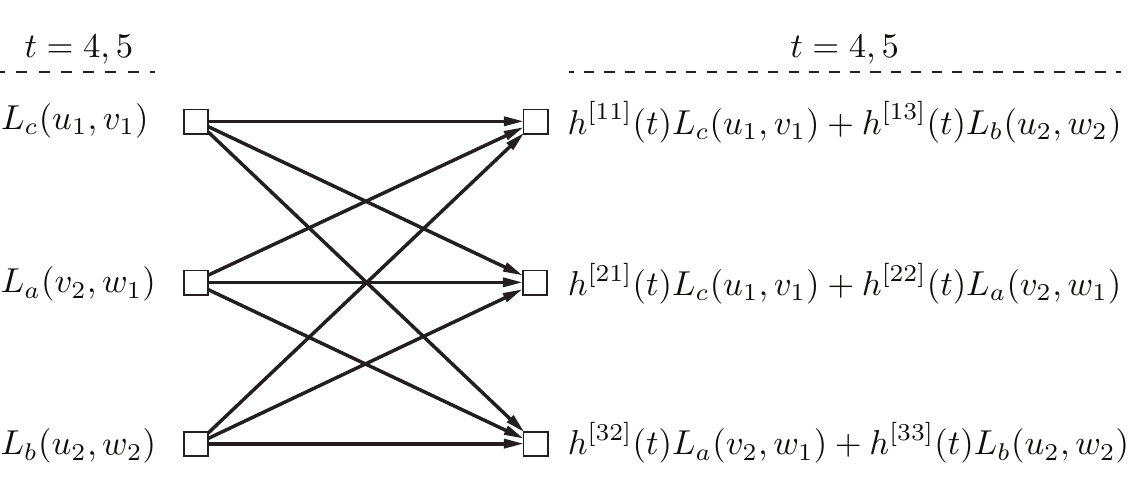}}\\
\subfloat[Output feedback.]{\label{Fig:3-IC-OF-Phase2}
\includegraphics[scale=.9]{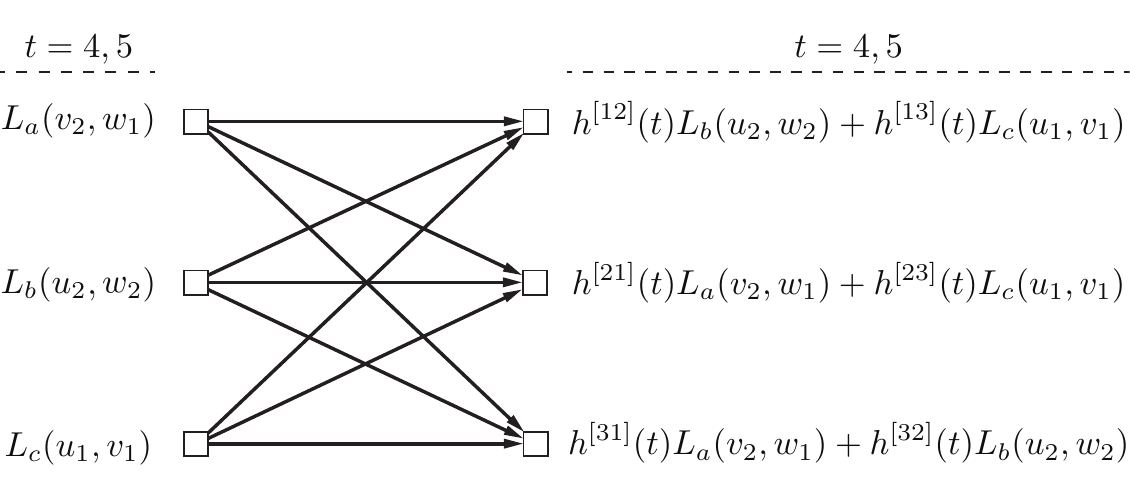}}\\
\subfloat[Shannon feedback.]{\label{Fig:3-IC-SF-Phase2}
\includegraphics[scale=.9]{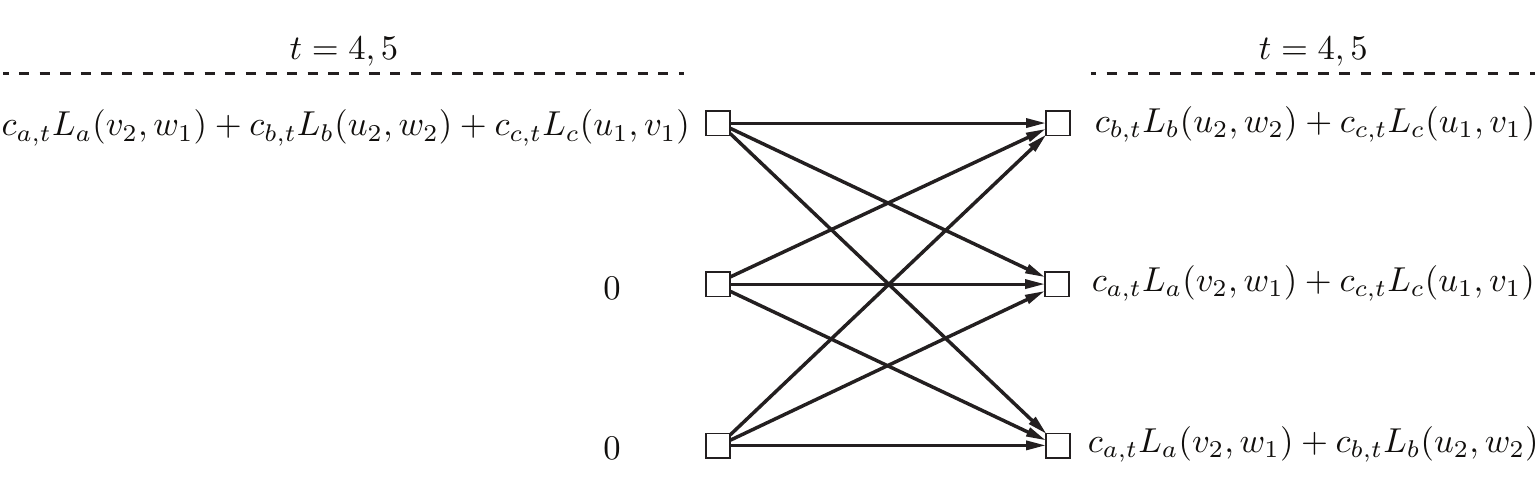}}
\caption{Phase $2$ of the transmission scheme for $3$-user IC.}
\label{Fig:3-IC-Scheme-Phase2}
\end{figure}

The above transmission scheme in phase $2$ is illustrated in \cref{Fig:3-IC-FD-Phase2}. This completes the delivery of the $6$ information symbols $\{u_1,u_2,v_1,v_2,w_1,w_2\}$ to their intended receivers in $5$ time slots, and thus, proves achievability of $6/5$ DoF with full-duplex delayed CSIT.
  
\phase{$2$}{$3$-user IC with Output Feedback}

With access to output feedback, the quantity $L_c(u_1,v_1)$ is available at TX$_3$ after the first time slot. Similarly, $L_a(v_2,w_1)$ and $L_b(u_2,w_2)$ are available at TX$_1$ and TX$_2$, respectively, after the second and third time slots. Hence, transmission of these symbols in phase $2$ can be done in two time slots using the same scheme explained above. The only difference is that here $L_a(v_2,w_1)$, $L_b(u_2,w_2)$, and  $L_c(u_1,v_1)$ are transmitted by TX$_1$, TX$_2$, and TX$_3$ respectively, as shown in \cref{Fig:3-IC-OF-Phase2}.

\phase{$2$}{$3$-user IC with Shannon Feedback}

Under the Shannon feedback assumption, we argue that $L_c(u_1,v_1)$ is available at all three transmitters after the first time slot as follows: TX$_3$ obtains $L_c(u_1,v_1)$ through the output feedback. On the other hand, TX$_1$, having access to output feedback, obtains $L_a(u_1,v_1)$ after this time slot. Then, since it also has access to delayed CSI and its own transmitted symbol $u_1$, it can cancel the effect of $u_1$ from $L_a(u_1,v_1)$ to obtain $v_1$. Therefore, it can reconstruct $L_c(u_1,v_1)$ using $u_1$, $v_1$, and the delayed CSI. Similarly, TX$_2$ can reconstruct $L_c(u_1,v_1)$. Using a similar argument, $L_a(v_2,w_1)$ and $L_b(u_2,w_2)$ will be available at all three transmitters after the second and third time slots, respectively.

Recall that under each of the assumptions of full-duplex delayed CSIT and output feedback, to deliver $L_a(v_2,w_1)$, $L_b(u_2,w_2)$, and $L_c(u_1,v_1)$ to their intended pairs of receiver in phase $2$, we delivered two random linear combinations of them to each receiver.  In those cases, each of these symbols was repeated by one of the transmitted in two time slots simultaneously. Here, we again deliver two random linear combinations of these three symbols to each receiver using another approach: two random linear combinations of $L_a(v_2,w_1)$, $L_b(u_2,w_2)$, and $L_c(u_1,v_1)$ are transmitted by one of the transmitters, say TX$_1$, in two time slots $t=4,5$, while the rest of transmitters are silent. The coefficients of these combinations are generated \emph{offline} and revealed to all receivers before the transmission begins. Hence, after two time slots, each receiver obtains two random linear combinations in terms of $L_a(v_2,w_1)$, $L_b(u_2,w_2)$, and $L_c(u_1,v_1)$, and will be able to remove its known undesired quantity and decode the other two desired quantities. Therefore, $6/5$ DoF is also achieved with Shannon feedback. The phase $2$ of the transmission scheme with Shannon feedback is depicted in \cref{Fig:3-IC-SF-Phase2}, where $\{c_{a,t}, c_{b,t}, c_{c,t}|t=1,2\}$ are the random coefficient.

\subsection{$4$-user Interference Channel}
\label{Subsec:4-user-IC}
Before proceeding with the $4$-user IC, let us summarize the common ingredients of the transmission schemes proposed for the $3$-user IC as follows:
\begin{enumerate}[(i)]
\item The transmission is accomplished in consecutive phases (two phases in case of the $3$-user IC).
\item In each time slot of phase $1$, fresh information symbols are transmitted by a subset $\xxS$ of transmitters (with $|\xxS|=2$ in case of the $3$-user IC). The set of all receivers is then partitioned into two subsets $\xxS$ and its complement $\xxS^\textup{c}$ (with $|\xxS^\textup{c}|=1$ in case of the $3$-user IC). Each receiver in $\xxS$ has a desired information symbol among the transmitted symbols, whereas the receivers in $\xxS^\textup{c}$ are not interested in decoding any transmitted symbol in this time slot.
\item \label{Item:S} Each receiver in $\xxS$ receives a piece of information (linear equation) in terms of its own desired symbol and $|\xxS|-1$ interference symbol(s). Since there are more than one unknowns in the received equation, the receiver cannot resolve the equation for its desired symbol. It requires another $|\xxS|-1$ linearly independent equations to resolve its own desired symbol (\emph{and} the interference symbols).
\item Each receiver in $\xxS^\textup{c}$ receives a piece of information (linear equation) in terms of $|\xxS|$ undesired (interference) symbols. The linear equations received by any arbitrary $|\xxS|-1$ receivers out of these $|\xxS^\textup{c}|$ receivers are, however, desired by all receivers in $\xxS$, in view of observation (\ref{Item:S}) and the fact of the channel coefficients are i.i.d. across the channel nodes.
\item Let RX$_{j^*}$ be one of these $|\xxS|-1$ receivers. The linear combination received by RX$_{j^*}$, if retransmitted, provides each receiver in $\xxS$ with a desired equation without causing any further interference at RX$_{j^*}$. In this sense, this linear combination can be considered as an ``aligned interference'' at RX$_{j^*}$ because it only occupies one dimension in the received equation space of RX$_{j^*}$. 
\item These $|\xxS|-1$ pieces of information are also available at a ``subset of transmitters'', depending on the channel feedback/cooperation assumption. These transmitters can cooperate to retransmit these $|\xxS|-1$ pieces of information in the subsequent phases of the transmission scheme (phase $2$ in case of the $3$-user IC).
\end{enumerate}

Along the direction highlighted by the above observations, we propose a $3$-phase transmission scheme for the $4$-user IC under each of the assumptions  (\ref{Item:FD-Assumption})-(\ref{Item:SF-Assumption}). As in the $3$-user case, the proposed schemes for the $4$-user IC have the same performance in terms of achievable DoF and achieve $24/19$ DoF under each assumption. We note that this is strictly greater than $45/38$ DoF which is the best known achievable DoF for the $4$-user IC with delayed CSIT \cite{Abdoli2011IC-X-Arxiv}. Since the three schemes are common in their phase $1$, we present phase $1$ only once and then present the remaining phases separately under each assumption:

\phase{$1$}{$4$-user IC}

\begin{figure}
\centering
\includegraphics[scale=.9]{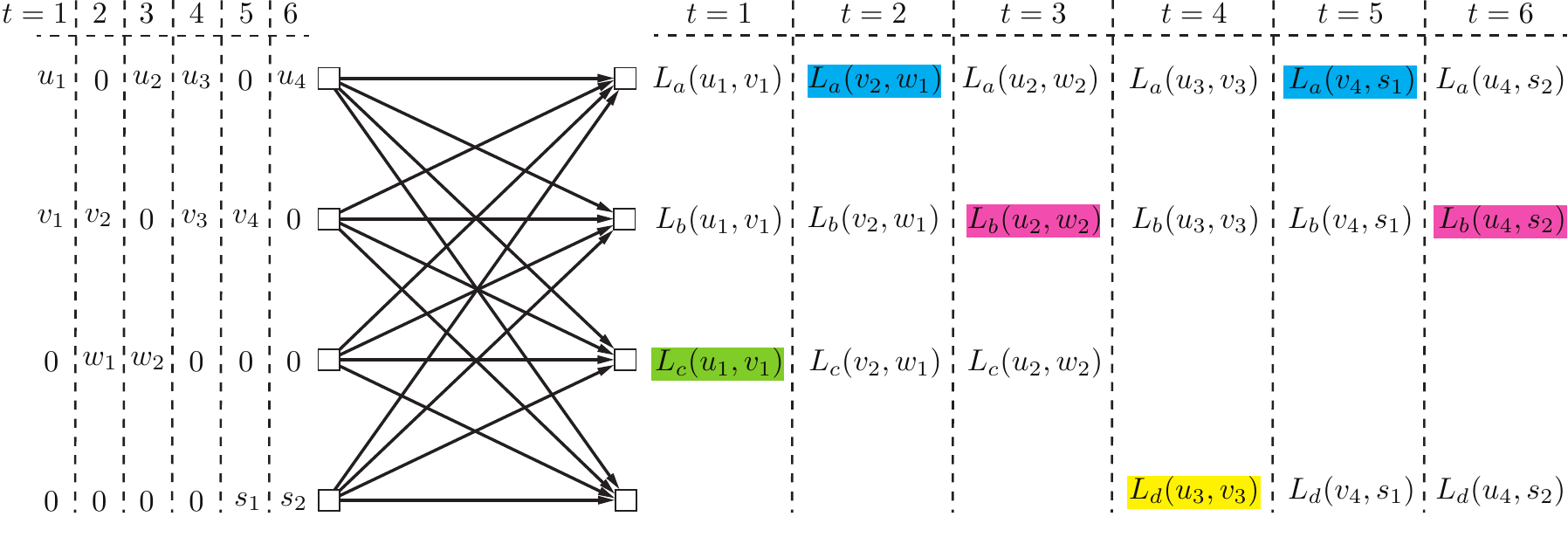}\\
\includegraphics[scale=.9]{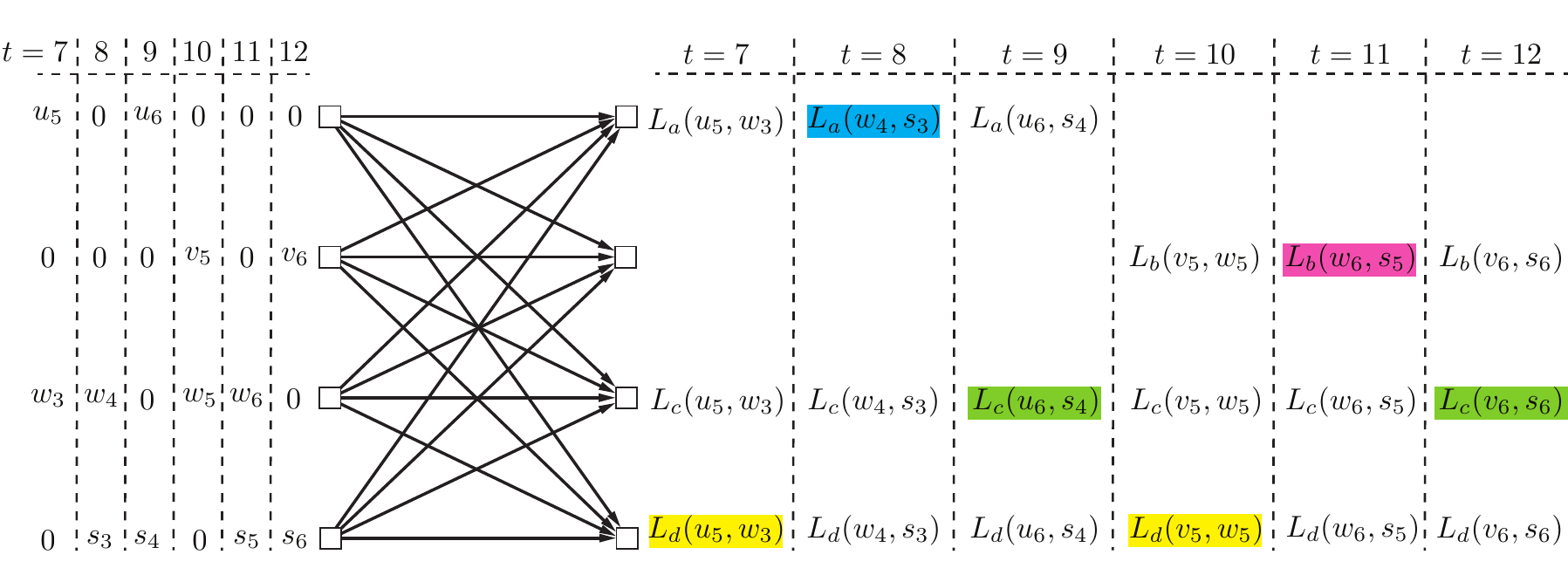}
\caption{Phase $1$ of the transmission scheme for $4$-user IC. Each coloured linear combination is the one which is (i) available at a receiver, (ii) not desired by that receiver, and (iii) desired by two of the other receivers.}
\label{Fig:4-IC-Scheme-Phase1}
\end{figure}

This phase takes $12$ time slots, during which $24$ information symbols 
\begin{align}
\{u_i,v_i,w_i,s_i|i=1,\cdots,6\}
\end{align}
are fed to the system by the transmitters in parallel with phase $1$ of the scheme for the $3$-user IC (see \cref{Subsec:3-user-IC}). \Cref{Fig:4-IC-Scheme-Phase1} illustrates the transmission in phase $1$ for the $4$-user IC.

$\rhd$ \emph{Time slots $t=1,\cdots, 3$}: TX$_1$, TX$_2$, and TX$_3$ transmit $6$ information symbols $\{u_1,u_2,v_1,v_2,w_1,w_2\}$ exactly as in the $3$-user case. TX$_4$ is silent during the first $3$ time slots. Consequently, the linear combinations $L_a(v_2,w_1)$, $L_b(u_2,w_2)$, and $L_c(u_1,v_1)$ which are respectively received by RX$_1$, RX$_2$, and RX$_3$ need to be delivered to their respective pairs of receivers during the remaining phases. The availability of these quantities at TX$_1$, TX$_2$ and TX$_2$ after the first $3$ time slots depends on the channel feedback/cooperation assumption and can be summarized as follows (see the corresponding phase $2$ in \cref{Subsec:3-user-IC} for a detailed discussion):
\begin{itemize}
\item \emph{Full-duplex delayed CSIT}: $L_a(v_2,w_1)$ is available at TX$_2$ and TX$_3$; $L_b(u_2,w_2)$ is available at TX$_1$ and TX$_3$; and $L_c(u_1,v_1)$ is available at TX$_1$ and TX$_2$.
\item \emph{Output feedback}: $L_a(v_2,w_1)$, $L_b(u_2,w_2)$, and $L_c(u_1,v_1)$ are available at TX$_1$, TX$_2$, and TX$_3$, respectively.
\item \emph{Shannon feedback}: $L_a(v_2,w_1)$, $L_b(u_2,w_2)$, and $L_c(u_1,v_1)$ are available at all three transmitters TX$_1$, TX$_2$, and TX$_3$.
\end{itemize}

The transmission in the remaining time slots of this phase is similarly proceeded by different subsets of $3$ out of the $4$ transmitters:

$\rhd$ \emph{Time slots $t=4,\cdots, 6$}: TX$_1$, TX$_2$, and TX$_4$ transmit fresh information symbols $\{u_3,u_4,v_3,v_4,s_1,s_2\}$, while TX$_3$ is silent. Similarly, $L_a(v_4,s_1)$, $L_b(u_4,s_2)$, and $L_d(u_3,v_3)$ which are respectively received by RX$_1$, RX$_2$, and RX$_4$ need to be delivered to their respective pairs of receivers during the remaining phases. These quantities are similarly available at subsets of $\{\textup{TX}_1,\textup{TX}_2,\textup{TX}_4\}$ based on the channel feedback/cooperation assumption.

$\rhd$ \emph{Time slots $t=7,\cdots, 9$}: TX$_1$, TX$_3$, and TX$_4$ transmit fresh information symbols $\{u_5,u_6,w_3,w_4,s_3,s_4\}$, while TX$_2$ is silent. Similarly, $L_a(w_4,s_3)$, $L_c(u_6,s_4)$, and $L_d(u_5,w_3)$ which are respectively received by RX$_1$, RX$_3$, and RX$_4$ need to be delivered to their respective pairs of receivers during the remaining phases. These quantities are similarly available at subsets of $\{\textup{TX}_1,\textup{TX}_3,\textup{TX}_4\}$ based on the channel feedback/cooperation assumption.

$\rhd$ \emph{Time slots $t=10,\cdots, 12$}: TX$_2$, TX$_3$, and TX$_4$ transmit fresh information symbols $\{v_5,v_6,w_5,w_6,s_5,s_6\}$, while TX$_1$ is silent. Similarly, $L_b(w_6,s_5)$, $L_c(v_6,s_6)$, and $L_d(v_5,w_5)$ which are respectively received by RX$_2$, RX$_3$, and RX$_4$ need to be delivered to their respective pairs of receivers during the remaining phases. These quantities are similarly available at subsets of $\{\textup{TX}_2,\textup{TX}_3,\textup{TX}_4\}$ based on the channel feedback/cooperation assumption.

Now, let us proceed with the remaining phases under each of the channel feedback/cooperation assumptions (\ref{Item:FD-Assumption})-(\ref{Item:SF-Assumption}):

\subsubsection{Full-duplex $4$-user IC with Delayed CSIT}

\phase{$2$}{Full-duplex $4$-user IC with  Delayed CSIT}

This phase takes $4$ time slots. In each time slot, $3$ transmitters simultaneously transmit three symbols generated during phase $1$ as follows:

$\rhd$ \emph{Time slot $t=13$}: TX$_1$, TX$_2$, and TX$_3$ respectively transmit $L_c(u_1,v_1)$, $L_a(v_2,w_1)$, and $L_b(u_2,w_2)$, while TX$_4$ is silent. RX$_1$ has $L_a(v_2,w_1)$ and wishes to decode $L_b(u_2,w_2)$ and $L_c(u_1,v_1)$. Hence, RX$_1$ can obtain a linear combination solely in terms of $L_b(u_2,w_2)$ and $L_c(u_1,v_1)$ by cancelling $L_a(v_2,w_1)$ from its received equation. Similarly, RX$_2$ and RX$_3$ each obtain a linear combination in terms of their desired pair of quantities. Thus, each of RX$_1$, RX$_2$, and RX$_3$ requires another linearly independent equation to resolve its both desired quantities. 

Now, consider the following linear combination received by RX$_4$ over this time slot:
\begin{align}
L'_d\left(L_a(v_2,w_1),L_b(u_2,w_2),L_c(u_1,v_1)\right)=h^{[41]}(13)L_c(u_1,v_1)+h^{[42]}(13)L_a(v_2,w_1)+h^{[43]}(13)L_b(u_2,w_2). \nonumber
\end{align}
If we somehow deliver the above linear combination to RX$_1$, it can obtain $h^{[41]}(13)L_c(u_1,v_1)+h^{[43]}(13)L_b(u_2,w_2)$ by cancelling $L_a(v_2,w_1)$. Since the channel coefficients are i.i.d. across the channel nodes, this linear combination is linearly independent of the equation RX$_1$ has received during this time slot, and hence, it will enable RX$_1$ to resolve its both desired quantities. Likewise, if we deliver $L'_d\left(L_a(v_2,w_1),L_b(u_2,w_2),L_c(u_1,v_1)\right)$ to RX$_2$ and RX$_3$, each of them will be able to decode its both desired quantities. Thus, it is desired by RX$_1$, RX$_2$, and RX$_3$, and will be delivered to them in phase $3$.

We now argue that this linear combination will be available at TX$_1$, TX$_2$ and TX$_3$ after this time slot. We indeed show that $L_c(u_1,v_1)$, $L_a(v_2,w_1)$, and $L_b(u_2,w_2)$ will be available at these three transmitters, which together with the delayed CSIT assumption yields the desired result. But this immediately follows from the fact that each of TX$_1$, TX$_2$, and TX$_3$ has two out of these three quantities, and thus, by the full-duplex operation, receives the third one during this time slot. 

The transmission in the remaining $3$ time slots of phase $2$ is similarly done by other subsets of $3$ out of the $4$ transmitters:

\begin{figure}
\centering
\includegraphics[scale=.9]{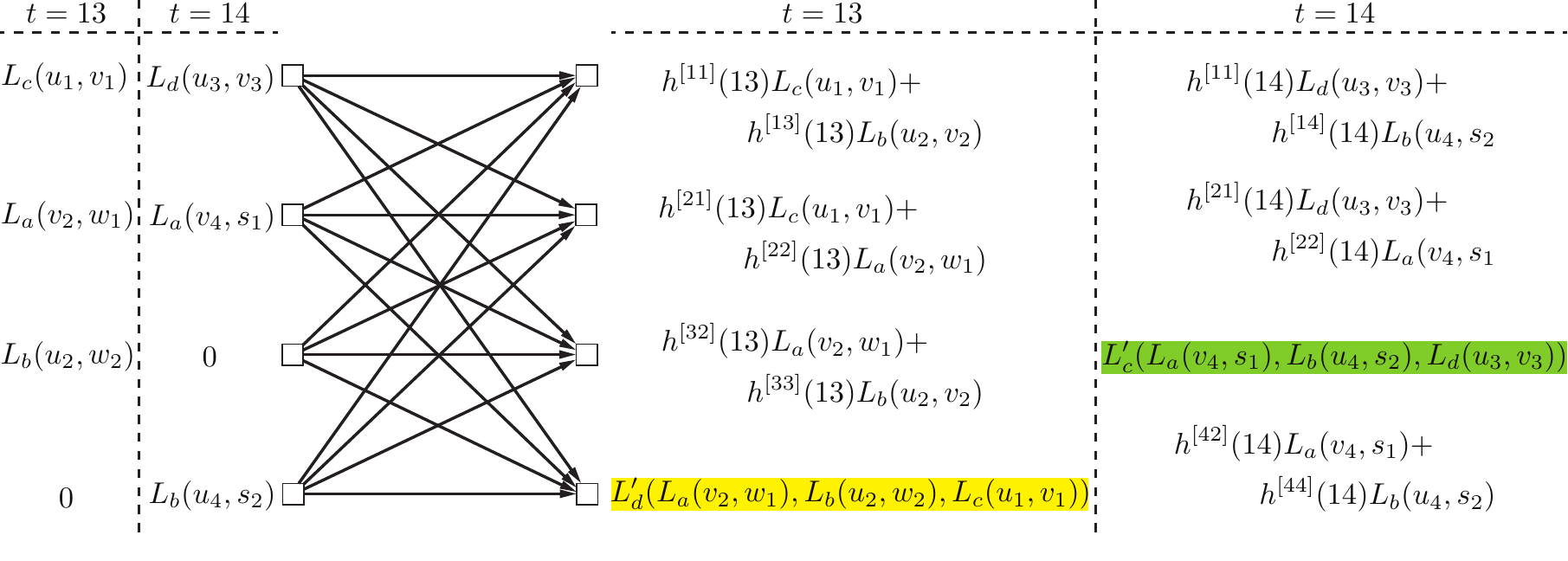}\\
\includegraphics[scale=.9]{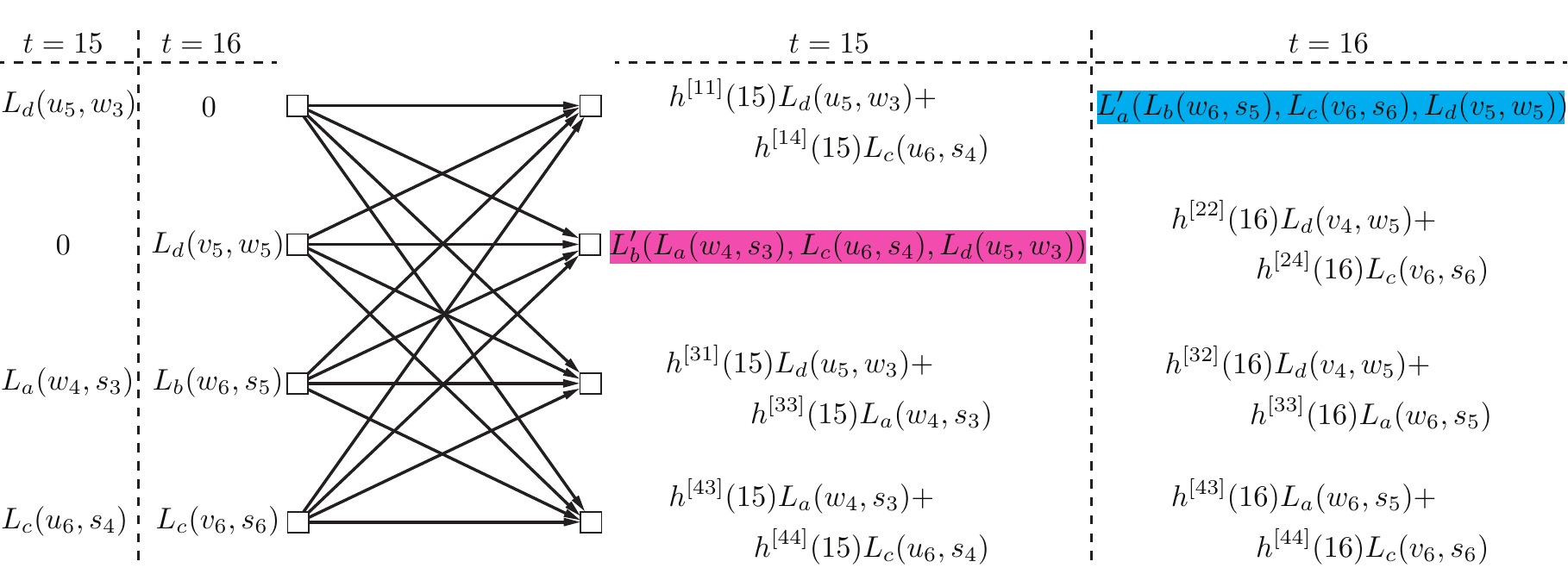}
\caption{Phase $2$ of the transmission scheme for full-duplex $4$-user IC with delayed CSIT. Each coloured linear combination is the one which is (i) available at a receiver, (ii) not desired by that receiver, and (iii) desired by the other receivers.}
\label{Fig:4-IC-FD-Scheme-Phase2}
\end{figure}

$\rhd$ \emph{Time slot $t=14$}: TX$_1$, TX$_2$, and TX$_4$ respectively transmit $L_d(u_3,v_3)$, $L_a(v_4,s_1)$, and $L_b(u_4,s_2)$ and the following linear combination which is received by RX$_3$ will be desired by RX$_1$, RX$_2$, and RX$_4$ and available at TX$_1$, TX$_2$, and TX$_4$:
\begin{align}
L'_c\left(L_a(v_4,s_1),L_b(u_4,s_2),L_d(u_3,v_3)\right)=h^{[31]}(14)L_d(u_3,v_3)+h^{[32]}(14)L_a(v_4,s_1)+h^{[34]}(14)L_b(u_4,s_2). \nonumber
\end{align}

$\rhd$ \emph{Time slot $t=15$}: TX$_1$, TX$_3$, and TX$_4$ respectively transmit $L_d(u_5,w_3)$, $L_a(w_4,s_3)$, and $L_c(u_6,s_4)$ and the following linear combination which is received by RX$_2$ will be desired by RX$_1$, RX$_3$, and RX$_4$ and available at TX$_1$, TX$_3$, and TX$_4$:
\begin{align}
L'_b\left(L_a(w_4,s_3),L_c(u_6,s_4),L_d(u_5,w_3)\right)=h^{[21]}(15)L_d(u_5,w_3)+h^{[23]}(15)L_a(w_4,s_3)+h^{[24]}(15)L_c(u_6,s_4). \nonumber
\end{align}

$\rhd$ \emph{Time slot $t=16$}: TX$_2$, TX$_3$, and TX$_4$ respectively transmit $L_d(v_5,w_5)$, $L_b(w_6,s_5)$, and $L_c(v_6,s_6)$ and the following linear combination which is received by RX$_1$ will be desired by RX$_2$, RX$_3$, and RX$_4$ and available at TX$_2$, TX$_3$, and TX$_4$:
\begin{align}
L'_a\left(L_b(w_6,s_5),L_c(v_6,s_6),L_d(v_5,w_5)\right)=h^{[12]}(16)L_d(v_5,w_5)+h^{[13]}(16)L_b(w_6,s_5)+h^{[14]}(16)L_c(v_6,s_6). \nonumber
\end{align}

\Cref{Fig:4-IC-FD-Scheme-Phase2} illustrates the transmission in phase $2$ for the $4$-user IC with full-duplex delayed CSIT. The symbols $L'_a$, $L'_b$, $L'_c$, and $L'_d$ will be delivered to their respective triples of receivers in phase $3$.

\phase{$3$}{Full-duplex $4$-user IC with Delayed CSIT}

This phase takes $3$ time slots. In each time slot, $L'_d$, $L'_c$, $L'_b$, and $L'_a$ are transmitted by TX$_1$, TX$_2$, TX$_3$, and TX$_4$, respectively. Each receiver has one of these quantities and requires the other three. By the end of this phase, each receiver will obtain three random linear combinations of its three desired quantities, and thus, will decode its desired quantities. 

\subsubsection{$4$-user IC with Output Feedback}

\phase{$2$}{$4$-user IC with Output Feedback}

The above scheme for the phase $2$ under full-duplex delayed CSIT assumption can be used under output feedback assumption as well. The only difference is that in each of the $4$ time slots, the three corresponding symbols are transmitted using a different permutation of the same transmitters as follows:

$\rhd$ \emph{Time slot $t=13$}: TX$_1$, TX$_2$, and TX$_3$ respectively transmit $L_a(v_2,w_1)$, $L_b(u_2,w_2)$, and $L_c(u_1,v_1)$, while TX$_4$ is silent. The linear combination $L'_d\left(L_a(v_2,w_1),L_b(u_2,w_2),L_c(u_1,v_1)\right)$ will then be desired by RX$_1$, RX$_2$, and RX$_3$ and will be available at TX$_4$ after this time slot via output feedback.

$\rhd$ \emph{Time slot $t=14$}: TX$_1$, TX$_2$, and TX$_4$ respectively transmit $L_a(v_4,s_1)$, $L_b(u_4,s_2)$, and $L_d(u_3,v_3)$, while TX$_3$ is silent. The linear combination $L'_c\left(L_a(v_4,s_1),L_b(u_4,s_2),L_d(u_3,v_3)\right)$ will be desired by RX$_1$, RX$_2$, and RX$_4$ and available at TX$_3$.

$\rhd$ \emph{Time slot $t=15$}: TX$_1$, TX$_3$, and TX$_4$ respectively transmit $L_a(w_4,s_3)$, $L_c(u_6,s_4)$, and $L_d(u_5,w_3)$, while TX$_2$ is silent. The linear combination $L'_b\left(L_a(w_4,s_3),L_c(u_6,s_4),L_d(u_5,w_3)\right)$ will be desired by RX$_1$, RX$_3$, and RX$_4$ and available at TX$_2$.

$\rhd$ \emph{Time slot $t=16$}: TX$_2$, TX$_3$, and TX$_4$ respectively transmit $L_b(w_6,s_5)$, $L_c(v_6,s_6)$, and $L_d(v_5,w_5)$, while TX$_1$ is silent. The linear combination $L'_a\left(L_b(w_6,s_5),L_c(v_6,s_6),L_d(v_5,w_5)\right)$ will be desired by RX$_2$, RX$_3$, and RX$_4$ and available at TX$_1$.

The symbols $L'_a$, $L'_b$, $L'_c$, and $L'_d$ will be delivered to their respective triples of receivers in phase $3$.

\phase{$3$}{$4$-user IC with output feedback}

This phase takes $3$ time slots. In each time slot, the symbols (linear combinations) $L'_a$, $L'_b$, $L'_c$, and $L'_d$ are transmitted by TX$_1$, TX$_2$, TX$_3$, and TX$_4$, respectively. Similar to the full-duplex delayed CSIT, by the end of this phase, each receiver will decode its desired symbols.

\subsubsection{$4$-user IC with Shannon Feedback}

\phase{$2$}{$4$-user IC with Shannon Feedback} 

This phase takes $4$ time slots as follows:

$\rhd$ \emph{Time slot $t=13$}: Recall from phase $1$ that $L_a(v_2,w_1)$, $L_b(u_2,w_2)$, and $L_c(u_1,v_1)$ are all available at each of TX$_1$, TX$_2$, and TX$_3$. We also note that if we deliver two random linear combinations of these three quantities to RX$_1$, RX$_2$, and RX$_3$, then each of them will be able to decode its two desired quantities out of these three quantities. Hence, two random linear combinations of them with offline generated coefficients are simultaneously transmitted by two transmitters out of TX$_1$, TX$_2$, and TX$_3$ (say, TX$_1$ and TX$_2$). Then, each of RX$_1$, RX$_2$, and RX$_3$ receives one linear equation in terms of $L_a(v_2,w_1)$, $L_b(u_2,w_2)$, and $L_c(u_1,v_1)$ and requires another random linear combination to resolve both desired quantities. Thus, the linear combination $L'_d\left(L_a(v_2,w_1),L_b(u_2,w_2),L_c(u_1,v_1)\right)$ which is received by RX$_4$ is desired by each of the other three receivers. Also, this linear combination is available at TX$_4$ through the output feedback and is available at the other transmitters, since they all have $L_a(v_2,w_1)$, $L_b(u_2,w_2)$, and $L_c(u_1,v_1)$.

The remaining $3$ time slots are similarly dedicated to transmission of other linear combinations as follows:

$\rhd$ \emph{Time slot $t=14$}: Two random linear combinations of $L_a(v_4,s_1)$, $L_b(u_4,s_2)$, and $L_d(u_3,v_3)$ are transmitted by TX$_4$ and TX$_1$. The linear combination $L'_c\left(L_a(v_4,s_1),L_b(u_4,s_2),L_d(u_3,v_3)\right)$ which is received by RX$_3$ is desired by each of the other three receivers, and is available at all $4$ transmitters.

$\rhd$ \emph{Time slot $t=15$}: Two random linear combinations of $L_a(w_4,s_3)$, $L_c(u_6,s_4)$, and $L_d(u_5,w_3)$ are transmitted by TX$_3$ and TX$_4$. The linear combination $L'_b\left(L_a(w_4,s_3),L_c(u_6,s_4),L_d(u_5,w_3)\right)$ which is received by RX$_2$ is desired by each of the other three receivers, and is available at all $4$ transmitters.

$\rhd$ \emph{Time slot $t=16$}: Two random linear combinations of $L_b(w_6,s_5)$, $L_c(v_6,s_6)$, and $L_d(v_5,w_5)$ are transmitted by TX$_2$ and TX$_3$. The linear combination $L'_a\left(L_b(w_6,s_5),L_c(v_6,s_6),L_d(v_5,w_5)\right)$ which is received by RX$_1$ is desired by each of the other three receivers, and is available at all $4$ transmitters.

\begin{figure}
\centering
\includegraphics[scale=.9]{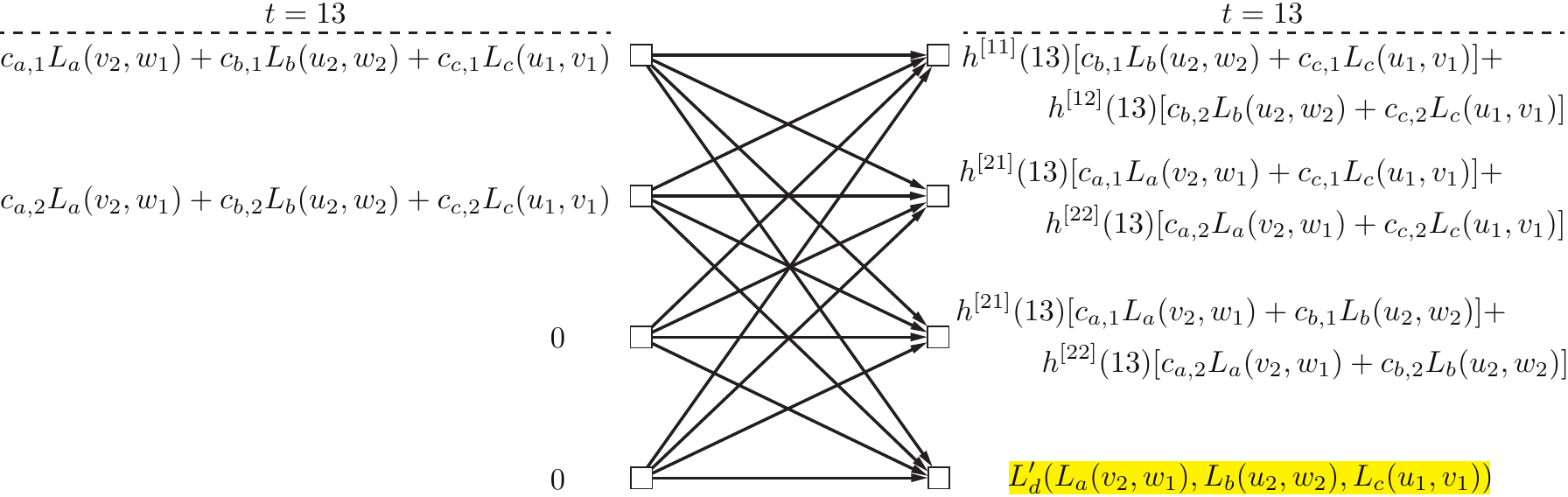}\\
\includegraphics[scale=.9]{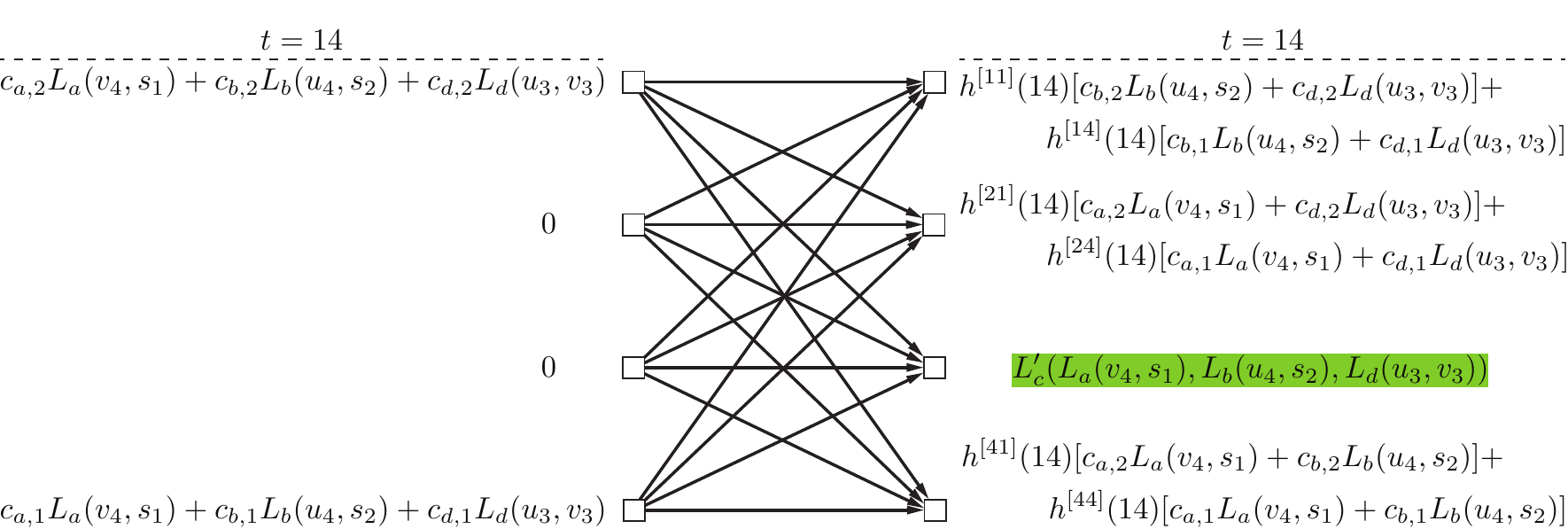}\\
\includegraphics[scale=.9]{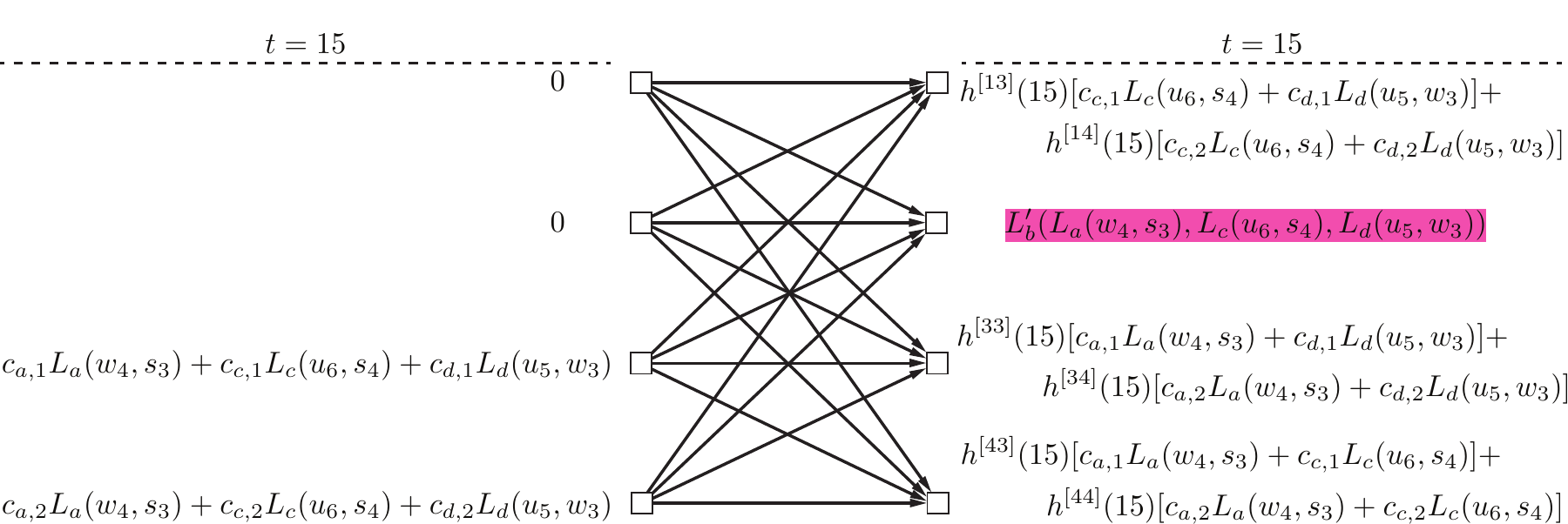}\\
\includegraphics[scale=.9]{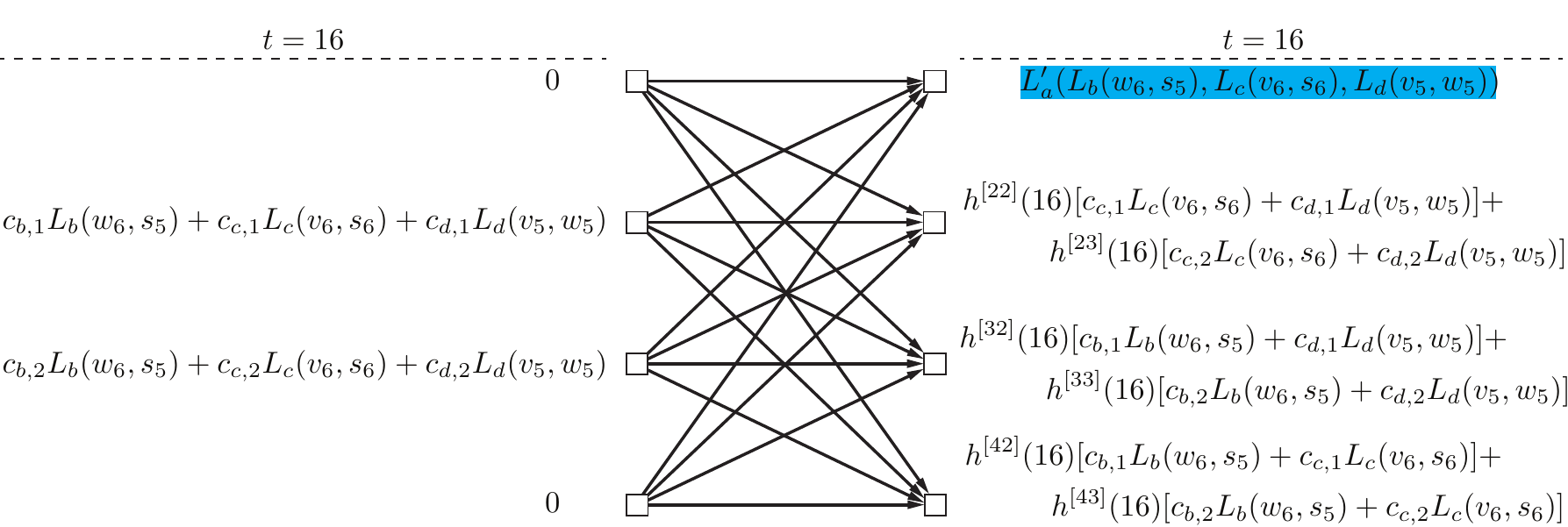}
\caption{Phase $2$ of the transmission scheme for $4$-user IC with Shannon feedback. Each coloured linear combination is the one which is (i) available at a receiver, (ii) not desired by that receiver, and (iii) desired by the other receivers.}
\label{Fig:4-IC-SF-Scheme-Phase2}
\end{figure}

The transmission in phase $2$ for the $4$-user IC with Shannon feedback is illustrated in \cref{Fig:4-IC-SF-Scheme-Phase2}.

\phase{$3$}{$4$-user IC with Shannon feedback}

Over $3$ time slots, one of the transmitters, say TX$_1$, transmits $3$ random linear combinations of $L'_a$, $L'_b$, $L'_c$, and $L'_d$, and thus, each receiver will decode its desired symbols by the end of this phase.

Our achievable DoF for the $4$-user IC under each of the feedback/cooperation assumptions will then be $24/(12+4+3)=24/19$.

\begin{remark}
Although the proposed transmission schemes for the $3$-user and $4$-user IC achieve the same DoF under each of the channel feedback/cooperation assumptions, this is not generally the case as it will be seen later. Indeed, for $K>6$, the proposed transmission schemes achieve strictly different DoFs under different feedback/cooperation assumptions.
\end{remark}

\section{Illustrative Examples: X Channel}
\label{Sec:X-Illustrative}
In this section, we illustrate our transmission schemes for the $2\times 2$ and $3\times 3$ X channel under each of the channel feedback/cooperation assumptions (\ref{Item:FD-Assumption})-(\ref{Item:SF-Assumption}). Before proceeding with the details of the transmission schemes, let us introduce a notation which is exclusively used in this section: 
\begin{notation}
The symbols $u^a$, $u^b$, and $u^c$ denote information symbols of TX$_1$, TX$_2$, and TX$_3$, respectively, all intended for RX$_1$. Similarly, $v^a$, $v^b$, and $v^c$ denote information symbols intended for RX$_2$, and $w^a$, $w^b$, and $w^c$ are all intended for RX$_3$.
\end{notation}
We also use the same notations for the linear combinations and their colouring as defined in \cref{Notation:L}.

\subsection{$2\times 2$ X Channel}
\label{Subsec:2-by-2-XC}
\begin{figure}
\centering
\includegraphics[scale=.9]{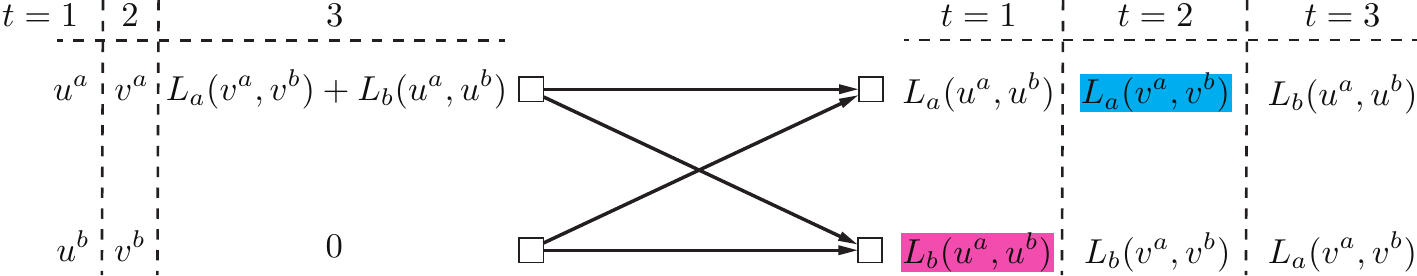}
\caption{The transmission scheme for full-duplex $2\times 2$ X channel with delayed CSIT. Each coloured linear combination is the one which is (i) available at a receiver, (ii) not desired by that receiver, and (iii) desired by the other receiver.}
\label{Fig:2-X-FD-Scheme}
\end{figure}

It is already known that $2\times 2$ X channel can achieve $4/3$ DoF with output feedback \cite{Maleki_Jafar_Retro}. This is indeed the DoF of $2$-user MISO broadcast channel with Shannon feedback \cite{maddah2010DoF_BCC_Delayed_Arxiv}, which is also an upper bound to the DoF of $2\times 2$ X channel under each of the assumptions (\ref{Item:FD-Assumption})-(\ref{Item:SF-Assumption}). Hence, the DoF of $2\times 2$ X channel with output feedback or with Shannon feedback is equal to $4/3$. In this section, we show that $2\times 2$ X channel has the same DoF under the full-duplex delayed CSIT assumption as well. The transmission scheme operates in parallel with scheme proposed in \cite{maddah2010DoF_BCC_Delayed_Arxiv} for the $2$-user MISO broadcast channel and employed in \cite{Maleki_Jafar_Retro} for the $2\times 2$ X channel with output feedback. It is a two-phase transmission scheme depicted in \cref{Fig:2-X-FD-Scheme}, wherein the fresh information symbols are transmitted over the channel in the first phase and delivery of the symbols to their intended receivers is completed in the second phase. In particular, $4$ information symbols are delivered in $3$ time slots as follows:

\phase{$1$}{Full-duplex $2\times 2$ X Channel with Delayed CSIT}

This phase takes $2$ time slots to transmit $4$ information symbols as follows:

$\rhd$ \emph{First time slot}: The symbols $u^a$ and $u^b$ are transmitted by TX$_1$ and TX$_2$, respectively. Ignoring the noise, RX$_1$ will receive a linear equation 
\begin{align}
L_a(u^a,u^b)=h^{[11]}(1)u^a+h^{[12]}(1)u^b,
\end{align}
in terms of $2$ desired information symbols, and hence, requires another linearly independent equation to resolve them. Simultaneously, RX$_2$ receives another linear equation, namely, 
\begin{align}
L_b(u^a,u^b)=h^{[21]}(1)u^a+h^{[22]}(1)u^b,
\end{align}
in terms of $u^a$ and $u^b$. Since the channel coefficients are i.i.d. across the channel nodes, $L_b(u^a,u^b)$ is linearly independent of $L_a(u^a,u^b)$ almost surely. Therefore, if we deliver $L_b(u^a,u^b)$ to RX$_1$ it will be able to decode both $u^a$ and $u^b$. On the other hand, according to full-duplex operation of the transmitters, both TX$_1$ and TX$_2$ will have both $u^a$ and $u^b$, and by the delayed CSIT assumption, they can reconstruct $L_b(u^a,u^b)$ after this time slot.

$\rhd$ \emph{Second time slot}: Similarly, $v^a$ and $v^b$ are transmitted respectively by TX$_1$ and TX$_2$. Then, the linear combination 
\begin{align}
L_a(v^a,v^b)=h^{[11]}(2)v^a+h^{[12]}(2)v^b,
\end{align}
which is received by RX$_1$ will be desired by RX$_2$ and available at both TX$_1$ and TX$_2$.

Therefore, it only remains to deliver $L_b(u^a,u^b)$ and $L_a(v^a,v^b)$ to RX$_1$ and RX$_2$ respectively. This is accomplished in one time slot in phase $2$:

\phase{$2$}{Full-duplex $2\times 2$ X Channel with Delayed CSIT}

$\rhd$ \emph{Third time slot}: One of the transmitters, say TX$_1$, transmits $L_b(u^a,u^b)+L_a(v^a,v^b)$, while the other transmitter is silent. RX$_1$ receives this linear combination, and it can cancel $L_a(v^a,v^b)$ which it already has, to obtain the desired quantity $L_b(u^a,u^b)$. Similarly, RX$_2$ can cancel $L_b(u^a,u^b)$ to obtain $L_a(v^a,v^b)$.

\subsection{$3\times 3$ X Channel}
\label{Subsec:3-by-3-XC}
For this channel, we achieve $24/17$ DoF with full-duplex delayed CSIT. We also achieve $3/2$ DoF and $27/17$ DoF with output feedback and Shannon feedback, respectively. In the following, we show the achievability of each of the above DoFs:

\subsubsection{Full-duplex $3\times 3$ X Channel with Delayed CSIT}
We propose a $3$-phase transmission scheme which delivers $72$ information symbols in $51$ time slots, and thus, achieves $24/17$ DoF as follows:

\phase{$1$}{Full-duplex $3\times 3$ X Channel with Delayed CSIT}

This phase takes $12$ times slots to transmit $24$ information symbols. 

$\rhd$ \emph{Time slots $t=1,\cdots,6$}: Only TX$_1$ and TX$_2$ transmit information symbols, and TX$_3$ is silent. In particular, for each pair of receivers, TX$_1$ and TX$_2$ use $2$ time slots to transmit $4$ information symbols exactly as in phase $1$ of the scheme proposed above for the full-duplex $2\times 2$ X channel with delayed CSIT. 

$\rhd$ \emph{Time slots $t=7,\cdots,12$}: Similarly, another $12$ information symbols are now transmitted by TX$_1$ and TX$_3$, while TX$_2$ is silent.

\begin{figure}
\centering
\includegraphics[scale=.9]{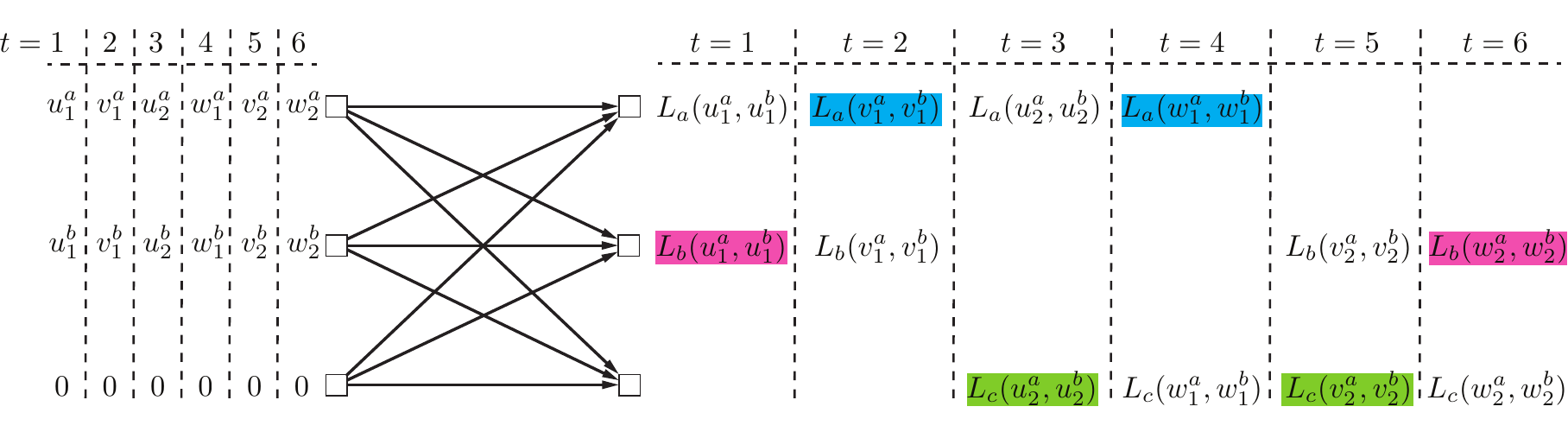}\\
\includegraphics[scale=.9]{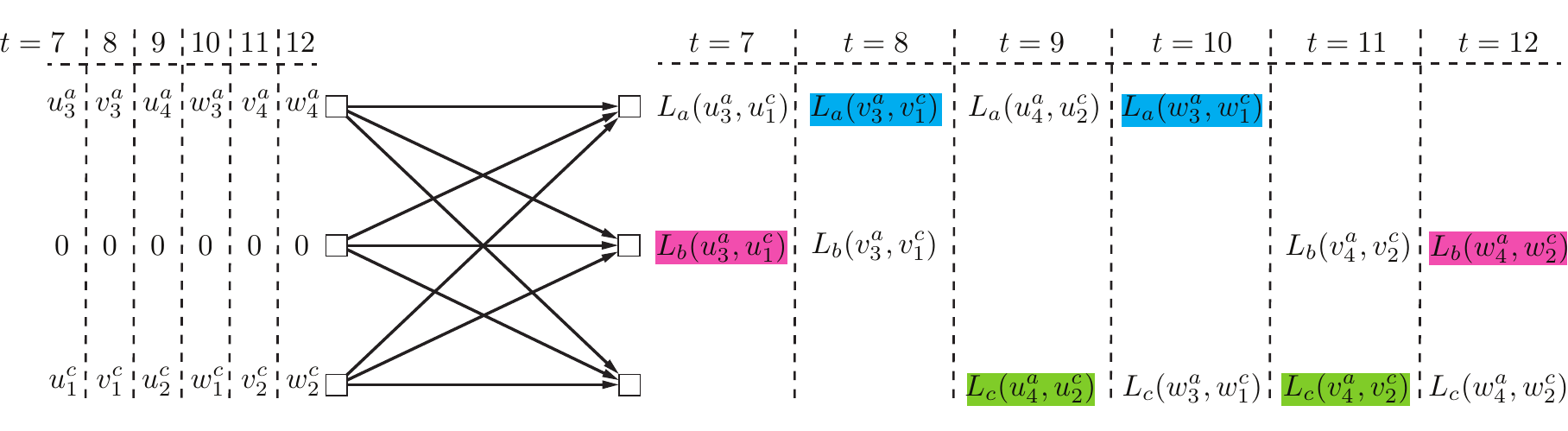}
\caption{Phase $1$ of the transmission scheme for full-duplex $3\times 3$ X channel with delayed CSIT. Each coloured linear combination is the one which is (i) available at a receiver, (ii) not desired by that receiver, and (iii) desired by one of the other receivers.}
\label{Fig:3-X-FD-Scheme-Phase1}
\end{figure}

The transmission in this phase is illustrated in \cref{Fig:3-X-FD-Scheme-Phase1}. Each coloured linear combination in the figure is available at one receiver and desired by another receiver, and will also be reconstructed by two of the transmitters after its corresponding time slot. For example,  $L_b(u^a_1,u^b_1)$ is available at RX$_2$ and desired by RX$_1$, and will be reconstructed by TX$_1$ and TX$_2$ after the first time slot. Now, it only remains to deliver the following $6$ linear combinations to their respective pairs of receivers (as discussed in phase $2$ of the full-duplex $2\times 2$ X channel with delayed CSIT):
\begin{align}
&\textup{TX}_1 \& \textup{TX}_2
\begin{cases}
L_b(u^a_1,u^b_1)+L_a(v^a_1,v^b_1) & \longrightarrow \textup{RX}_1 \& \textup{RX}_2\\
L_c(u^a_2,u^b_2)+L_a(w^a_1,w^b_1) & \longrightarrow \textup{RX}_1 \& \textup{RX}_3\\
L_c(v^a_2,v^b_2)+L_b(w^a_2,w^b_2) & \longrightarrow \textup{RX}_2 \& \textup{RX}_3
\end{cases}, \label{Eq:TX1-TX2_FD_Phase2}\\
&\textup{TX}_1 \& \textup{TX}_3
\begin{cases}
L_b(u^a_3,u^c_1)+L_a(v^a_3,v^c_1) & \longrightarrow \textup{RX}_1 \& \textup{RX}_2\\
L_c(u^a_4,u^c_2)+L_a(w^a_3,w^c_1) & \longrightarrow \textup{RX}_1 \& \textup{RX}_3\\
L_c(v^a_4,v^c_2)+L_b(w^a_4,w^c_2) & \longrightarrow \textup{RX}_2 \& \textup{RX}_3
\end{cases}. \label{Eq:TX1-TX3_FD_Phase2}
\end{align}

This will be accomplished during the remaining phases of the transmission scheme.

\phase{$2$}{Full-duplex $3\times 3$ X Channel with Delayed CSIT}

This phase takes $3$ time slots to transmit the linear combinations indicated in \cref{Eq:TX1-TX2_FD_Phase2,Eq:TX1-TX3_FD_Phase2} by TX$_1$ and TX$_2$ as follows. TX$_3$ is silent in this phase.

$\rhd$ \emph{Time slot $t=13$}: TX$_1$ and TX$_2$ transmit $L_b(u^a_3,u^c_1)+L_a(v^a_3,v^c_1)$ and $L_b(u^a_1,u^b_1)+L_a(v^a_1,v^b_1)$, respectively, while TX$_3$ is silent. By the end of this time slot, RX$_1$ obtains a linear combination in terms of the (desired) $L_b$ quantities (after cancelling the known $L_a$ quantities). Hence, it requires another linearly independent combination of the $L_b$ quantities to decode both of them. Similarly, RX$_2$ obtains a linear combination of the (desired) $L_a$ quantities and needs another linearly independent combination of them to decode both. Now, one can easily verify that the linear combination 
\begin{align}
L'_c = h^{[31]}(13)[L_b(u^a_3,u^c_1)+L_a(v^a_3,v^c_1)]+h^{[32]}(13)[L_b(u^a_1,u^b_1)+L_a(v^a_1,v^b_1)],
\end{align}
received by RX$_3$ during this time slot, is linearly independent of the linear combination received by each of RX$_1$ and RX$_2$. Therefore, if we deliver this linear combination to both RX$_1$ and RX$_2$, each of them will be able to decode its both desired $L_b$ or $L_a$ quantities. On the other hand, by the delayed CSIT assumption, $L'_c$ is available at TX$_1$ as well (note that TX$_1$ has both transmitted linear combinations).

The next two time slots are similarly dedicated to the other pairs of receivers:

$\rhd$ \emph{Time slot $t=14$}: TX$_1$ and TX$_2$ transmit $L_c(u^a_4,u^c_2)+L_a(w^a_3,w^c_1)$ and $L_c(u^a_2,u^b_2)+L_a(w^a_1,w^b_1)$, respectively. Now, each of RX$_1$ and RX$_3$ receives a desired linear combination and the linear combination
\begin{align}
L'_b= h^{[21]}(14)[L_c(u^a_4,u^c_2)+L_a(w^a_3,w^c_1)]+h^{[22]}(14)[L_c(u^a_2,u^b_2)+L_a(w^a_1,w^b_1)],
\end{align}
received by RX$_2$ during this time slot, will be desired by both RX$_1$ and RX$_3$. This linear combination is also available at TX$_1$ after this time slot.

$\rhd$ \emph{Time slot $t=15$}: TX$_1$ and TX$_2$ transmit $L_c(v^a_4,v^c_2)+L_b(w^a_4,w^c_2)$ and $L_c(v^a_2,v^b_2)+L_b(w^a_2,w^b_2)$, respectively. Each of RX$_2$ and RX$_3$ receives a desired linear combination and the linear combination
\begin{align}
L'_a=h^{[11]}(15)[L_c(v^a_4,v^c_2)+L_b(w^a_4,w^c_2)]+h^{[12]}(15)[L_c(v^a_2,v^b_2)+L_b(w^a_2,w^b_2)],
\end{align}
received by RX$_1$ during this time slot, will be desired by both RX$_2$ and RX$_3$. This linear combination is also available at TX$_1$ after this time slot.

In summary, the linear combinations $L'_a$, $L'_b$, and $L'_c$ each are available at one receiver and desired by the other two receivers, and all of them are available at TX$_1$. They will be delivered to their respective pairs of receivers in phase $3$.

\phase{$3$}{Full-duplex $3\times 3$ X Channel with Delayed CSIT}

$\rhd$ \emph{Time slots $t=16,17$}: In each time slot, a random linear combination of $L'_a$, $L'_b$, and $L'_c$ is transmitted by TX$_1$, while the rest of transmitters are silent. It can be easily verified that after these two time slots, each receiver will be able to decode its both desired quantities.

\subsubsection{$3\times 3$ X Channel with Output Feedback}
Our transmission scheme for this channel is a $2$-phase scheme wherein $9$ information symbols are delivered to the receivers in $6$ time slots, yielding $3/2$ DoF, as illustrated in \cref{Fig:3-X-OF-Scheme} and elaborated on in the following:

\begin{figure}
\centering
\subfloat[Phase $1$]{\includegraphics[scale=.9]{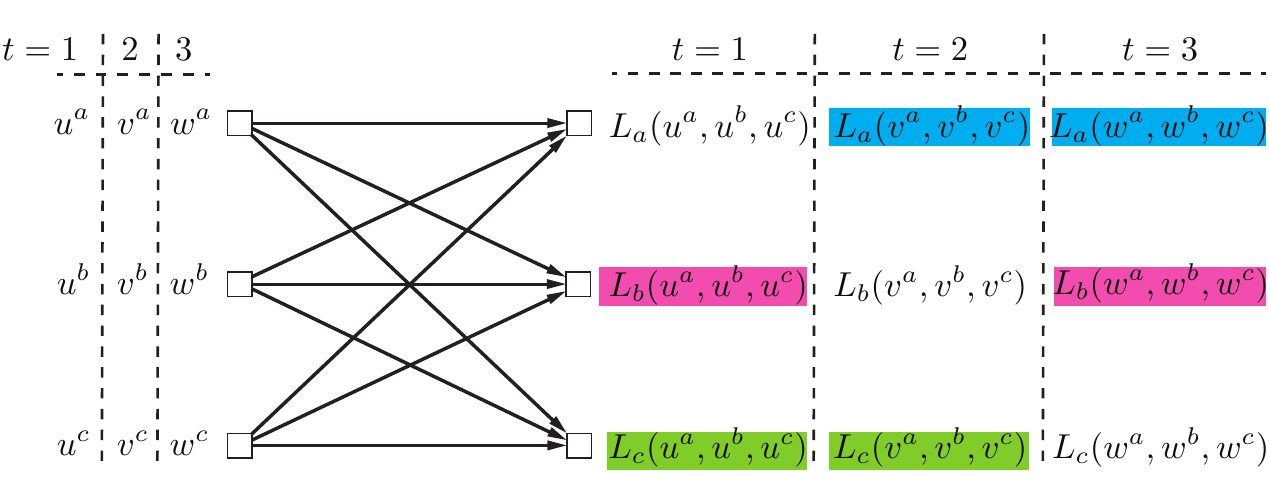}}\\
\subfloat[Phase $2$]{\includegraphics[scale=.9]{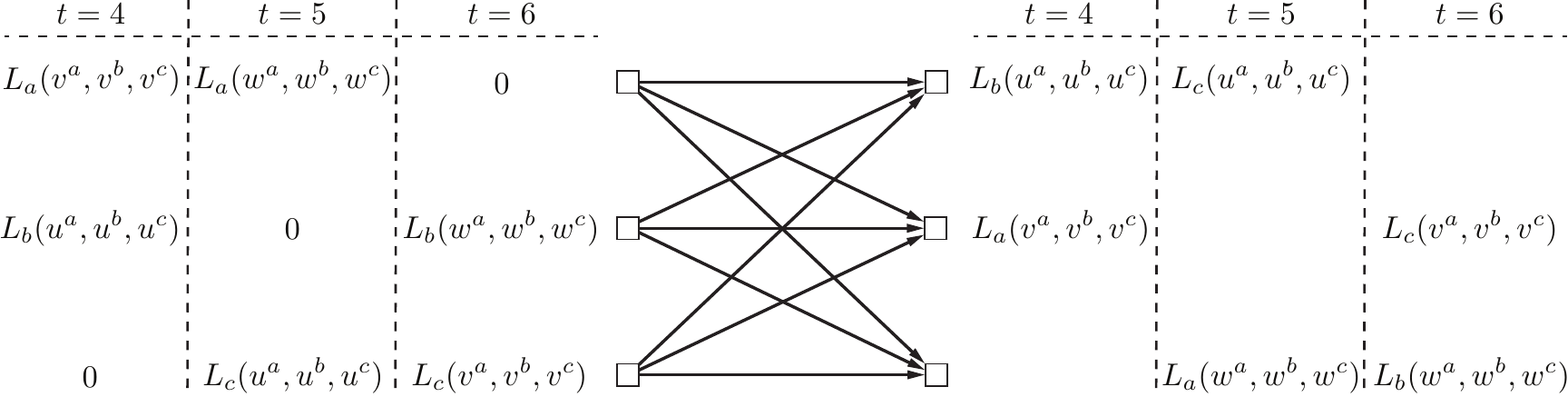}}
\caption{Transmission scheme for $3\times 3$ X channel with output feedback. Each coloured linear combination is the one which is (i) available at a receiver, (ii) not desired by that receiver, and (iii) desired by one of the other receivers.}
\label{Fig:3-X-OF-Scheme}
\end{figure}

\phase{$1$}{$3\times 3$ X Channel with Output Feedback} 
This phase has $3$ time slots. Each time slot is dedicated to transmission of information symbols intended for one of the receivers:

$\rhd$ \emph{First time slot}: The information symbols $u^a$, $u^b$, and $u^c$, all intended for RX$_1$, are transmitted by TX$_1$, TX$_2$ and TX$_3$, respectively. By the end of this time slot, RX$_1$ receives linear combination $L_a(u^a,u^b,u^c)$ of the three desired symbols and requires two extra linearly independent equations to resolve all three symbols. RX$_2$ receives the linear combination $L_b(u^a,u^b,u^c)$ which is linearly independent of $L_a(u^a,u^b,u^c)$, and thus, is desired by RX$_1$. Similarly, the linear combination $L_c(u^a,u^b,u^c)$ received by RX$_3$ is desired by RX$_1$. On the other hand, $L_b(u^a,u^b,u^c)$ (resp.\ $L_c(u^a,u^b,u^c)$) will be also available at TX$_2$ (resp.\ TX$_3$) through the output feedback.

The second and third time slots are similarly dedicated to RX$_2$ and RX$_3$, respectively:

$\rhd$ \emph{Second time slot}: The information symbols $v^a$, $v^b$, and $v^c$, all intended for RX$_2$, are transmitted by TX$_1$, TX$_2$ and TX$_3$, respectively. Similarly, $L_a(v^a,v^b,v^c)$ and $L_c(v^a,v^b,v^c)$, received by RX$_1$ and RX$_3$ and available at TX$_1$ and TX$_3$ through the output feedback, will be desired by RX$_1$ after this time slot.

$\rhd$ \emph{Third time slot}: The information symbols $w^a$, $w^b$, and $w^c$, all intended for RX$_3$, are transmitted by TX$_1$, TX$_2$ and TX$_3$, respectively. Similarly, $L_a(w^a,w^b,w^c)$ and $L_b(w^a,w^b,w^c)$, received by RX$_1$ and RX$_2$ and available at TX$_1$ and TX$_2$ through the output feedback, will be desired by RX$_3$ after this time slot.

Therefore, to deliver the transmitted information symbols to their intended receivers, it suffices to 
\begin{itemize}
\item[(i)] deliver $L_b(u^a,u^b,u^c)$ and $L_c(u^a,u^b,u^c)$ to RX$_1$;
\item[(ii)] deliver $L_a(v^a,v^b,v^c)$ and $L_c(v^a,v^b,v^c)$ to RX$_2$;
\item[(iii)] deliver $L_a(w^a,w^b,w^c)$ and $L_b(w^a,w^b,w^c)$ to RX$_3$.
\end{itemize}

This will be done in phase $2$.

\phase{$2$}{$3\times 3$ X Channel with Output Feedback} 

This phase takes $3$ time slots. Each time slot is dedicated to a pair of receivers as follows:

$\rhd$ \emph{Fourth time slot}: Over this time slot, which is dedicated to RX$_1$ and RX$_2$, $L_a(v^a,v^b,v^c)$ and $L_b(u^a,u^b,u^c)$ are respectively transmitted by TX$_1$ and TX$_2$, while TX$_3$ is silent. After this time slot, RX$_1$ obtains the desired linear combination $L_b$ by cancelling the known undesired linear combination $L_a$. Similarly, RX$_2$ obtains its own desired linear combination $L_a$ by cancelling $L_b$.

$\rhd$ \emph{Fifth time slot}: The quantities $L_a(w^a,w^b,w^c)$ and $L_c(u^a,u^b,u^c)$ are transmitted by TX$_1$ and TX$_3$, while TX$_2$ is silent. Then, each of RX$_1$ and RX$_3$ similarly obtains its desired quantity.

$\rhd$ \emph{Sixth time slot}: The quantities $L_b(w^a,w^b,w^c)$ and $L_c(v^a,v^b,v^c)$ are transmitted by TX$_2$ and TX$_3$, while TX$_1$ is silent. Then, each of RX$_2$ and RX$_3$ similarly obtains its desired quantity.

\subsubsection{$3\times 3$ X Channel with Shannon Feedback}
Our transmission scheme for this channel has two rounds of operation, during which $27$ information symbols are delivered to the receivers in $17$ time slots as follows:

\round{$1$}{$3\times 3$ X Channel with Shannon Feedback}

The first round consists of two phases. Phase $1$ takes $3$ time slots to transmit $9$ information symbols $\{u^a_1,u^b_1,u^c_1,v^a_1,v^b_1,v^c_1,w^a_1,w^b_1,w^c_1\}$ exactly as in phase $1$ of the scheme proposed above for the same channel with output feedback. Before proceeding with phase $2$, one notes that TX$_1$ after the first time slot will obtain the linear combination
\begin{align}
L_a(u^a_1,u^b_1,u^c_1)=h^{[11]}(1)u^a_1+h^{[12]}(1)u^b_1+h^{[13]}(1)u^c_1,
\end{align}
through the output feedback. Since TX$_1$ has access to delayed CSI as well (Shannon feedback assumption), it can cancel its own transmitted symbols $u^a_1$ to obtain 
\begin{align}
\label{Eq:u^bu^c_1}
h^{[12]}(1)u^b_1+h^{[13]}(1)u^c_1,
\end{align}
which is a linear combination of $u^b_1$ and $u^c_1$. TX$_1$ knows the coefficients $h^{[12]}(1)$ and $h^{[13]}(1)$ of this linear combination. Similarly, TX$_2$ will obtain $h^{[21]}(2)v^a_1+h^{[23]}(2)v^c_1$ after the second time slot using Shannon feedback.

In phase $2$, over one time slot, TX$_1$ and TX$_2$ transmit $L_a(v^a_1,v^b_1,v^c_1)$ and $L_b(u^a_1,u^b_1,u^c_1)$, while TX$_3$ is silent. Hence, $L_a(v^a_1,v^b_1,v^c_1)$ and $L_b(u^a_1,u^b_1,u^c_1)$ are delivered to RX$_2$ and RX$_1$, respectively (as in the phase $2$ of the scheme proposed with output feedback). Now, TX$_1$ will obtain $L_b(u^a_1,u^b_1,u^c_1)$ since it has access to Shannon feedback and its own transmitted quantity \ie $L_a(v^a_1,v^b_1,v^c_1)$. Therefore, by cancelling $u^a_1$ from $L_b(u^a_1,u^b_1,u^c_1)$, TX$_1$ will obtain
\begin{align}
\label{Eq:u^bu^c_2}
h^{[22]}(1)u^b_1+h^{[23]}(1)u^c_1,
\end{align}
which is another linear combination of $u^b_1$ and $u^c_1$. Hence, using \cref{Eq:u^bu^c_1,Eq:u^bu^c_2}, TX$_1$ will be able to decode both $u^b_1$ and $u^c_1$. Thereby, having access to delayed CSI, TX$_1$ can reconstruct $L_c(u^a_1,u^b_1,u^c_1)$. Likewise, TX$_2$ will be able to decode both $v^a_1$ and $v^c_1$, and hence, can reconstruct $L_c(v^a_1,v^b_1,v^c_1)$.

In summary, after these $4$ time slots, it only remains to 
\begin{itemize}
\item[(i)] deliver $L_c(u^a_1,u^b_1,u^c_1)$ to RX$_1$;
\item[(ii)] deliver $L_c(v^a_1,v^b_1,v^c_1)$ to RX$_2$;
\item[(iii)] deliver $L_a(w^a_1,w^b_1,w^c_1)$ and $L_b(w^a_1,w^b_1,w^c_1)$ to RX$_3$.
\end{itemize}
On the other hand, TX$_1$ has access to $L_c(u^a_1,u^b_1,u^c_1)$ by above argument and has access to $L_a(w^a_1,w^b_1,w^c_1)$ using output feedback. Similarly, TX$_2$ has access to $L_c(v^a_1,v^b_1,v^c_1)$ and $L_b(w^a_1,w^b_1,w^c_1)$. Hence, it suffices to deliver the following two linear combinations to their respective pairs of receivers:
\begin{align}
&\textup{TX}_1: L_c(u^a_1,u^b_1,u^c_1)+L_a(w^a_1,w^b_1,w^c_1) &\longrightarrow \hspace{2mm} \textup{RX}_1 \& \textup{RX}_3, \label{Eq:Shannon_R13_1}\\
&\textup{TX}_2: L_c(v^a_1,v^b_1,v^c_1)+L_b(w^a_1,w^b_1,w^c_1) &\longrightarrow \hspace{2mm} \textup{RX}_2 \& \textup{RX}_3. \label{Eq:Shannon_R23_1}
\end{align}

Before proceeding with the second round, we repeat the above procedure two more times and transmit another $2\times 9 =18$ \emph{fresh} information symbols, namely $\{u^a_i,u^b_i,u^c_i,v^a_i,v^b_i,v^c_i,w^a_i,w^b_i,w^c_i\}_{i=2,3}$, in another $2\times 4=8$ time slots. However, in the first repetition, $L_a(w^a_2,w^b_2,w^c_2)$ and $L_c(u^a_2,u^b_2,u^c_2)$ are transmitted by TX$_1$ and TX$_3$ in phase $2$, and it will suffice to deliver the following two linear combinations to their respective pairs of receivers:
\begin{align}
&\textup{TX}_1: L_b(u^a_2,u^b_2,u^c_2)+L_a(v^a_2,v^b_2,v^c_2) &\longrightarrow \hspace{2mm}\textup{RX}_1 \& \textup{RX}_2, \label{Eq:Shannon_R12_1}\\
&\textup{TX}_3: L_b(w^a_2,w^b_2,w^c_2)+L_c(v^a_2,v^b_2,v^c_2) &\longrightarrow \hspace{2mm}\textup{RX}_2 \& \textup{RX}_3. \label{Eq:Shannon_R23_2}
\end{align}
Similarly, in the second repetition, $L_b(w^a_3,w^b_3,w^c_3)$ and $L_c(v^a_3,v^b_3,v^c_3)$ are transmitted by TX$_2$ and TX$_3$ in phase $2$, and it will suffice to deliver the following two linear combinations to their respective pairs of receivers:
\begin{align}
&\textup{TX}_2: L_a(v^a_3,v^b_3,v^c_3)+L_b(u^a_3,u^b_3,u^c_3) &\longrightarrow \hspace{2mm}\textup{RX}_1 \& \textup{RX}_2, \label{Eq:Shannon_R12_2} \\
&\textup{TX}_3: L_a(w^a_3,w^b_3,w^c_3)+L_c(u^a_3,u^b_3,u^c_3) &\longrightarrow \hspace{2mm}\textup{RX}_1 \& \textup{RX}_3. \label{Eq:Shannon_R13_2}
\end{align}

Up to this point, we have spent $12$ time slots, transmitted $27$ information symbols. Now, we need to to deliver the above $6$ linear combinations to their respective pairs of receivers. This will be done in the second round. 

\round{$2$}{$3\times 3$ X Channel with Shannon Feedback}

This round takes $5$ time slots, \ie $t=13,\cdots,17$. During the first $3$ time slots the above $6$ linear combinations are transmitted over the channel. Each time slot is dedicated to a pair of receivers as follows:

$\rhd$ \emph{Time slot $t=13$}: TX$_1$ and TX$_2$ respectively transmit $L_b(u^a_2,u^b_2,u^c_2)+L_a(v^a_2,v^b_2,v^c_2)$ and $L_a(v^a_3,v^b_3,v^c_3)+L_b(u^a_3,u^b_3,u^c_3)$ (both to be delivered to RX$_1$ and RX$_2$ according to \cref{Eq:Shannon_R12_1,Eq:Shannon_R12_2}), while TX$_3$ is silent. Then, using an argument similar to the phase $2$ of the transmission scheme proposed for the full-duplex $3\times 3$ X channel with delayed CSIT, RX$_1$ (resp.\ RX$_2$) receives an equation in terms of the (desired) linear combinations $L_b(u^a_2,u^b_2,u^c_2)$ and $L_b(u^a_3,u^b_3,u^c_3)$ (resp.\ $L_a(v^a_2,v^b_2,v^c_2)$ and $L_a(v^a_3,v^b_3,v^c_3)$). Also, the equation 
\begin{align}
L'_c=h^{[31]}(13)[L_b(u^a_2,u^b_2,u^c_2)+L_a(v^a_2,v^b_2,v^c_2)]+h^{[32]}(13)[L_a(v^a_3,v^b_3,v^c_3)+L_b(u^a_3,u^b_3,u^c_3)],
\end{align}
received by RX$_3$ in this time slot will be desired by both RX$_1$ and RX$_2$. It can also be easily verified that $L'_c$ can be reconstructed by TX$_1$ due to Shannon feedback.

$\rhd$ \emph{Time slot $t=14$}: TX$_1$ and TX$_3$ respectively transmit $L_c(u^a_1,u^b_1,u^c_1)+L_a(w^a_1,w^b_1,w^c_1)$ and $L_a(w^a_3,w^b_3,w^c_3)+L_c(u^a_3,u^b_3,u^c_3)$, both desired by RX$_1$ and RX$_3$, while TX$_2$ is silent. Then, the linear combination
\begin{align}
L'_b=h^{[21]}(14)[L_c(u^a_1,u^b_1,u^c_1)+L_a(w^a_1,w^b_1,w^c_1)]+h^{[23]}(14)[L_a(w^a_3,w^b_3,w^c_3)+L_c(u^a_3,u^b_3,u^c_3)],
\end{align}
received by RX$_2$ will be desired by both RX$_1$ and RX$_3$ and can be reconstructed by TX$_1$ using Shannon feedback.

$\rhd$ \emph{Time slot $t=15$}: TX$_2$ and TX$_3$ respectively transmit $L_b(w^a_2,w^b_2,w^c_2)+L_c(v^a_2,v^b_2,v^c_2)$ and $L_b(w^a_2,w^b_2,w^c_2)+L_c(v^a_2,v^b_2,v^c_2)$, both desired by RX$_2$ and RX$_3$, while TX$_1$ is silent. Then, the linear combination
\begin{align}
L'_a=h^{[12]}(15)[L_b(w^a_2,w^b_2,w^c_2)+L_c(v^a_2,v^b_2,v^c_2)]+h^{[13]}(15)[L_b(w^a_2,w^b_2,w^c_2)+L_c(v^a_2,v^b_2,v^c_2) ],
\end{align}
received by RX$_1$ will be desired by both RX$_2$ and RX$_3$ and is received by TX$_1$ using Shannon feedback (output feedback).

During the last $2$ time slots of this round, the linear combinations $L'_a$, and $L'_b$, and $L'_c$ are delivered to their intended pairs of receivers:

$\rhd$ \emph{Time slots $t=16,17$}: Two random linear combinations of $L'_a$, and $L'_b$, and $L'_c$ are transmitted by TX$_1$, while the rest of transmitters are silent. Each receiver will then be able to decode its two desired linear combinations. 

The achieved DoF is therefore equal to $27/(12+3+2)=27/17$.

\section{Main Results}
\label{Sec:MainResults}
The main results of this paper are summarized in the following six theorems. The proof of each theorem is provided in its respective section.

\subsection{Full-duplex Transmitter Cooperation and Delayed CSIT}
\label{Subsec:ICX-DCSIT-FD}
\begin{theorem}
\label{Th:ICFD}
The $K$-user ($K\geq 3$) SISO Gaussian interference channel with delayed CSIT and full-duplex transmitters can achieve $\DoFa_1^\textup{ICFD}(K)$ degrees of freedom almost surely, where $\DoFa_1^\textup{ICFD}(K)$ is given by
\begin{align}
\label{Eq:DoF_ICFD}
\DoFa_1^\textup{ICFD}(K) &=\frac{4}{3-\frac{2}{\lceil \frac{K}{2}\rceil(\lceil \frac{K}{2}\rceil-1)}+\frac{4}{\lfloor \frac{K}{2}\rfloor(\lceil \frac{K}{2}\rceil-1)}\sum_{\ell=\lceil \frac{K}{2}\rceil+1}^K\frac{1}{\ell}}.
\end{align}
\end{theorem}
\begin{IEEEproof}
See \cref{Sec:IC-DCSIT-FD}.
\end{IEEEproof}

\begin{theorem}
\label{Th:XFD}
The $M\times K$ SISO Gaussian X channel with delayed CSIT and full-duplex transmitters can achieve $\DoFa_1^\textup{XFD}(M,K)$ degrees of freedom almost surely, where $\DoFa_1^\textup{XFD}(M,K)$ is given by
\begin{multline}
\label{Eq:DoF_XFD}
\DoFa_1^\textup{XFD}(M,K)\!=\!
\begin{cases}
\left(\frac{1}{\lceil \frac{K}{2}\rceil}-1+\!\sum_{\ell_1=1}^{\lceil \frac{K}{2}\rceil-1}\frac{1}{\ell_1^2}+\frac{1}{\lceil \frac{K}{2}\rceil(\lfloor \frac{K}{2}\rfloor+1)}\sum_{\ell_2=\lceil \frac{K}{2}\rceil}^K\frac{1}{\ell_2}\right)^{-1}\!\!, & M> \lceil\frac{K}{2}\rceil\\
\left( \frac{1}{M-1}-1+\!\sum_{\ell_1=1}^{M-2}\frac{1}{\ell_1^2}+\frac{1}{M^2}\sum_{\ell_2=M-1}^{K}\frac{1}{\ell_2}(\frac{M-1}{M})^{\min(\ell_2,K-M+1)-M}\right)^{-1}\!\!, & M \leq \lceil\frac{K}{2}\rceil
\end{cases}.
\end{multline}
\end{theorem}
\begin{IEEEproof}
See \cref{Sec:X-DCSIT-FD}.
\end{IEEEproof}

\subsection{Output Feedback}
\label{Subsec:ICX-OF}
\begin{theorem}
\label{Th:ICOF}
The $K$-user ($K\geq 3$) SISO Gaussian interference channel with output feedback can achieve $\DoFa_1^\textup{ICOF}(K)$ degrees of freedom almost surely, where $\DoFa_1^\textup{ICOF}(K)$ is given by
\begin{align}
\DoFa_1^\textup{ICOF}(K)=\max_{w\in \{\lfloor w^*_K\rfloor, \lceil w^*_K\rceil \}}\frac{w}{a(K)w(w-1)^2+(w+1)/2}, \label{Eq:DoF_ICOF}
\end{align}
with $w^*_K$ and $a(K)$ defined as
\begin{align}
w^*_K&\Def \frac{1}{3}+\frac{1}{6}\left(\frac{8a(K)+3\sqrt{48a(K)+81}+27}{a(K)}\right)^{\frac{1}{3}}+\frac{1}{6}\left(\frac{8a(K)-3\sqrt{48a(K)+81}+27}{a(K)}\right)^{\frac{1}{3}}, \label{Eq:w*(K)_def} \\
a(K)&\Def\frac{1}{\lceil \frac{K}{2}\rceil-1}\left(-\frac{1}{2\lceil \frac{K}{2}\rceil}+\frac{1}{\lfloor \frac{K}{2}\rfloor}\sum_{\ell=\lceil \frac{K}{2}\rceil+1}^K\frac{1}{\ell}\right). \label{Eq:a(K)_def}
\end{align}
\end{theorem}
\begin{IEEEproof}
See \cref{Sec:IC-OF}.
\end{IEEEproof}

\begin{theorem}
\label{Th:XOF}
The $K\times K$ SISO Gaussian X channel with output feedback can achieve $\DoFa_1^\textup{XOF}(K,K)=\frac{2K}{K+1}$ degrees of freedom almost surely\symbolfootnote[2]{The result of this theorem has been simultaneously and independently reported in \cite{tandon2012x}}.
\end{theorem}
\begin{IEEEproof}
See \cref{Sec:X-OF}.
\end{IEEEproof}

\subsection{Shannon Feedback}
\label{Subsec:ICX-SF}
\begin{theorem}
\label{Th:ICSF}
The $K$-user ($K\geq 3$) SISO Gaussian interference channel with Shannon feedback can achieve $\DoFa_1^\textup{ICSF}(K)$ degrees of freedom almost surely, where $\DoFa_1^\textup{ICSF}(K)$ is given by
\begin{align}
\label{Eq:DoF_ICSF}
\DoFa_1^\textup{ICSF}(K) = \max_{\substack{2\leq w \leq \lceil K/2 \rceil\\ w \in \bbZ^+}} \frac{w}{1+\frac{w-2}{\DoFa_w^\textup{ICOF}(K)}+\frac{w}{(w+1)\DoFa_{w+1}^\textup{ICSF}(K)}},
\end{align}
with $\DoFa_m^\textup{ICOF}(K)$ given by \cref{Eq:ICOF-DoF_m}, and $\DoFa_m^\textup{ICSF}(K)$ given by
\begin{align}
\DoFa_m^\textup{ICSF}(K)=
\begin{cases}
\left(\frac{1}{m}+m(m-1)\!\left[\frac{1}{m}-\frac{1}{\lfloor \frac{K}{2}\rfloor}-\sum_{\ell_1=m+1}^{\lfloor \frac{K}{2}\rfloor}\frac{1}{\ell_1^2}+\frac{1}{\lfloor \frac{K}{2}\rfloor \lceil \frac{K}{2} \rceil}\sum_{\ell_2=\lfloor \frac{K}{2} \rfloor+1}^K\frac{1}{\ell_2}\right]\right)^{-1}\!\!, & 2 \leq m \leq \lfloor \frac{K}{2} \rfloor\\
\left(\frac{m}{K-m+1}\sum_{\ell=m}^K\frac{1}{\ell}\right)^{-1}\!\!, & \lfloor \frac{K}{2} \rfloor < m \leq K
\end{cases}. \label{Eq:ICSF-DoF_m}
\end{align}
\end{theorem}
\begin{IEEEproof}
See \cref{Sec:IC-SF}.
\end{IEEEproof}

\begin{theorem}
\label{Th:XSF}
The $K\times K$ SISO Gaussian X channel with Shannon feedback can achieve $\DoFa_1^\textup{XSF}(K,K)$ degrees of freedom almost surely, where $\DoFa_1^\textup{XSF}(K,K)$ is given by
\begin{align}
\label{Eq:DoF_XSF}
\DoFa_1^\textup{XSF}(K,K) =\frac{K^2}{\frac{K^2+7K-6}{2}-\frac{2(K-1)}{\lfloor \frac{K}{2}\rfloor}-2(K-1) \sum_{\ell_1=1}^{\lfloor \frac{K}{2}\rfloor}\frac{1}{\ell_1^2}+\frac{2(K-1)}{\lfloor \frac{K}{2}\rfloor \lceil \frac{K}{2}\rceil}\sum_{\ell_2=\lfloor \frac{K}{2}\rfloor+1}^K\frac{1}{\ell_2}}.
\end{align}
\end{theorem}
\begin{IEEEproof}
See \cref{Sec:X-SF}.
\end{IEEEproof}

\subsection{Some Comments}
\label{Subsec:Comments}
Before proceeding with the proof details, we highlight some key features of our proposed transmission schemes through the following observations:

\begin{enumerate}
\item For each of IC and X channel and under each of the feedback/cooperation assumptions, a ``multi-phase'' transmission scheme is proposed.
\item During phase $1$, in each time slot, fresh information symbols are transmitted by a subset of transmitters such that:
\begin{enumerate}[(i)]
\item \label{Comment:EnoughEq1} Each receiver receives a number of linear combinations of its own desired information symbols (and possibly some interference symbols). The received linear combinations are not enough to resolve all desired symbols (possibly including some interference symbols). 
\item \label{Comment:EnoughEq2} Each receiver also receives some linear combinations solely in terms of undesired information symbols. However, these linear combinations are desired by some other receivers in view of observation (\ref{Comment:EnoughEq1}). On the other hand, by the end of phase $1$, each of these linear combinations will be also available at a subset of transmitters based on the feedback/cooperation assumption.
\end{enumerate}
\item During the remaining transmission phases, the transmitters deliver the linear combinations mentioned in observation (\ref{Comment:EnoughEq2}) to the receivers where they are desired: 
\begin{enumerate}[(i)]
\item Phase $m$, $m\geq 2$, takes some linear combinations as its inputs. Each of these linear combinations is available at a subset of transmitters and is desired by a subset of cardinality $m$ of receivers (and is \emph{at most} available at one unintended receiver as well). 
\item During phase $m$, the input linear combinations are transmitted over the channel such that each intended receiver obtains ``part'' of the information required to decode the input linear combinations. The rest of information required by each intended receiver (to decode all its desired linear combinations) is obtained by a subset of  unintended receivers. These pieces of information will be delivered to the intended receivers during phases $m+1, m+2, \cdots$.
\item In specific, the mentioned pieces of information (or a mixture of them) is now desired by a subset of cardinality $m+1$ of receivers, and is available at a subset of transmitters and \emph{at most} one unintended receiver. These linear combinations constitute the inputs of phase $m+1$. 
\item The transmission continues until the last phase. The input of the last phase is the linear combinations which are desired by all receivers (except for \emph{at most} one unintended receiver where the linear combination is already available). These linear combinations are delivered to their intended receivers by an appropriate number of transmissions.
\end{enumerate}

\item Under the full-duplex delayed CSIT assumption, for both IC and X channel, only two transmitters are simultaneously active in each time slot of phase $1$.
\item Under the output feedback and Shannon feedback assumptions, in each time slot of phase $1$,
\begin{enumerate}[(i)]
\item for the X channel, all transmitters are simultaneously active.
\item for the IC, the number of active transmitters is a function of the number of users.
 \end{enumerate}
\item Under the Shannon feedback assumption, the schemes proposed for both IC and X channel operate in two rounds: The first round follows the scheme proposed for the output feedback. However, as the scheme proceeds, each transmitter obtains more information about the symbols of the other transmitters using Shannon feedback. Eventually, each transmitter will be able to decode some information symbols of the other transmitters. Then, the transmission scheme will move on to the second round, where more transmitters can cooperate in the rest of transmissions.
\end{enumerate}

Here, we introduce some notations which are widely used in the subsequent proof sections, namely, \cref{Sec:ICX-DCSIT-FD,Sec:ICX-OF,Sec:ICX-SF}:
\begin{notation}  
\label{Notation:K-user-Symbols}
In the $K$-user IC and $K\times K$ X channel, $\xxS_m \subseteq \{1,2,\cdots,K\}$ denotes a subset of cardinality $m$ of transmitters (or receivers), $m\leq K$, where $\xxS_K=\{1,2,\cdots,K\}$ is the index set of all transmitters (or receivers). In the $M\times K$ X channel, subsets of cardinality $m_1$ and $m_2$ of transmitters and receivers are denoted by $\xxSt_{m_1}\subseteq \xxSt_M$ and $\xxSr_{m_2}\subseteq \xxSr_K$, respectively, where $\xxSt_M=\{1,2,\cdots,M\}$ and $\xxSr_K=\{1,2,\cdots,K\}$ are respectively the index sets of all transmitters and all receivers and $m_1\leq M$, $m_2\leq K$. A coded symbol which is available at all transmitters TX$_i$, $i\in \xxSt_{m_1}$, and all receivers RX$_{j'}$, $j'\in \xxSr_{m_3}$, and is intended to be decoded at all receivers RX$_j$, $j\in \xxSr_{m_2}$, is denoted by $u^{[\xxSt_{m_1}|\xxSr_{m_2};\xxSr_{m_3}]}$. The superscripts ``(t)'' and ``(r)'' may be omitted whenever it is clear. If $\xxSr_{m_3}=\{\}$, the mentioned symbol is denoted by $u^{[\xxSt_{m_1}|\xxSr_{m_2}]}$ and is called an order-$m_2$ symbol.
\end{notation}

\section{SISO Interference and X Channels with Full-duplex Transmitter Cooperation and Delayed CSIT}
\label{Sec:ICX-DCSIT-FD}
In this section, we investigate the impact of full-duplex transmitter cooperation on the DoF of the $K$-user IC and $M\times K$ X channel with delayed CSIT. We will demonstrate how transmitters can exploit their knowledge about each other's messages (attained through the full-duplex cooperation) combined with the delayed CSIT to achieve a higher DoF compared to the non-cooperative delayed CSIT. In specific, we prove \cref{Th:ICFD,Th:XFD} as follows:

\subsection{Proof of \Cref{Th:ICFD}}
\label{Sec:IC-DCSIT-FD}
Our transmission scheme for the $K$-user IC consists of $K-1$ phases as follows:

\phase{$1$}{Full-duplex $K$-user IC with Delayed CSIT}
In this phase, fresh information symbols are fed to the channel as follows: For every subset ${\xxS_3=\{i_1,i_2,i_3\} \subseteq \xxS_K}$, spend $3$ time slots to transmit $6$ \emph{fresh} information symbols $\{u^{[i_1]}_1,u^{[i_1]}_2,u^{[i_2]}_1,u^{[i_2]}_2,u^{[i_3]}_1,u^{[i_3]}_2\}$ by $\{\textrm{TX}_{i_1},\textrm{TX}_{i_2},\textrm{TX}_{i_3}\}$ as follows:

In the first time slot, TX$_{i_1}$ and TX$_{i_2}$ transmit $u_1^{[i_1]}$ and $u_1^{[i_2]}$, respectively, the rest of transmitters are silent. Hence, ignoring the noise, RX$_{i_1}$ and RX$_{i_2}$ each receive one linear equation in terms of $u_1^{[i_1]}$ and $u_1^{[i_2]}$ by the end of the first time slot. Therefore, if we deliver a linearly independent equation in terms of $u_1^{[i_1]}$ and $u_1^{[i_2]}$ to both RX$_{i_1}$ and RX$_{i_2}$, each of them will be able to decode both transmitted symbols (desired and interference). This linearly independent equation is indeed the linear combination $h^{[i_3i_1]}(1)u_1^{[i_1]}+h^{[i_3i_2]}(1)u_1^{[i_2]}$ received by RX$_{i_3}$ during this time slot. On the other hand, according to full-duplex operation of the transmitters, both TX$_{i_1}$ and TX$_{i_2}$ will have both $u_1^{[i_1]}$ and $u_1^{[i_2]}$ by the end of the first time slot. This along with the delayed CSIT assumption enables both TX$_{i_1}$ and TX$_{i_2}$ to reconstruct $h^{[i_3i_1]}(1)u_1^{[i_1]}+h^{[i_3i_2]}(1)u_1^{[i_2]}$. Thus, according to \cref{Notation:K-user-Symbols}, one can define 
\begin{align}
u^{[i_1,i_2|i_1,i_2;i_3]} \Def h^{[i_3i_1]}(1)u_1^{[i_1]}+h^{[i_3i_2]}(1)u_1^{[i_2]}.
\end{align}

Similarly, the second and third time slots are described as follows:
\begin{itemize}
\item \emph{Second time slot}: TX$_{i_2}$ and TX$_{i_3}$ transmit $u_2^{[i_2]}$ and $u_1^{[i_3]}$, respectively. The symbol $u^{[i_2,i_3|i_2,i_3;i_1]}$ will be accordingly generated after this time slot.
\item \emph{Third time slot}: TX$_{i_3}$ and TX$_{i_1}$ transmit $u_2^{[i_3]}$ and $u_2^{[i_1]}$, respectively. The symbol $u^{[i_3,i_1|i_3,i_1;i_2]}$ will be accordingly generated after this time slot.
\end{itemize}

Therefore, $6\binom{K}{3}$ information symbols are transmitted in $3\binom{K}{3}$ time slots and $3\binom{K}{3}$ symbols of type $u^{[\xxS_2|\xxS_2;j]}$, $j\in \xxS_K \backslash \xxS_2$, are generated by the end of phase $1$. We denote by $\DoFa_m^\textrm{ICFD}(K)$, $2\leq m \leq K-1$, our achievable DoF for transmission of symbols of type $u^{[\xxS_m|\xxS_m;j]}$, $j\in \xxS_K \backslash \xxS_m$, over the $K$-user IC with full-duplex delayed CSIT. The achieved DoF is then calculated as
\begin{align}
\DoFa_1^\textrm{ICFD}(K) = \frac{6\binom{K}{3}}{3\binom{K}{3}+\frac{3\binom{K}{3}}{\DoFa_2^\textrm{ICFD}(K)}} = \frac{2}{1+\frac{1}{\DoFa_2^\textrm{ICFD}(K)}}. \label{Eq:ICFD_Phase1_DoF}
\end{align}

\phase{$m$, $2\leq m\leq K-2$}{Full-duplex $K$-user IC with Delayed CSIT}
For $m,n\in \mathbb{Z}$, define
\begin{align}
L_m(n) &\Def \lcm\{n-m,m\} \\
Q_m(n) &\Def \min\{n-m,m\}, \label{Eq:Q_m-def}
\end{align}
where $\lcm\{x,y\}$, $x,y\in \mathbb{Z}$, is the least common multiplier of $x$ and $y$. This phase takes $\frac{m+1}{m}\alpha_m(K)$ symbols $u^{[\xxS_m|\xxS_m;j]}$, $j\in \xxS_K \backslash \xxS_m$, transmits them over the channel in $\frac{\alpha_m(K)}{Q_m(K)}$ time slots, and generates $\frac{Q_m(K)-1}{Q_m(K)}\alpha_m(K)$ symbols of type $u^{[\xxS_{m+1}|\xxS_{m+1};j]}$, $j\in \xxS_K \backslash \xxS_{m+1}$, where $\alpha_m(K)$ is defined as
\begin{align}
\alpha_m(K) \Def \binom{K}{m+1}\binom{K-m-1}{Q_m(K)-1}L_m(K).
\end{align}

Fix a subset ${\xxS_{m+1}=\{i_1,i_2,\cdots,i_{m+1}\}\subset \xxS_K}$, and a subset ${\xxS_{Q_m(K)-1}\subseteq \xxS_K\backslash \xxS_{m+1}}$. During $\frac{L_m(K)}{Q_m(K)}$ time slots, each TX$_{i_n}$, ${1\leq n\leq m+1}$, transmits a random linear combination of $u_k^{[\xxS_{m+1}\backslash \{i_{n-1}\}|\xxS_{m+1}\backslash \{i_{n-1}\};i_{n-1}]}$, $1\leq k\leq L_m(K)/m$, (with $i_0\Def i_{m+1}$) in each time slot. Therefore, a total of $(m+1)\frac{L_m(K)}{m}$ symbols are transmitted in $\frac{L_m(K)}{Q_m(K)}$ time slots. We note that the random coefficients of these linear combinations are generated offline and shared with all nodes. Now, the following observations are important:
\begin{enumerate}[(i)]
\item \label{FD-Observ1} RX$_j$, ${j\in \xxS_{m+1}}$, wishes to decode the $L_m(K)$ symbols ${\{u_k^{[\xxS_{m+1}\backslash \{j'\}|\xxS_{m+1}\backslash \{j'\};j']}\}_{k=1}^{L_m(K)/m}}\!\!$, ${j'\in \xxS_{m+1}\backslash \{j\}}$. Since it has all the symbols $\{u_k^{[\xxS_{m+1}\backslash \{j\}|\xxS_{m+1}\backslash \{j\};j]}\}_{k=1}^{L_m(K)/m}$, by canceling them, it will obtain $\frac{L_m(K)}{Q_m(K)}$ equations out of its received equations, solely in terms of its desired symbols.

\item \label{FD-Observ2} TX$_i$, $i\in \xxS_{m+1}$, has all the transmitted symbols except for $\{u_k^{[\xxS_{m+1}\backslash \{i\}|\xxS_{m+1}\backslash \{i\};i]}\}_{k=1}^{L_m(K)/m}$. According to the full-duplex operation, it will obtain $\frac{L_m(K)}{Q_m(K)}$ random linear combinations of these symbols after canceling its known symbols, and since $\frac{L_m(K)}{Q_m(K)}\geq \frac{L_m(K)}{m}$, it can decode all of them.

\item \label{FD-Observ3} RX$_{j'}$, ${j'}\in \xxS_{Q_m(K)-1}$, receives $\frac{L_m(K)}{Q_m(K)}$ linear equations in terms of all transmitted symbols. If we deliver these linear combinations to RX$_j$, $j\in \xxS_{m+1}$, it will be able to cancel its undesired part as argued in observation (\ref{FD-Observ1}) and obtain $\frac{L_m(K)}{Q_m(K)}$ equations solely in terms of its desired symbols. On the other hand, in view of observation (\ref{FD-Observ2}) and according to the delayed CSIT assumption, TX$_i$, $i\in \xxS_{m+1}$, can reconstruct all these linear combinations by the end of the $\frac{L_m(K)}{Q_m(K)}$ time slots. Thus, the $\frac{L_m(K)}{Q_m(K)}$ linear combinations received by RX$_{j'}$, $j'\in \xxS_{Q_m(K)-1}$, are denoted by $\{u_k^{[\xxS_{m+1}|\xxS_{m+1};j']}\}_{k=1}^{L_m(K)/Q_m(K)}$. After delivering these $(Q_m(K)-1)\times \frac{L_m(K)}{Q_m(K)}$ symbols to RX$_j$, $j\in\xxS_{m+1}$, it will be provided with a total of $L_m(K)$ linear combinations in terms of its $L_m(K)$ desired symbols. Also, it is easy to show that these linear combinations are linearly independent almost surely, and hence, can be solved for the desired symbols.
\end{enumerate}
Since there are $\binom{K}{m+1}$ choices of $\xxS_{m+1}$ and $\binom{K-m-1}{Q_m(K)-1}$ choices of $\xxS_{Q_m(K)-1}$ for each $\xxS_{m+1}$, the achieved DoF equals
\begin{align}
\DoFa_m^\textrm{ICFD}(K) &= \frac{(m+1)\alpha_m(K)/m}{\frac{\alpha_m(K)}{Q_m(K)}+\frac{(Q_m(K)-1)\alpha_m(K)/Q_m(K)}{\DoFa_{m+1}^\textrm{ICFD}(K)}} \nonumber \\
&=\frac{m+1}{m}\times \frac{Q_m(K)}{1+\frac{Q_m(K)-1}{\DoFa_{m+1}^\textrm{ICFD}(K)}},\quad\quad  2\leq m \leq K-2. \label{Eq:DoF_ICFD_phasem}
\end{align}

\phase{$K-1$}{Full-duplex $K$-user IC with Delayed CSIT}
During $K-1$ consecutive time slots, TX$_i$, $i\in \xxS_K$, repeats the symbol $u^{[\xxS_K\backslash \{i-1\}|\xxS_K\backslash \{i-1\};i-1]}$ (with $u^{[\xxS_K\backslash \{0\}|\xxS_K\backslash \{0\};0]}\Def u^{[\xxS_K\backslash \{K\}|\xxS_K\backslash \{K\};K]}$). It is easily verified that, in each time slot, each receiver obtains a linear combination of its $K-1$ desired symbols. Hence, after $K-1$ time slots, every receiver will be able to decode all its $K-1$ desired symbols. One then can write
\begin{align}
\label{Eq:DoF_ICFD_phaseK-1}
\DoFa_{K-1}^\textup{ICFD}(K)=\frac{K}{K-1}.
\end{align}

At the end, following Appendix \ref{App:ClosedForm-ICFD-ICOF}, it can be shown that \cref{Eq:DoF_ICFD} is indeed the closed form solution to the recursive \cref{Eq:ICFD_Phase1_DoF,Eq:DoF_ICFD_phasem} with initial condition \eqref{Eq:DoF_ICFD_phaseK-1}.

\subsection{Proof of \Cref{Th:XFD}}
\label{Sec:X-DCSIT-FD}
For the general $M\times K$ SISO X channel, a $K$-phase transmission scheme is proposed wherein the information symbols are transmitted in the first phase towards generation of higher order symbols during the subsequent phases. The order-$K$ symbols will be finally delivered to all receivers in phase $K$.

\phase{$1$}{Full-duplex $M\times K$ X Channel with Delayed CSIT}
Fix $i_1, i_2\in \xxSt_M$. For any $\{j_1,j_2\}\in \xxSr_K$, TX$_{i_1}$ and TX$_{i_2}$ transmit four \emph{fresh} information symbols $u^{[i_1|j_1]}$, $u^{[i_2|j_1]}$, $u^{[i_1|j_2]}$, and $u^{[i_2|j_2]}$ in two time slots as follows (we have ignored the indices of symbols for ease of notations): over the first time slot, TX$_{i_1}$ and TX$_{i_2}$ respectively transmit $u^{[i_1|j_1]}$, $u^{[i_2|j_1]}$, both intended for RX$_{j_1}$. After this time slot, the linear combination $h^{[j_2i_1]}u^{[i_1|j_1]}+h^{[j_2i_2]}u^{[i_2|j_1]}$, which has been received by RX$_{j_2}$, is available at TX$_{i_1}$ and TX$_{i_2}$ due to full-duplex operation of the transmitters and delayed CSIT, and is desired by RX$_{j_1}$ to be able to decode $u^{[i_1|j_1]}$ and $u^{[i_2|j_1]}$. Hence, it is denoted as $u^{[i_1,i_2|j_1;j_2]}$. Similarly, over the second time slot, TX$_{i_1}$ and TX$_{i_2}$ respectively transmit $u^{[i_1|j_2]}$, $u^{[i_2|j_2]}$, both intended now for RX$_{j_2}$, and the symbol $u^{[i_1,i_2|j_2;j_1]}$ is generated. It is easily verified that $u^{[i_1,i_2|j_1;j_2]} + u^{[i_1,i_2|j_2;j_1]}$ is desired by both RX$_{j_1}$ and RX$_{j_2}$. Hence, one can define the following order-$2$ symbol:
\begin{align}
u^{[i_1,i_2|j_1,j_2]} \Def u^{[i_1,i_2|j_1;j_2]} + u^{[i_1,i_2|j_2;j_1]}.
\end{align}

By the end of this phase, $4\binom{M}{2}\binom{K}{2}$ fresh information symbols are transmitted in $2\binom{M}{2}\binom{K}{2}$ time slots and $\binom{M}{2}\binom{K}{2}$ order-$2$ symbols are generated, which will be delivered to their corresponding pairs of receivers during the rest of the transmission scheme. The achieved DoF is then calculated as
\begin{align}
\DoFa_1^\textup{XFD}(M,K)=\frac{4\binom{M}{2}\binom{K}{2}}{2\binom{M}{2}\binom{K}{2}+\frac{\binom{M}{2}\binom{K}{2}}{\DoFa^\textup{XFD}_2(M,K)}}=\frac{4}{2+\frac{1}{\DoFa^\textup{XFD}_2(M,K)}}, \label{Eq:DoF_XFD_Phase1}
\end{align}
where $\DoFa_2^\textup{XFD}(M,K)$ denotes our achievable DoF for transmission of order-$2$ symbols of type $u^{[\xxSt_2|\xxSr_2]}$ over the full-duplex $M\times K$ SISO X channel with delayed CSIT.

\phase{$m$, $2\leq m \leq K-1$}{Full-duplex $M\times K$ X Channel with Delayed CSIT}
Consider the following distinct cases:
\begin{enumerate}[(i)]
\item \label{Item:Case1} $M> \frac{K}{2}, \,\,\, 2\leq m\leq \frac{K}{2}$: 

In this case, order-$m$ symbols of type $u^{[\xxSt_m|\xxSr_m]}$ are transmitted over the channel. Fix a subset $\xxSr_{2m}\subseteq \xxSr_K$, and a subset $\xxSt_{m+1}=\{i_1,i_2,\cdots,i_{m+1}\}\subseteq \xxSt_M$. Note that since $m\leq K/2< M$, both subsets exist. All transmitters TX$_j$, $j\in \xxSt_M \backslash \xxSt_{m+1}$, are silent, while the transmitters TX$_{i_n}$, $1\leq n \leq m+1$, simultaneously transmit as follows: For every subset $\xxSr_m\subset \xxSr_{2m}$, spend one time slot to transmit $u^{[i_n,i_{n+1},\cdots,i_{n+m-1}|\xxSr_m]}$ by $\textup{TX}_{i_n}$, $n=1,\cdots,m+1$, where $i_k\Def i_{k-m-1}$ for $m+1<k\leq2m$. Every RX$_j$, $j\in \xxSr_m$, receives one linear equation in terms of $m+1$ desired symbols, and thus, requires $m$ extra independent equations to resolve all the $m+1$ symbols. It is easy to see that the equation received by RX$_j$, $j\in \xxSr_{2m} \backslash \xxSr_m$, is linearly independent of the equation received by each RX$_j$, $j\in \xxSr_m$, and hence, is desired by all of them. On the other hand, every TX$_{i_n}$, $1\leq n \leq m+1$, knows exactly $m$ symbols out of the $m+1$ transmitted symbols, and thus, obtains the last one using the full-duplex operation by the end of this time slot. Hence, TX$_{i_n}$, $1\leq n \leq m+1$, having access to all the $m+1$ transmitted symbols and the delayed CSI, can reconstruct the linear combinations received by all receivers by the end of this time slot. In particular, one can denote the linear combination received by RX$_j$, $j\in \xxSr_{2m}\backslash \xxSr_{m}$, as $u^{[\xxSt_{m+1}|\xxSr_{m};j]}$.

Now, we have the following observation: For any subset $\xxSr_{m+1}\subset \xxSr_{2m}$, consider the $m+1$ symbols $u^{[\xxSt_{m+1}|\xxSr_{m+1}\backslash \{j\};j]}$, $j\in \xxSr_{m+1}$, as defined above. Each receiver RX$_j$, $j\in \xxSr_{m+1}$, has exactly one of these symbols and requires the other $m$. Therefore, if we deliver $m$ random linear combinations of these $m+1$ symbols to all receivers RX$_j$, $j\in \xxSr_{m+1}$, each of them will be provided with $m$ random linear combinations of $m$ desired unknowns, and thus, will resolve all of them. Hence, these $m$ random linear combinations can be denoted as $\{u^{[\xxSt_{m+1}|\xxSr_{m+1}]}_k\}_{k=1}^m$. These order-$(m+1)$ symbols will be delivered to their corresponding receivers during the rest of the transmission scheme. We denote by $\DoFa_m^\textup{XFD}(M,K)$, $2\leq m\leq K/2<M$, our achievable DoF for transmission of order-$m$ symbols of type $u^{[\xxSt_m|\xxSr_m]}$ over the full-duplex $M\times K$ SISO X channel with delayed CSIT. Since there are $\binom{K}{2m}$ choices for $\xxSr_{2m}$, $\binom{M}{m+1}$ choices for $\xxSt_{m+1}$, and $\binom{2m}{m}$ choices for $\xxSr_m$, the achieved DoF is calculated as
\begin{align}
\DoFa_m^\textup{XFD}(M,K) &= \frac{\binom{M}{m+1}\binom{K}{2m}\binom{2m}{m}(m+1)}{\binom{M}{m+1}\binom{K}{2m}\binom{2m}{m}+\frac{\binom{M}{m+1}\binom{K}{2m}\binom{2m}{m+1}m}{\DoFa_{m+1}^\textup{XFD}(M,K)}} \nonumber \\
&=\frac{(m+1)^2}{m+1+\frac{m^2}{\DoFa_{m+1}^\textup{XFD}(M,K)}}, \hspace{2cm} M> \frac{K}{2}, \quad 2\leq m\leq \frac{K}{2}. \label{Eq:DoF_XFD_Case1}
\end{align}
\item $M> \frac{K}{2}, \,\,\, \frac{K}{2}<m\leq K-1$:

In this case, order-$m$ symbols of type $u^{[\xxSt_{\lfloor K/2\rfloor+1}|\xxSr_m]}$ are transmitted over the channel. Since $K/2 < M$ and $K/2<m$, we have $K-m+1\leq \lfloor K/2\rfloor+1\leq M$, and thus, these symbols can be optimally transmitted over the channel using the scheme proposed in \cite{maddah2010DoF_BCC_Delayed_Arxiv} for transmission of order-$m$ symbols over an $M\times K$ MISO broadcast channel with delayed CSIT when $K-m+1\leq M$ (cf.\ \cite{maddah2010DoF_BCC_Delayed_Arxiv} Section III-C). We note that, in this case, the transmitters do not use their full-duplex capabilities, since enough number of transmitters already know the order-$m$ symbols which are going to be transmitted over the channel. Here, we denote by $\DoFa_m^\textup{XFD}(M,K)$, $0\leq K-m<K/2 <M$, the achievable DoF for transmission of order-$m$ symbols of type $u^{[\xxSt_{\lfloor K/2\rfloor+1}|\xxSr_m]}$ over the full-duplex $M\times K$ SISO X channel with delayed CSIT. Hence, the following recursion holds (cf.\ Eq. (27) in \cite{maddah2010DoF_BCC_Delayed_Arxiv}):
\begin{align}
\DoFa_m^\textup{XFD}(M,K) =\frac{(m+1)(K-m+1)}{m+1+\frac{m(K-m)}{\DoFa_{m+1}^\textup{XFD}(M,K)}}, \hspace{2cm} M>\frac{K}{2}, \quad \frac{K}{2}<m\leq K-1. \label{Eq:DoF_XFD_Case2}
\end{align}

We emphasize here that the achievable DoF given in \cref{Eq:DoF_XFD_Case2} is indeed tight, since it meets the upper bound of the MISO broadcast channel with delayed CSIT (cf.\ \cite{maddah2010DoF_BCC_Delayed_Arxiv}). 
\item $2\leq M\leq \frac{K}{2}, \,\,\, 2\leq m < M$:

In this case, order-$m$ symbols of type $u^{[\xxSt_m|\xxSr_m]}$ are transmitted over the channel. Since in this case we have $m<M\leq K/2$, the transmission scheme proposed for case (\ref{Item:Case1}) works for this case as well and the achieved DoF is given by \cref{Eq:DoF_XFD_Case1}.

\item $2\leq M\leq \frac{K}{2}, \,\,\,  M\leq m \leq K-1$:
\end{enumerate}

In this case, order-$m$ symbols of type $u^{[\xxSt_M|\xxSr_m]}$ are transmitted over the channel without operating in the full-duplex mode using the scheme proposed in \cite{maddah2010DoF_BCC_Delayed_Arxiv} for transmission of order-$m$ symbols over an $M\times K$ MISO broadcast channel with delayed CSIT (see \cite{maddah2010DoF_BCC_Delayed_Arxiv} Section VI-B), and the following DoF is achieved (cf.\ Eq. (39) in \cite{maddah2010DoF_BCC_Delayed_Arxiv}):
\begin{align}
\DoFa_m^\textup{XFD}(M,K)=\frac{(m+1)(\min\{M-1,K-m\}+1)}{m+1+\frac{m\times\min\{M-1,K-m\}}{\DoFa_{m+1}^\textup{XFD}(M,K)}}, \hspace{2cm} M\leq \frac{K}{2}, \quad M\leq m \leq K-1, \label{Eq:DoF_XFD_Case4}
\end{align}
where $\DoFa_m^\textup{XFD}(M,K)$ (resp.\ $\DoFa_{m+1}^\textup{XFD}(M,K)$), in this case, denotes our achievable DoF for transmission of symbols of type $u^{[\xxSt_M|\xxSr_m]}$ (resp.\ $u^{[\xxSt_M|\xxSr_{m+1}]}$) over the full-duplex $M\times K$ SISO X channel with delayed CSIT.

To summarize our achievable results for the above cases, for $m,M,K\in \mathbb{Z}$, we define
\begin{align}
Q_m(M,K)&\Def \min \{M-1,K-m,m\},\\
\Theta_m(M,K)&\Def \min \{M,\lfloor K/2\rfloor+1,m\},
\end{align}
and denote by $\DoFa_m^\textup{XFD}(M,K)$ our achievable DoF for transmission of order-$m$ symbols of type $u^{[\xxSt_{\Theta_m(M,K)}|\xxSr_m]}$ over the full-duplex $M\times K$ SISO X channel with delayed CSIT. Then, it is easy to see from \cref{Eq:DoF_XFD_Phase1,Eq:DoF_XFD_Case1,Eq:DoF_XFD_Case2,Eq:DoF_XFD_Case4} that our achievable DoF satisfies the following recursive equation:
\begin{align}
\DoFa_m^\textup{XFD}(M,K) =\frac{(m+1)(Q_m(M,K)+1)}{m+1+\frac{m\times Q_m(M,K)}{\DoFa_{m+1}^\textup{XFD}(M,K)}}, \hspace{2cm} \quad 1\leq m\leq K-1. \label{Eq:DoF_XFD_Recursive}
\end{align}

\phase{$K$}{Full-duplex $M\times K$ X Channel with Delayed CSIT}

In this phase, the symbols of type $u^{[\xxSt_{\Theta_K(M,K)}|\xxSr_K]}$ are delivered to all $K$ receivers by simple transmission of one symbol per time slot by one of the transmitters (which has access to that symbol). Therefore,
\begin{align}
\DoFa_{K}^\textup{XFD}(M,K) = 1. \label{Eq:DoF_XFD_phaseK}
\end{align}

It is shown in Appendix \ref{App:ClosedForm-XFD} that \cref{Eq:DoF_XFD} is indeed the closed form solution to the recursive Eq. \eqref{Eq:DoF_XFD_Recursive} together with the initial condition \eqref{Eq:DoF_XFD_phaseK}.

\section{SISO Interference and X Channels with Output Feedback}
\label{Sec:ICX-OF}
In this section, we investigate the impact of output feedback on the DoF of the $K$-user IC and $K\times K$ X channel. As defined in \cref{Sec:ProblemFormulation}, we assume that output of each receiver is fed back to its paired transmitter. This provides each transmitter with ``some'' information about the other transmitters' messages, which enables the transmitters to cooperate in their subsequent transmissions. Recall that in our achievable schemes for the full-duplex IC and X channel with delayed CSIT, described in \cref{Sec:ICX-DCSIT-FD}, each transmitter acquired \emph{pure} symbols of the other transmitters via full-duplex cooperation in order to \emph{reconstruct} the linear combinations received by the receivers. The number of simultaneously active transmitters was restricted in each time slot such that each active transmitter can obtain a pure symbol transmitted by one of the others. For instance, in phase $1$ of the scheme, only two transmitters per time slot were allowed to simultaneously transmit over the channel. In contrast, when the output feedback is available, the linear combination received by each receiver will become readily available at one of the transmitters, and thus, the restriction on the number of simultaneously active transmitters is relaxed, providing for a higher level of transmitter cooperation and interference alignment. The rest of this section presents proofs of \cref{Th:ICOF,Th:XOF}.

\subsection{Proof of \Cref{Th:ICOF}}
\label{Sec:IC-OF}
Our transmission scheme for the $K$-user IC with output feedback consists of $K-\mu(K)+1$ phases as follows, where the integer $\mu(K)$, $2\leq \mu(K)\leq\lceil K/2\rceil$, will be determined later:

\phase{$1$}{$K$-user IC with Output Feedback}
For every subset ${\xxS_{\mu(K)}\subset \xxS_K}$, and every subset $\xxS_{\mu(K)-1}\subseteq \xxS_K \backslash \xxS_{\mu(K)}$, in one time slot, each TX$_i$, $i\in \xxS_{\mu(K)}$, transmits a \emph{fresh} information symbol $u^{[i]}$. Then, if we deliver $\mu(K)-1$ linearly independent combinations of the $\mu(K)$ transmitted symbols to RX$_i$, $i\in \xxS_{\mu(K)}$, it will be able to decode all the transmitted symbols. Thus, the equation received by RX$_j$, $j\in \xxS_{\mu(K)-1}$, which will be available at TX$_j$ via the output feedback, is desired by all the receivers RX$_i$, $i\in \xxS_{\mu(K)}$. Hence, they can be denoted as $u^{[j|\xxS_{\mu(K)};j]}$, $j\in \xxS_{\mu(K)-1}$.

Therefore, $\mu(K)\binom{K}{\mu(K)}\binom{K-\mu(K)}{\mu(K)-1}$ information symbols are transmitted in $\binom{K}{\mu(K)}\binom{K-\mu(K)}{\mu(K)-1}$ time slots and $(\mu(K)-1)\binom{K}{\mu(K)}\binom{K-\mu(K)}{\mu(K)-1}$ symbols $u^{[j|\xxS_{\mu(K)};j]}$ are generated by the end of phase $1$. Denoting by $\DoFa_m^\textrm{ICOF}(K)$ our achievable DoF for transmission of symbols $u^{[j|\xxS_m;j]}$, $j\in \xxS_K\backslash \xxS_m$, over the $K$-user IC with output feedback, the achieved DoF is equal to
\begin{align}
\DoFa_1^\textrm{ICOF}(K)=\frac{\mu(K)}{1+\frac{\mu(K)-1}{\DoFa_{\mu(K)}^\textrm{ICOF}(K)}}. \label{Eq:ICOF-DoF_1}
\end{align}

\phase{$m$, $2\leq m\leq K-2$}{$K$-user IC with Output Feedback}
This phase feeds $\frac{m+1}{m}\alpha_m(K)$ symbols of type $u^{[j|\xxS_m;j]}$, $j\in \xxS_K \backslash \xxS_m$, to the channel in $\frac{\alpha_m(K)}{Q_m(K)}$ time slots, and generates $\frac{Q_m(K)-1}{Q_m(K)}\alpha_m(K)$  symbols of type 
$u^{[j|\xxS_{m+1};j]}$, $j\in \xxS_K \backslash \xxS_{m+1}$. In specific, for every subset ${\xxS_{m+1}\subset \xxS_K}$, and every subset ${\xxS_{Q_m(K)-1}\subseteq \xxS_K\backslash \xxS_{m+1}}$, during $\frac{L_m(K)}{Q_m(K)}$ time slots, every TX$_i$, ${i\in \xxS_{m+1}}$, transmits $\frac{L_m(K)}{Q_m(K)}$ random linear combinations of symbols $\{u_k^{[i|\xxS_{m+1}\backslash \{i\};i]}\}_{k=1}^{L_m(K)/m}$. Each RX$_j$, ${j\in \xxS_{m+1}}$, wishes to decode the $L_m(K)$ symbols $\{u_k^{[j'|\xxS_{m+1}\backslash \{j'\};j']}\}_{k=1}^{L_m(K)/m}$, ${j'\in \xxS_{m+1}\backslash \{j\}}$. Also, RX$_j$, ${j\in \xxS_{m+1}}$, after removing $\{u_k^{[j|\xxS_{m+1}\backslash \{j\};j]}\}_{k=1}^{L_m(K)/m}$ from its received equations, obtains $\frac{L_m(K)}{Q_m(K)}$ linear equations solely in terms of its desired symbols. If we deliver the $\frac{L_m(K)}{Q_m(K)}$ linear equations received by RX$_{j'}$, $j'\in \xxS_{Q_m(K)-1}$, to RX$_j$, $j\in \xxS_{m+1}$, it will obtain another $(Q_m(K)-1)\times \frac{L_m(K)}{Q_m(K)}$ linear equations solely in terms of its desired symbols. Since these equations will be available at TX$_{j'}$, ${j'\in \xxS_{Q_m(K)-1}}$, via the output feedback, they are denoted as $\{u_k^{[j'|\xxS_{m+1};j']}\}_{k=1}^{L_m(K)/Q_m(K)}$. Therefore, RX$_j$, $j\in\xxS_{m+1}$, will have $L_m(K)$ (linearly independent) equations in terms of its $L_m(K)$ desired symbols, and can solve them for its desired symbols.

Finally, since the number of input symbols, spent time slots, and output symbols of this phase are equal to those of phase $m$ in the proposed transmission scheme for the full-duplex $K$-user IC with delayed CSIT described in proof of \cref{Th:ICFD}, the achieved DoF for phase $m$ satisfies the same recursive equation, \ie \cref{Eq:DoF_ICFD_phasem}:
\begin{align}
\DoFa_m^\textup{ICOF}(K)=\frac{m+1}{m}\times \frac{Q_m(K)}{1+\frac{Q_m(K)-1}{\DoFa_{m+1}^\textup{ICOF}(K)}},\quad\quad 2\leq m\leq K-2. \label{Eq:DoF_ICOF_phasem}
\end{align}

\phase{$K-1$}{$K$-user IC with Output Feedback}
During $K-1$ consecutive time slots, TX$_i$, $i\in \xxS_K$, repeats the symbol $u^{[i|\xxS_K\backslash \{i\};i]}$. Therefore, each receiver receives $K-1$ linear combination of its $K-1$ desired symbols, and thus, will be able to decode all its $K-1$ desired symbols. Hence, 
\begin{align}
\DoFa_{K-1}^\textup{ICOF}=\frac{K}{K-1}. \label{Eq:DoF_ICOF_phaseK-1}
\end{align}

\begin{figure}
\centering
\includegraphics{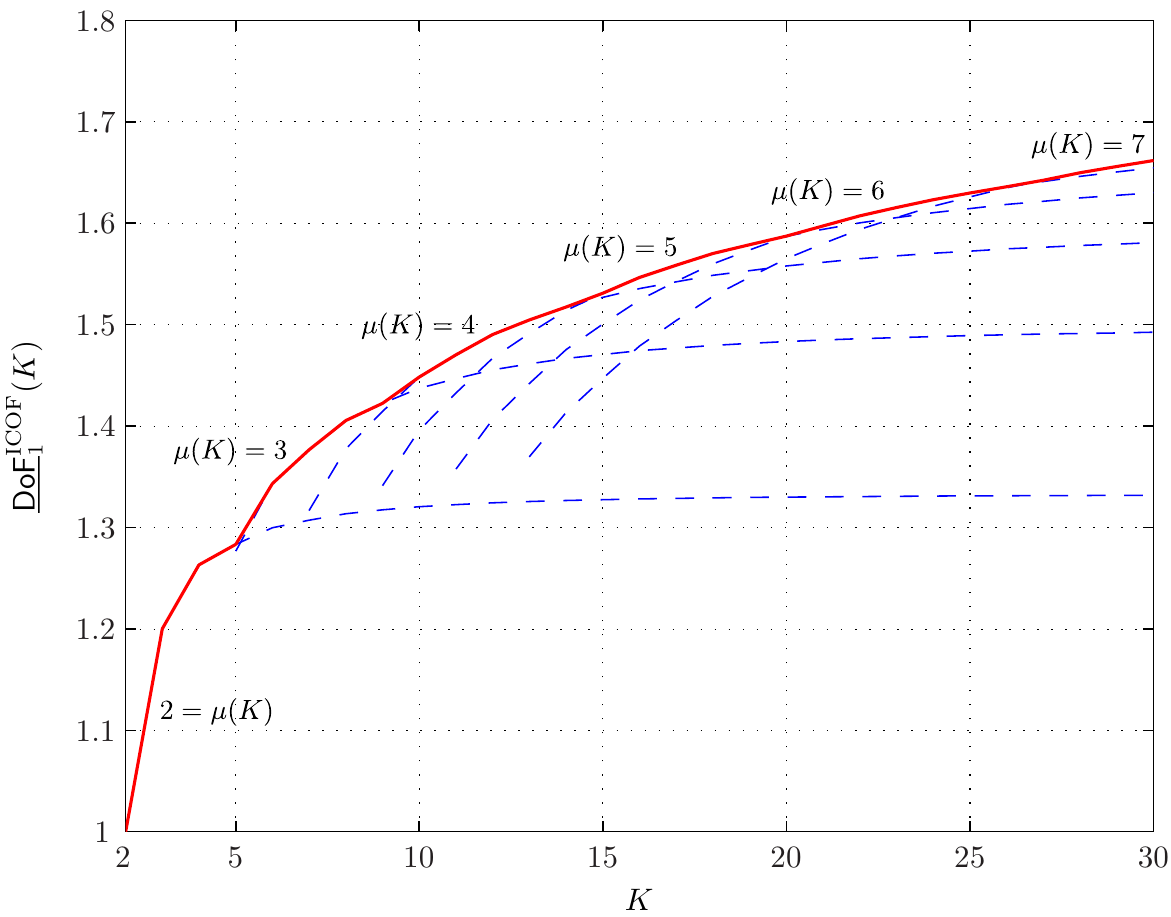}
\caption{Achievable DoFs for the $K$-user IC with output feedback.}
\label{Fig:ICOF_DoF}
\end{figure}

It is shown in Appendix \ref{App:ClosedForm-ICFD-ICOF} that the solution to recursive Eq. \eqref{Eq:DoF_ICOF_phasem} with initial condition \eqref{Eq:DoF_ICOF_phaseK-1} is given by
\begin{align}
\DoFa_m^\textup{ICOF}(K)=
\begin{cases}
\left(\frac{1}{2}-\frac{m(m-1)}{2\lceil \frac{K}{2}\rceil(\lceil \frac{K}{2}\rceil-1)}+\frac{m(m-1)}{\lfloor \frac{K}{2}\rfloor(\lceil \frac{K}{2}\rceil-1)}\sum_{\ell=\lceil \frac{K}{2}\rceil+1}^K\frac{1}{\ell}\right)^{-1},  &\hspace{5mm}2\leq m \leq \lceil \frac{K}{2}\rceil\\
\left(\frac{m}{K-m}\sum_{\ell=m+1}^K\frac{1}{\ell}\right)^{-1}, & \hspace{-3.5mm}\lceil \frac{K}{2} \rceil < m \leq K-1
\end{cases}.
\label{Eq:ICOF-DoF_m}
\end{align}
Substituting \cref{Eq:ICOF-DoF_m} for $\DoFa_{\mu(K)}^\textup{ICOF}(K)$ in \cref{Eq:ICOF-DoF_1}, we get
\begin{align}
\DoFa_1^\textup{ICOF}(K)=\frac{\mu(K)}{a(K)\mu(K)\left(\mu(K)-1\right)^2+(\mu(K)+1)/2}, \label{Eq:ICOF-DoF_1'}
\end{align}
where $a(K)$ is defined by \cref{Eq:a(K)_def}. Now, we choose $\mu(K)$ such that $\DoFa_1^\textup{ICOF}(K)$ given in \cref{Eq:ICOF-DoF_1'} is maximized. In other words,
\begin{align}
\mu(K)= \argmax_{\substack{2\leq w \leq \lceil K/2 \rceil\\ w \in \bbZ^+}} f_K^\textup{ICOF}(w), \label{Eq:Delta_ICOF_maximization}
\end{align}
where $f_K^\textup{ICOF}(w)$ is defined as
\begin{align}
f_K^\textup{ICOF}(w)\Def\frac{w}{a(K)w(w-1)^2+(w+1)/2}. \label{Eq:f_K^ICOF(w)_def}
\end{align}
By taking the derivative of  $f_K^\textup{ICOF}(w)$ with respect to $w$, it can be shown that the solution $w^*_K$ to the maximization problem $w^*_K= \argmax_{2\leq w \leq \lceil K/2 \rceil} f_K^\textup{ICOF}(w)$ is given by \cref{Eq:w*(K)_def}. Thus, since  $f_K^\textup{ICOF}(w)$ is a continuous and concave function of $w$, the solution $\mu(K)$ to the maximization problem \eqref{Eq:Delta_ICOF_maximization} is either $\lfloor w^*_K\rfloor$ or $\lceil w^*_K\rceil$, depending on which yields a greater $f_K^\textup{ICOF}(w)$, \ie
\begin{align}
\mu(K)=\argmax_{w\in \{\lfloor w^*_K\rfloor, \lceil w^*_K\rceil \}}f_K^\textup{ICOF}(w), \label{Eq:Delta_ICOF_final}
\end{align}
which in view of \cref{Eq:f_K^ICOF(w)_def,Eq:ICOF-DoF_1'} completes the proof. \Cref{Fig:ICOF_DoF} shows the achievable DoF for different values of $\mu(K)$ together with the optimized achievable DoF, \ie $\DoFa_1^\textup{ICOF}(K)$, for $3\leq K \leq 30$.

\subsection{Proof of \Cref{Th:XOF}}
\label{Sec:X-OF}
We propose a transmission scheme which consists of $2$ main phases as follows:

\phase{$1$}{$K\times K$ X Channel with Output Feedback}
For every $j\in \xxS_K$, spend one time slot to transmit the fresh information symbols $u^{[1|j]}$, $u^{[2|j]}$, $\cdots$, $u^{[K|j]}$ respectively by TX$_1$, TX$_2$, $\cdots$, TX$_K$, all intended for RX$_j$. By the end of this time slot, RX$_j$ has received one linear combination of all $K$ desired symbols. Therefore, if the linear combinations received by RX$_{j'}$, $j'\in \xxS_K \backslash \{j\}$, are delivered to RX$_j$, it can decode all the $K$ symbols. On the other hand, according to the output feedback, the linear combination received by RX$_{j'}$, $j'\in \xxS_K \backslash \{j\}$, will be available at TX$_{j'}$ after this time slot. Hence, they can be denoted as $u^{[j'|j;j']}$, $j'\in \xxS_K \backslash \{j\}$. Therefore, after $K$ time slots, $K(K-1)$ symbols $u^{[j'|j;j']}$, $j\in \xxS_K$, $j'\in \xxS_K \backslash \{j\}$, will be generated. These symbols will be delivered to their respective receiver during the next phase.

\phase{$2$}{$K\times K$ X Channel with Output Feedback}
This phase takes $K(K-1)/2$ time slots to deliver the $K(K-1)$ symbols generated in phase $1$ as follows: For any subset $\{j,j'\}\subseteq \xxS_K$, spend one time slot to transmit $u^{[j|j';j]}$ and $u^{[j'|j;j']}$ by TX$_j$ and TX$_{j'}$, respectively, while the other transmitters are silent. After this time slot, each of RX$_j$ and RX$_{j'}$ can decode its desired symbol by canceling the interference symbol which it already has. The achieved DoF is then equal to
\begin{align}
\DoFa_1^\textup{XOF}(K,K)=\frac{K^2}{K+K(K-1)/2}=\frac{2K}{K+1},
\end{align}
completing the proof.

\section{SISO Interference and X Channels with Shannon Feedback}
\label{Sec:ICX-SF}
With Shannon feedback, each transmitter has access to \emph{all} observations made by its paired receiver, \ie the channel output and all the channel coefficients, with some delay. Moreover, it has access to its own transmitted symbols. If a receiver wants to decode, say, $n$ symbols (some of which might be interference), it requires $n$ linearly independent equations in terms of the $n$ symbols. However, the key observation is that after delivering $n-1$ required equations to a receiver, its paired transmitter having access to Shannon feedback and its own transmitted symbol (which is one of the $n$ symbols), will be able to decode all the remaining $n-1$ symbols. Then, using the delayed CSIT, it will be able to reconstruct the last (yet undelivered) linear combination, and hence, to cooperate for its delivery. This allows for achieving higher DoFs compared to what we achieved in \cref{Sec:ICX-DCSIT-FD,Sec:ICX-OF}. The following two subsections offer proofs of \cref{Th:ICSF,Th:XSF}.

\subsection{Proof of \Cref{Th:ICSF}}
\label{Sec:IC-SF}
Our achievable scheme for the $K$-user IC with Shannon feedback has two rounds of operation:

\round{$1$}{$K$-user IC with Shannon Feedback}
In this round, the transmitters use only the output feedback in parallel with the scheme proposed in proof of \cref{Th:ICOF}. In specific, during phase $1$, for every subset $\xxS_{\nu(K)}\subset \xxS_K$, every subset $\xxS_{\nu(K)-1}\subseteq \xxS_K \backslash \xxS_{\nu(K)}$, and every $j_0\in \xxS_{\nu(K)-1}$, in one time slot, each TX$_i$, $i\in \xxS_{\nu(K)}$, transmits a \emph{fresh} information symbol $u^{[i]}$. The integer $\nu(K)$, $2\leq \nu(K)\leq\lceil K/2\rceil$, will be determined later. The linear combination received by RX$_j$, $j\in \xxS_{\nu(K)-1}$, which will be available at TX$_j$ via the output feedback, is desired by every RX$_i$, $i\in \xxS_{\nu(K)}$. 

Now, TX$_i$, $i\in \xxS_{\nu(K)}$, using Shannon feedback and having $u^{[i]}$, obtains an equation in terms of the symbols $u^{[i']}$, $i'\in \xxS_{\nu(K)}\backslash \{i\}$. We deliver the $\nu(K)-2$ linear combinations available at the receivers RX$_j$, $j\in \xxS_{\nu(K)-1} \backslash \{j_0\}$, to every RX$_i$, $i\in \xxS_{\nu(K)}$, using the scheme proposed in proof of \cref{Th:ICOF}. Meanwhile, TX$_i$ using Shannon feedback and having $u^{[i]}$, will obtain another $\nu(K)-2$ linearly independent combinations of $u^{[i']}$, $i'\in \xxS_{\nu(K)}\backslash \{i\}$, and hence, can decode all of them. Thereby, it can reconstruct the linear combination available at RX$_{j_0}$, which is still required by every RX$_i$, $i\in \xxS_{\nu(K)}$. Hence, this linear combination will be denoted as $u^{[\xxS_{\nu(K)}\cup \{j_0\}|\xxS_{\nu(K)};j_0]}$. 

We note that, for every subset ${\xxS_{\nu(K)+1}\subseteq \xxS_K}$, and every subset ${\xxS_{\nu(K)-2}\subseteq \xxS_K \backslash \xxS_{\nu(K)+1}}$, we have generated ${\nu(K)+1}$ symbols $u^{[\xxS_{\nu(K)+1}|\xxS_{\nu(K)+1}\backslash \{j_0\};j_0]}$, $j_0 \in \xxS_{\nu(K)+1}$. Since every RX$_i$, ${i\in \xxS_{\nu(K)+1}}$, needs exactly $\nu(K)$ out of these $\nu(K)+1$ symbols, $\nu(K)$ random linear combinations of these symbols are desired by each RX$_i$, $i\in \xxS_{\nu(K)+1}$, and can be denoted as $\{u^{[\xxS_{\nu(K)+1}|\xxS_{\nu(K)+1}]}_k\}_{k=1}^{\nu(K)}$. They will be delivered during round $2$ of the transmission scheme. The achieved DoF is therefore given by
\begin{align}
\DoFa_1^\textrm{ICSF}(K)&{=}\frac{\nu(K) \beta(K)}{\beta(K)+\frac{(\nu(K)-2)\beta(K)}{\DoFa_{\nu(K)}^\textrm{ICOF}(K)}+\frac{\binom{K}{\nu(K)+1}\binom{K-\nu(K)-1}{\nu(K)-2}\nu(K)}{\DoFa_{\nu(K)+1}^\textrm{ICSF}(K)}} \nonumber \\
&=\frac{\nu(K)}{1+\frac{\nu(K)-2}{\DoFa_{\nu(K)}^\textrm{ICOF}(K)}+\frac{\nu(K)}{(\nu(K)+1)\DoFa_{\nu(K)+1}^\textrm{ICSF}(K)}}, \label{Eq:ICSF-DoF_1}
\end{align}
where
\begin{align}
\beta(K)\Def\binom{K}{\nu(K)}\binom{K-\nu(K)}{\nu(K)-1}(\nu(K)-1),
\end{align}
and $\DoFa_m^\textrm{ICSF}(K)$ denotes our achievable DoF for transmission of the symbols of type $u^{[\xxS_m|\xxS_m]}$ over the $K$-user IC with Shannon feedback.

\round{$2$}{$K$-user IC with Shannon Feedback}
This round consists of $K-\nu(K)$ phases described as follows:

\phase{$m$, $\nu(K)+1\leq m\leq K-1$}{$K$-user IC with Shannon Feedback}
In this phase, symbols of type $u^{[\xxS_m|\xxS_m]}$ are fed to the channel and symbols of type $u^{[\xxS_{m+1}|\xxS_{m+1}]}$ are generated as follows: Fix a subset ${\xxS_{Q_m(K+1)+m-1}\subseteq \xxS_K}$, where $Q_m(n)$, ${n\in \mathbb{Z}}$, is defined in \cref{Eq:Q_m-def}. For any ${\xxS_m\subset \xxS_{Q_m(K+1)+m-1}}$, spend one time slot to transmit $\{u_k^{[\xxS_m|\xxS_m]}\}_{k=1}^{Q_m(K+1)}$ by $Q_m(K+1)$ arbitrary transmitters out of $\{\textrm{TX}_j:j\in \xxS_m\}$. Then, RX$_j$, $j\in \xxS_m$, requires $Q_m(K+1)-1$ extra equations to resolve all the transmitted symbols. Thus, the linear combination received by RX$_{j'}$, $j'\in \xxS_{Q_m(K+1)+m-1} \backslash \xxS_m$, which will be available at TX$_{j'}$ via the output feedback, is desired by every RX$_j$, $j\in \xxS_m$. On the other hand, every TX$_j$, $j\in \xxS_m$, having access to all the transmitted symbols and delayed CSI, can reconstruct this linear combination. Therefore, it is denoted as $u^{[\xxS_m \cup \{j'\}|\xxS_m;j']}$.

Now, for any subset ${\xxS_{m+1}\subseteq \xxS_{Q_m(K+1)+m-1}}$, consider ${m+1}$ symbols $u^{[\xxS_{m+1}|\xxS_{m+1}\backslash \{j\};j]}$, ${j\in \xxS_{m+1}}$. It is easy to see that $m$ random linear combinations of these symbols are desired by each RX$_i$, ${i\in \xxS_{m+1}}$, and can be denoted as $\{u^{[\xxS_{m+1}|\xxS_{m+1}]}_k\}_{k=1}^m$. The achieved DoF equals
\begin{align}
\DoFa_m^\textrm{ICSF}(K)&=\frac{Q_m(K+1)\binom{Q_m(K+1)+m-1}{m}\binom{K}{Q_m(K+1)+m-1}}{ \binom{K}{Q_m(K+1)+m-1}\binom{Q_m(K+1)+m-1}{m}+\frac{m\binom{Q_m(K+1)+m-1}{m+1}\binom{K}{Q_m(K+1)+m-1}}{\DoFa_{m+1}^\textrm{ICSF}(K)}} \nonumber \\
&=\frac{(m+1)Q_m(K+1)}{m+1+\frac{m\times(Q_m(K+1)-1)}{\DoFa_{m+1}^\textrm{ICSF}(K)}}, \hspace{10mm}2\leq m \leq K-1. \label{Eq:DoF_ICSF_phasem}
\end{align}

\phase{$K$}{$K$-user IC with Shannon Feedback}
In this phase, one symbol $u^{[\xxS_K|\xxS_K]}$ per time slot is transmitted by an arbitrary transmitter. Hence,
\begin{align}
\DoFa_K^\textrm{ICSF}(K)=1. \label{Eq:DoF_ICSF_phaseK}
\end{align}

\begin{figure}
\centering
\includegraphics{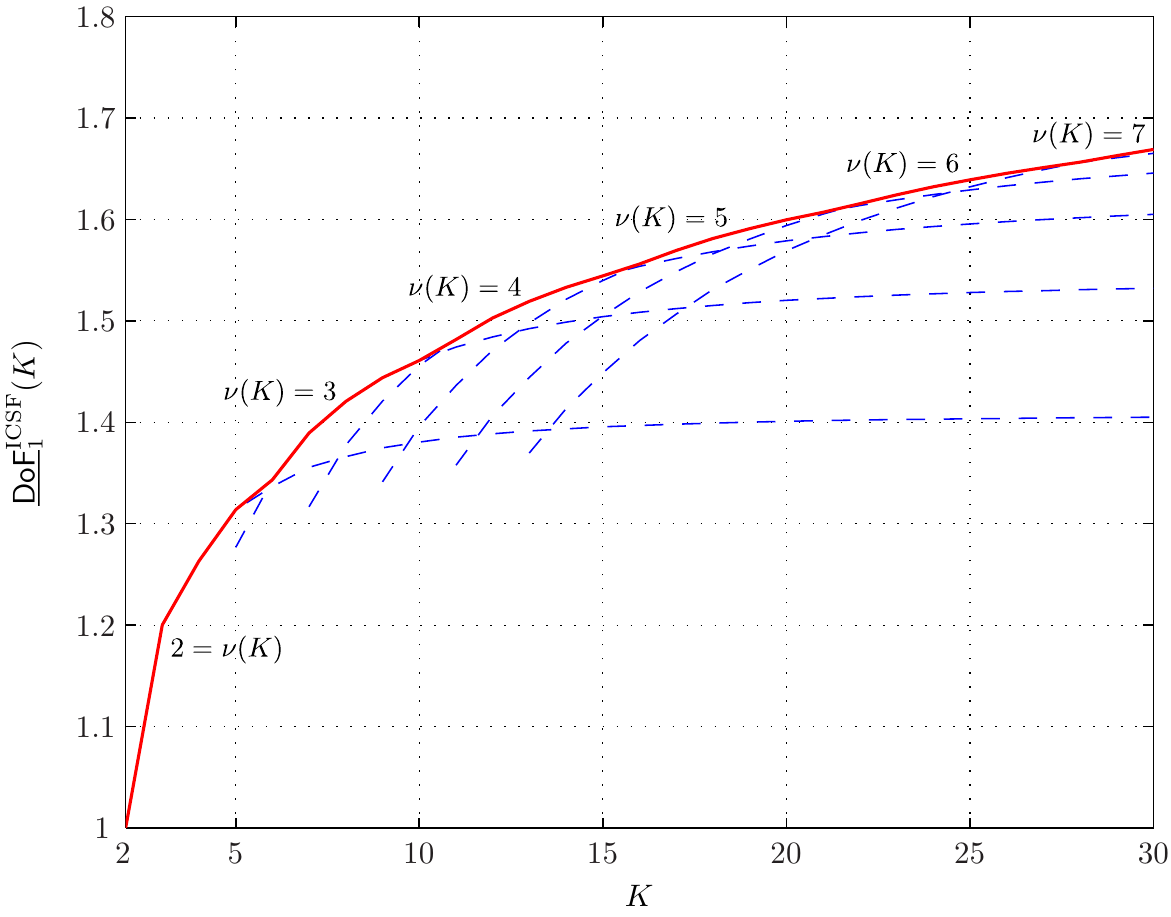}
\caption{Achievable DoFs for the $K$-user IC with Shannon feedback.}
\label{Fig:ICSF_DoF}
\end{figure}

It is shown in Appendix \ref{App:ClosedForm-ICSF} that the solution $\DoFa_m^\textup{ICSF}(K)$ to the recursive Eq. \eqref{Eq:DoF_ICSF_phasem} with initial condition \eqref{Eq:DoF_ICSF_phaseK} is given by \cref{Eq:ICSF-DoF_m}. Therefore, the proof is complete in view of \cref{Eq:ICSF-DoF_1} and the fact that $\nu(K)$ is chosen to maximize $\DoFa_1^\textup{ICSF}(K)$. The achievable DoF for different values of $\nu(K)$ and the optimized achieved DoF are plotted in \cref{Fig:ICSF_DoF} for $2\leq K \leq 30$.

\subsection{Proof of \Cref{Th:XSF}}
\label{Sec:X-SF}
Our transmission scheme for the $K\times K$ X channel with output feedback operates in $2$ rounds:

\round{$1$}{$K\times K$ X Channel with Shannon Feedback}
This round has $2$ phases in parallel with the scheme proposed in proof of \cref{Th:XOF} for the same channel with output feedback. In particular, in phase $1$, $K^2$ fresh information symbols $u^{[i|j]}$, $1\leq i,j\leq K$, are transmitted over the channel during $K$ time slots in the same way as the phase $1$ of the scheme proposed in proof of \cref{Th:XOF}, and $K(K-1)$ symbols $u^{[j'|j;j']}$, $\{j,j'\}\subseteq \xxS_K$, are generated correspondingly. After time slot $j$, TX$_j$, having access to its own transmitted symbol and Shannon feedback, will obtain a linear combination of the $K-1$ symbols $u^{[i|j]}$, $i\in \xxS_K\backslash \{j\}$. Therefore, if TX$_j$ is provided with extra $K-2$ linearly independent combinations of these $K-1$ symbols (with known coefficients), it will be able to decode all of them. 

In phase $2$, the symbols $u^{[j'|j;j']}$ are transmitted in the same way as in the phase $2$ of the scheme presented in proof of \cref{Th:XOF}. However, here, according to the Shannon feedback, each TX$_i$ obtains more linear combinations of the symbols $u^{[j|i]}$, $j\in \xxS_K\backslash \{i\}$, as we proceed with the transmissions. In specific, fix an index $j_0$, $j_0\in \xxS_K$. Then, for any $\{j,j'\}\in \xxS_K\backslash \{j_0\}$, spend one time slot to transmit $u^{[j|j';j]}$ and $u^{[j'|j;j']}$ respectively by TX$_j$ and TX$_{j'}$, while the other transmitters are silent. By the end of this time slot, $u^{[j|j';j]}$ and $u^{[j'|j;j']}$ are delivered to RX$_j$ and RX$_{j'}$, respectively. Also, TX$_j$ will obtain $u^{[j'|j;j']}$ through Shannon feedback, which is a linear combination of $u^{[i|j]}$, $i\in \xxS_K \backslash \{j\}$. Similarly, TX$_{j'}$ will obtain $u^{[j|j';j]}$ which is a linear combination of $u^{[i|j']}$, $i\in \xxS_K \backslash \{j'\}$. Therefore, one can verify that, after the $\binom{K-1}{2}$ time slots of this phase,
\begin{enumerate}[(i)]
\item \label{Observ1} each RX$_j$, $j\in \xxS_K\backslash \{j_0\}$, will receive all the symbols $u^{[j'|j;j']}$, $j'\in \xxS_K \backslash \{j_0,j\}$;
\item \label{Observ2} each TX$_j$, $j\in \xxS_K\backslash \{j_0\}$, will obtain $u^{[j'|j;j']}$, $j'\in \xxS_K \backslash \{j_0,j\}$, which are $K-2$ linear combinations of the symbols $u^{[i|j]}$, $i\in \xxS_K \backslash \{j\}$. These linear combinations together with the linear combination obtained during phase $1$, constitute $K-1$ linearly independent combinations of $K-1$ unknowns, and thus, can be solved for the symbols $u^{[i|j]}$, $i\in \xxS_K \backslash \{j\}$.
\end{enumerate}

By observation (\ref{Observ1}), it only remains to deliver the $2(K-1)$ symbols $u^{[j|j_0,j]}$, $u^{[j_0|j;j_0]}$, $j\in \xxS_K \backslash \{j_0\}$, to their respective receivers. On the other hand, by observation (\ref{Observ2}), the symbol $u^{[j_0|j;j_0]}$, $j\in \xxS_K \backslash \{j_0\}$, can now be reconstructed by TX$_j$, and thus, can be denoted as $u^{[j,j_0|j;j_0]}$. Consequently, one can define the following order-$2$ symbol which is available at TX$_j$:
\begin{align}
u^{[j|j,j_0]}\Def u^{[j|j_0;j]}+u^{[j,j_0|j;j_0]}, \quad\quad j\in \xxS_K \backslash \{j_0\}.
\end{align}
Therefore, it only remains to deliver the above $K-1$ order-$2$ symbols to their respective pairs of receivers. Before proceeding with the next round, we point out here that by $K$ times repetition of phase $1$, each time with $K^2$ fresh information symbols and a new $j_0$, $1\leq j_0\leq K$, we will generate $K(K-1)$ order-$2$ symbols $u^{[j|j,j_0]}$, $j_0 \in \xxS_K, j\in \xxS_K \backslash \{j_0\}$, as above. The achieved DoF will then be given by
\begin{align}
\DoFa_1^\textup{XSF}(K,K)&=\frac{K\times K^2}{K\times K + K\times \binom{K-1}{2}+\frac{K\times (K-1)}{\DoFa_2^\textup{XSF}(K,K)}} \nonumber \\
&=\frac{K^2}{K + \frac{(K-1)(K-2)}{2}+\frac{K-1}{\DoFa_2^\textup{XSF}(K,K)}}, \label{Eq:DoF_1-XSF}
\end{align}
where $\DoFa_2^\textup{XSF}(K,K)$ represents our achievable DoF for transmission of symbols $u^{[i|i,j]}$ and $u^{[j|i,j]}$, $\{i,j\}\subseteq \xxS_K$, over the $K\times K$ SISO X channel with Shannon feedback. These symbols will be delivered to their respective pairs of receivers during the next round.

\round{$2$}{$K\times K$ X Channel with Shannon Feedback}
This round has $K-1$ phases (\ie phases $2$ to $K$). If $K=2$, then the symbols $u^{[1|1,2]}$ and $u^{[2|1,2]}$ are transmitted respectively by TX$_1$ and TX$_2$ in $2$ time slots, by the end of which both receivers will obtain both symbols. If $K>2$, then the $K(K-1)$ order-$2$ symbols of type $u^{[i|i,j]}$ and $u^{[j|i,j]}$, $\{i,j\}\subseteq \xxS_K$, are transmitted over the channel in phase $2$ as follows: For each $\xxS_3=\{i_1,i_2,i_3\}\subseteq \xxS_K$, spend three time slots to transmit $u^{[i_k|i_k,i_\ell]}$ and $u^{[i_\ell|i_k,i_\ell]}$, $\{k,\ell\}\subset \{1,2,3\}$. In specific, over the first time slot, $u^{[i_1|i_1,i_2]}$ and $u^{[i_2|i_1,i_2]}$ are respectively transmitted by TX$_{i_1}$ and TX$_{i_2}$ while the other transmitters are silent. Then, RX$_{i_1}$ and RX$_{i_2}$ each require an extra linear equation to decode both symbols. Hence, after this time slot, the linear combination $h^{[i_3i_1]}u^{[i_1|i_1,i_2]}+h^{[i_3i_2]}u^{[i_2|i_1,i_2]}$ received by RX$_{i_3}$, which is now available at TX$_{i_3}$ via the output feedback, is desired by both RX$_{i_1}$ and RX$_{i_2}$, where the time indices have been omitted for brevity. On the other hand, TX$_{i_1}$ and TX$_{i_2}$ having access to their own transmitted symbol and Shannon feedback, can decode each other's symbol. Therefore, using delayed CSIT, they can reconstruct $h^{[i_3i_1]}u^{[i_1|i_1,i_2]}+h^{[i_3i_2]}u^{[i_2|i_1,i_2]}$. Thus, we can define $u^{[\xxS_3|i_1,i_2;i_3]}\Def h^{[i_3i_1]}u^{[i_1|i_1,i_2]}+h^{[i_3i_2]}u^{[i_2|i_1,i_2]}$. 

Similarly, the second and third time slots are dedicated respectively to transmission of $\{u^{[i_1|i_1,i_3]},u^{[i_3|i_1,i_3]}\}$ and $\{u^{[i_2|i_2,i_3]},u^{[i_3|i_2,i_3]}\}$, and generation of $u^{[\xxS_3|i_1,i_3;i_2]}$ and $u^{[\xxS_3|i_2,i_3;i_1]}$. Now, if we deliver two random linear combinations of $u^{[\xxS_3|i_1,i_2;i_3]}$, $u^{[\xxS_3|i_1,i_3;i_2]}$, and $u^{[\xxS_3|i_2,i_3;i_1]}$ to RX$_{i_1}$, RX$_{i_2}$, and RX$_{i_3}$, each of them will be able to decode its desired symbols. Therefore, we can define the following order-$3$ symbols:
\begin{align}
u_1^{[\xxS_3|\xxS_3]}\Def \alpha_1u^{[\xxS_3|i_2,i_3;i_1]}+\alpha_2u^{[\xxS_3|i_1,i_3;i_2]}+\alpha_3u^{[\xxS_3|i_1,i_2;i_3]}, \\
u_2^{[\xxS_3|\xxS_3]}\Def \alpha'_1u^{[\xxS_3|i_2,i_3;i_1]}+\alpha'_2u^{[\xxS_3|i_1,i_3;i_2]}+\alpha'_3u^{[\xxS_3|i_1,i_2;i_3]},
\end{align}
where $\alpha_k$, $\alpha'_k$, $k=1,2,3$, are random coefficients. The achieved DoF is thus given by 

\begin{align}
\DoFa_2^\textup{XSF}(K,K)=\frac{6\binom{K}{3}}{3\binom{K}{3}+\frac{2\binom{K}{3}}{\DoFa_3^\textup{XSF}(K,K)}}=\frac{6}{3+\frac{2}{\DoFa_3^\textup{XSF}(K,K)}}, \label{Eq:DoF_2-XSF}
\end{align}
where $\DoFa_3^\textup{XSF}(K,K)$ denotes our achievable DoF for transmission of symbols of type $u^{[\xxS_3|\xxS_3]}$ over the $K\times K$ SISO X channel with Shannon feedback.

Since the $K\times K$ SISO X channel has the same input-output relationship as the $K$-user SISO IC, the problem of transmission of order-$3$ symbols of type $u^{[\xxS_3|\xxS_3]}$ over the $K\times K$ X channel with Shannon feedback is equivalent to that of the IC with Shannon feedback. Hence, phase $m$, $3\leq m \leq K$, of round $2$ the scheme proposed in proof of \cref{Th:ICSF} can be used for transmission of the order-$3$ symbols and generation of higher order symbols up to order-$K$ symbols which will be delivered to all receivers in phase $K$. Therefore, the same recursive equation, \ie \cref{Eq:DoF_ICSF_phasem}, holds for $\DoFa_m^\textup{XSF}(K,K)$, $3\leq m \leq K-1$, with $\DoFa_K^\textup{XSF}(K,K)=1$, and thus, $\DoFa_m^\textup{XSF}(K,K)$, $3\leq m \leq K$, is given by \cref{Eq:ICSF-DoF_m}. Finally, \cref{Eq:DoF_XSF} results from \cref{Eq:DoF_1-XSF,Eq:DoF_2-XSF,Eq:ICSF-DoF_m}.

\section{Comparison and Discussion}
\label{Sec:Comparison}
\begin{figure}
\centering
\includegraphics{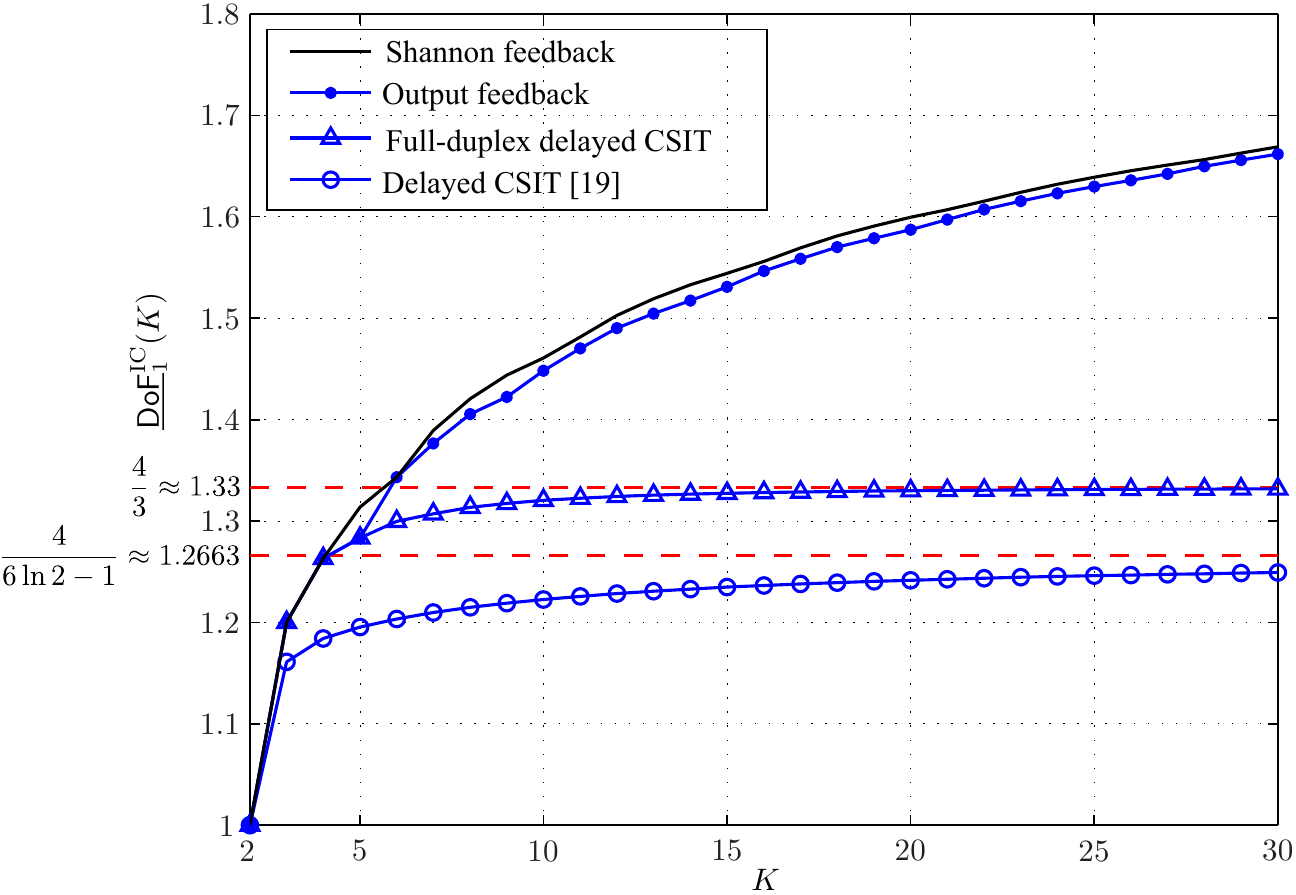}
\caption{Achievable DoFs for the $K$-user IC with Shannon feedback, output feedback, full-duplex delayed CSIT, and delayed CSIT.}
\label{Fig:IC_DoF}
\end{figure}

We compare our results with the best known achievable results on the DoF of both channels with delayed CSIT by the authors of this paper in \cite{Abdoli2011IC-X-Arxiv}. \Cref{Fig:IC_DoF} plots our achievable DoF for the $K$-user SISO IC with delayed CSIT and full-duplex transmitter cooperation, given by \cref{Eq:DoF_ICFD}, together with our achievable DoFs for the $K$-user IC with output and Shannon feedback, respectively given by \cref{Eq:DoF_ICOF,Eq:DoF_ICSF}, and compares them with the achievable DoF for the same channel with delayed CSIT \cite{Abdoli2011IC-X-Arxiv} for $2\leq K \leq 30$. It is seen from the figure that all our achievable DoFs for the $K$-user IC are strictly increasing in $K$, and for $K\geq 3$, they are greater than the achievable DoF for the same channel with delayed CSIT. Also, for $K\geq 6$, we achieve greater DoF with output feedback than with full-duplex delayed CSIT. Our achievable DoF with Shannon feedback is greater than that with output feedback for $K=5$ and $K\geq 7$. One can also verify from \cref{Eq:DoF_ICFD} that
\begin{align}
\lim_{K\to \infty}\DoFa_1^\textup{ICFD}(K)=\frac{4}{3}.
\end{align}

Regarding \cref{Eq:a(K)_def,Eq:w*(K)_def} and the fact that $\mu(K)$ is either $\lfloor w^*_K\rfloor$ or $\lceil w^*_K\rceil$, one can show $\mu(K)=o(K)$, which in view of \cref{Eq:ICOF-DoF_m} yields $\lim_{K\to \infty} \DoFa_{\mu(K)}^\textup{ICOF}(K)=2$. This together with \cref{Eq:ICOF-DoF_1}, and the fact that $\lim_{K\to \infty}\mu(K)=\infty$, implies that
\begin{align}
\lim_{K\to \infty}\DoFa_1^\textup{ICOF}(K)=2. \label{Eq:ICOF-limit}
\end{align}

We now show that $\lim_{K\to \infty}\DoFa_1^\textup{ICSF}(K)=2$. To do so, it suffices to show that $\DoFa_1^\textup{ICSF}(K)<2$. Then, an application of the Squeeze theorem regarding \cref{Eq:ICOF-limit} and the fact that $\DoFa_1^\textup{ICOF}(K)\leq\DoFa_1^\textup{ICSF}(K)$ will yield the desired result. Using \cref{Eq:DoF_ICSF}, we have
\begin{align}
\DoFa_1^\textup{ICSF}(K) &= \max_{\substack{2\leq w \leq \lceil K/2 \rceil\\ w \in \bbZ^+}} \frac{w}{1+\frac{w-2}{\DoFa_w^\textup{ICOF}(K)}+\frac{w}{(w+1)\DoFa_{w+1}^\textup{ICSF}(K)}} \nonumber \\
&< \max_{\substack{2\leq w \leq \lceil K/2 \rceil\\ w \in \bbZ^+}}  \frac{w}{1+\frac{w-2}{\DoFa_w^\textup{ICOF}(K)}} \nonumber \\
&\stackrel{\textup{(a)}}{=}\max_{\substack{2\leq w \leq \lceil K/2 \rceil\\ w \in \bbZ^+}} \frac{1}{a(K)(w-1)(w-2)+\frac{1}{2}} \nonumber \\
&\stackrel{\textup{(b)}}{=}2,
\end{align}
where (a) follows from \cref{Eq:ICOF-DoF_m,Eq:a(K)_def}, and (b) uses the fact that the denominator is strictly increasing in $w$ for $w\geq2$, and thus, is minimized at $w=2$.

\begin{figure}
\centering
\includegraphics[scale=1.08]{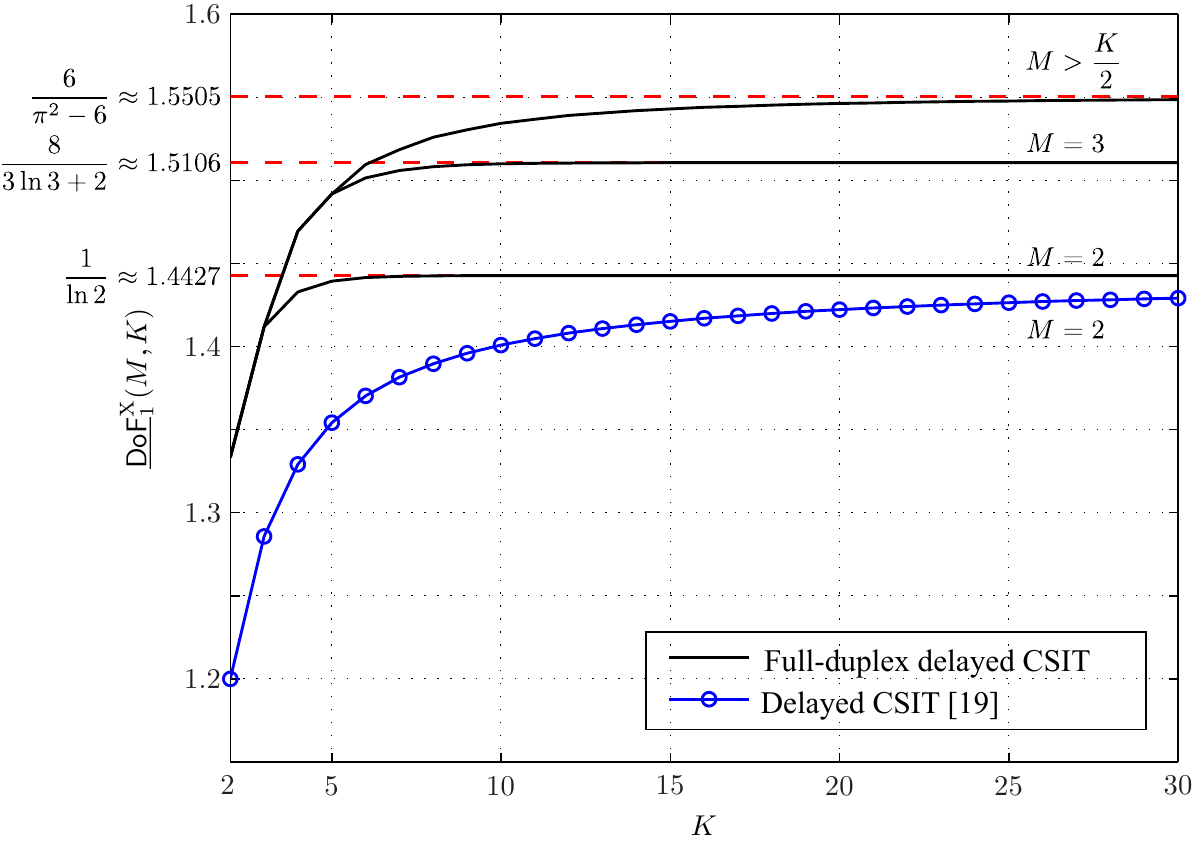}
\caption{Achievable DoFs for the $M\times K$ X channel with delayed CSIT, with and without full-duplex transmitter cooperation.}
\label{Fig:XFD_DoF}
\end{figure}

\Cref{Fig:XFD_DoF} plots our achievable DoFs for the $M\times K$ SISO X channel with delayed CSIT and full-duplex transmitter cooperation, given by \cref{Eq:DoF_XFD}, for $M=2, 3$, and $M>\frac{K}{2}$, and $2\leq K \leq 30$, and compares them with the achievable DoF reported in \cite{Abdoli2011IC-X-Arxiv} for the $2\times K$ X channel with delayed CSIT. For all values of $M$, our achievable DoF for the full-duplex $M\times K$ X channel with delayed CSIT is strictly increasing in $K$ and greater than that of the $2\times K$ X channel with delayed CSIT. Also, it can be shown using \cref{Eq:DoF_XFD} that for a fixed $M$:
\begin{align}
\label{Eq:DoF_XFD_lim_fixedM}
\lim_{K\to \infty}\DoFa_1^\textup{XFD}(M,K)=\frac{1}{\sum_{\ell_1=2}^{M-2}\frac{1}{\ell_1^2}+\frac{1}{M-1}+\frac{1}{(M-1)^2}\left[\left(\frac{M}{M-1}\right)^{M-2}\ln M-\sum_{\ell_2=1}^{M-2}\left(\frac{M}{M-1}\right)^{M-2-\ell_2}\frac{1}{\ell_2}\right]}.
\end{align}
For instance, $\lim_{K\to \infty}\DoFa_1^\textup{XFD}(2,K)=\frac{1}{\ln 2}$ and $\lim_{K\to \infty}\DoFa_1^\textup{XFD}(3,K)=\frac{8}{3\ln 3+2}$, as indicated in \cref{Fig:XFD_DoF}. Moreover, it follows from \cref{Eq:DoF_XFD} and $\sum_{n=1}^{\infty}\frac{1}{n^2}=\frac{\pi^2}{6}$ that, if $M>K/2$ for sufficiently large $K$, then
\begin{align}
\lim_{K\to \infty}\DoFa_1^\textup{XFD}(M,K)=\frac{6}{\pi^2-6}.
\end{align}

\begin{figure}
\centering
\includegraphics[scale=1]{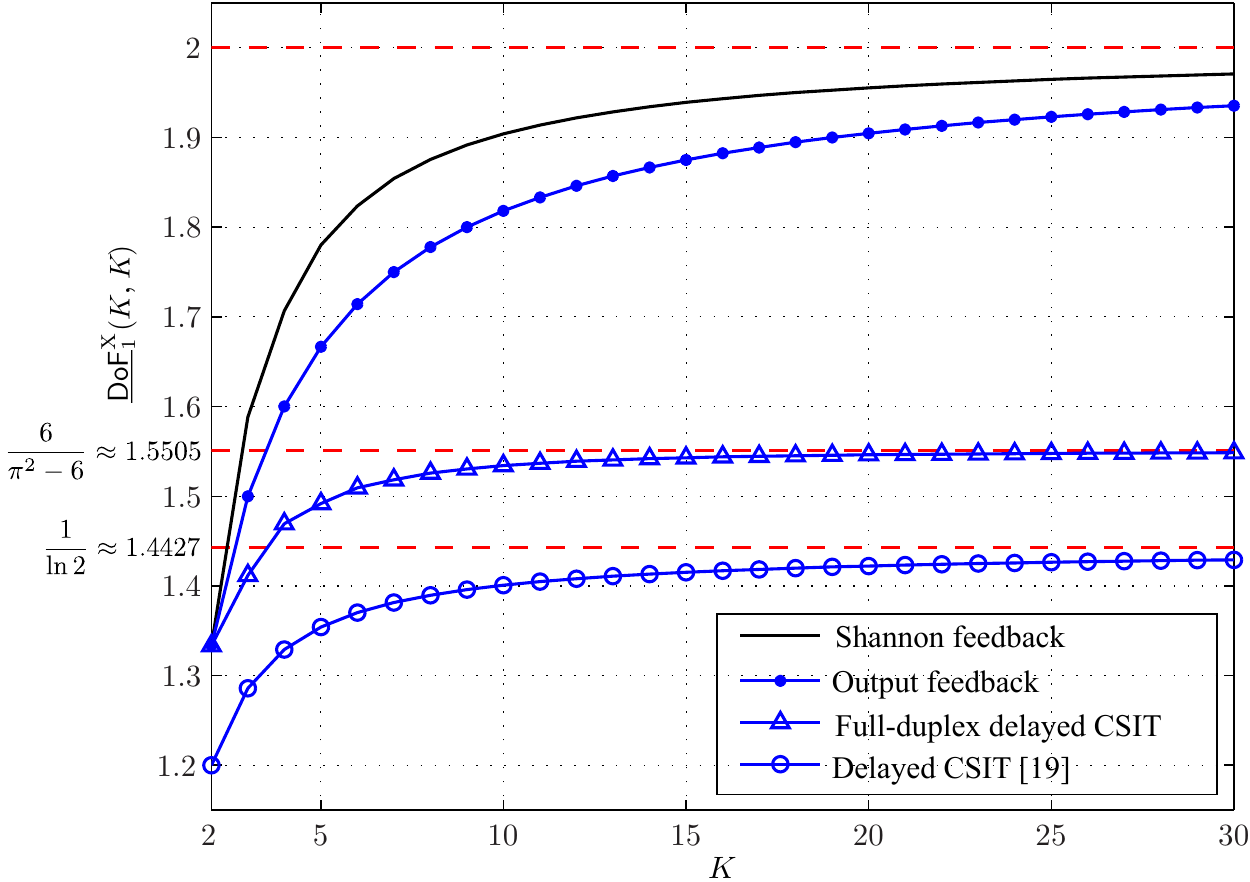}
\caption{Achievable DoFs for the $K\times K$ X channel with Shannon feedback, output feedback, full-duplex delayed CSIT, and delayed CSIT.}
\label{Fig:X_DC_FD_OF_SF_DoF}
\end{figure}

\Cref{Fig:X_DC_FD_OF_SF_DoF} compares our achievable DoF for the $K\times K$ X channel with Shannon feedback (given by \cref{Eq:DoF_XSF}), output feedback (which is $2K/(K+1)$ by \cref{Th:XOF}), full-duplex delayed CSIT (given by \cref{Eq:DoF_XFD}), and delayed CSIT \cite{Abdoli2011IC-X-Arxiv} for $2\leq K \leq 30$. It is observed that for $K>2$, 
\begin{align}
\DoFa_1^\textup{XFD}(K,K)<\DoFa_1^\textup{XOF}(K,K)<\DoFa_1^\textup{XSF}(K,K).
\end{align} 
Also, one can easily verify using \cref{Eq:DoF_XSF} and $\DoFa_1^\textup{XOF}(K,K)=2K/(K+1)$ that
\begin{align}
\lim_{K\to \infty}\DoFa_1^\textup{XOF}(K,K)= \lim_{K\to \infty}\DoFa_1^\textup{XSF}(K,K)=2.
\end{align}

\section{Conclusions}
\label{Sec:Conclusions}
The SISO AWGN interference and X channels with arbitrary number of users were investigated in this paper, where it was assumed that the CSI is not instantaneously available at the transmitters. Achievable results were obtained on the DoF of these channels under three different assumptions, namely, \emph{full-duplex delayed CSIT} (where the transmitters access the delayed CSI \emph{and} can operate in full-duplex mode), \emph{output feedback} (where each transmitter causally accesses the output of its paired receiver), and \emph{Shannon feedback} (where each transmitter accesses both the output feedback and delayed CSI). Under each assumption, the transmitters, obtaining side information about each other's messages through full-duplex or feedback links, could cooperate to align the interference at the receivers in a multi-phase fashion. 

For each channel, the transmitters enjoyed a different level of cooperation under each assumption, and hence, different values of DoF were achieved. The achieved DoFs are greater than the best previously reported achievable DoFs for both channels with delayed CSIT, and are strictly increasing with the number of receivers, though approaching limiting values not greater than $2$ for asymptotically large networks. Our DoF results under the full-duplex delayed CSIT assumption are the first to demonstrate the potential of full-duplex transmitter cooperation to yield DoF gains in multi-user networks. The problem of DoF characterization of both channels under each of the considered assumptions remains open in the lack of tight upper bounds.

\appendices

\section{Closed Form Expression for the Recursive \cref{Eq:DoF_ICFD_phasem,Eq:DoF_ICOF_phasem}}
\label{App:ClosedForm-ICFD-ICOF}
Consider the following recursive equation:
\begin{align}
\DoFa_m(K) =\frac{m+1}{m}\times \frac{Q_m(K)}{1+\frac{Q_m(K)-1}{\DoFa_{m+1}(K)}},\quad\quad 2\leq m \leq K-2,
\end{align}
with $Q_m(K)=\min\{K-m,m\}$ and the initial condition $\DoFa_{K-1}(K)=K/(K-1)$. We treat two different cases separately:

(i) $\lceil K/2 \rceil \leq m \leq K-1$: In this case, we have $Q_m(K)=K-m$, and hence,
\begin{align}
\frac{K-m}{m\DoFa_m(K)}=\frac{1}{m+1}+\frac{K-m-1}{(m+1)\DoFa_{m+1}(K)}.
\end{align}
Then, defining $\gamma_m(K)\Def \frac{K-m}{m\DoFa_m(K)}$, one can write $\gamma_m(K)=\frac{1}{m+1}+\gamma_{m+1}(K)$, which implies that $\gamma_m(K)=\sum_{\ell=m+1}^K\frac{1}{\ell}$, or equivalently,
\begin{align}
\DoFa_m(K)=\left(\frac{m}{K-m}\sum_{\ell=m+1}^K\frac{1}{\ell}\right)^{-1},\hspace{2cm} \lceil K/2 \rceil \leq m \leq K-1.
\label{Eq:Case(i)-DoF-ICFD-ICOF}
\end{align}

(ii) $2 \leq m < \lceil K/2 \rceil $: In this case, we have
\begin{align}
\frac{1}{\DoFa_m(K)}=\frac{1}{m+1}+\left(\frac{m-1}{m+1}\right)\frac{1}{\DoFa_{m+1}(K)},
\end{align}
which can be rewritten as
\begin{align}
\frac{2}{\DoFa_m(K)}-1=\frac{m-1}{m+1}\left(\frac{2}{\DoFa_{m+1}(K)}-1\right).
\end{align}
It immediately follows that
\begin{align}
\frac{2}{\DoFa_m(K)}-1&=\frac{m(m-1)}{\lceil \frac{K}{2}\rceil(\lceil \frac{K}{2}\rceil-1)}\left(\frac{2}{\DoFa_{\lceil \frac{K}{2}\rceil}(K)}-1\right) \nonumber \\
&\stackrel{\textup{(a)}}{=}\frac{m(m-1)}{\lceil \frac{K}{2}\rceil(\lceil \frac{K}{2}\rceil-1)}\left(\frac{2\lceil \frac{K}{2}\rceil\sum_{\ell=\lceil \frac{K}{2}\rceil+1}^K\frac{1}{\ell}}{\lfloor \frac{K}{2}\rfloor}-1\right), \hspace{2cm} 2\leq m < \lceil K/2 \rceil, \label{Eq:Case(ii)-DoF-ICFD-ICOF}
\end{align}
where (a) uses \cref{Eq:Case(i)-DoF-ICFD-ICOF} with $m=\lceil \frac{K}{2}\rceil$, and the fact that $K-\lceil \frac{K}{2}\rceil=\lfloor \frac{K}{2} \rfloor$.

It finally follows from \cref{Eq:Case(i)-DoF-ICFD-ICOF,Eq:Case(ii)-DoF-ICFD-ICOF} that
\begin{align}
\DoFa_m(K)=
\begin{cases}
\left(\frac{1}{2}-\frac{m(m-1)}{2\lceil \frac{K}{2}\rceil(\lceil \frac{K}{2}\rceil-1)}+\frac{m(m-1)}{\lfloor \frac{K}{2}\rfloor(\lceil \frac{K}{2}\rceil-1)}\sum_{\ell=\lceil \frac{K}{2}\rceil+1}^K\frac{1}{\ell}\right)^{-1},  &  \hspace{5mm} 2\leq m \leq \lceil \frac{K}{2}\rceil\\
\left(\frac{m}{K-m}\sum_{\ell=m+1}^K\frac{1}{\ell}\right)^{-1}, & \hspace{-3.5mm} \lceil \frac{K}{2} \rceil < m \leq K-1
\end{cases}.
\end{align}

\section{Closed Form Expression for the Recursive Eq. \eqref{Eq:DoF_XFD_Recursive}}
\label{App:ClosedForm-XFD}
Consider the recursive equation
\begin{align}
\DoFa_m(M,K) =\frac{(m+1)(Q_m(M,K)+1)}{m+1+\frac{m\times Q_m(M,K)}{\DoFa_{m+1}(M,K)}}, \hspace{2cm} \quad 1\leq m\leq K-1,
\end{align}
with $Q_m(M,K)=\min\{M-1,K-m,m\}$ and initial condition $\DoFa_K(M,K)=1$. The following distinct cases can be differentiated:

(i) $M-1 \geq \lceil K/2 \rceil$: In this case, $Q_m(M,K)=Q_m(K)=\min \{K-m,m\}$, and hence,
\begin{align}
\frac{Q_m(K)+1}{m\DoFa_m(M,K)}=\frac{1}{m}+\frac{Q_m(K)}{(m+1)\DoFa_{m+1}(M,K)}. \label{Eq:XFD_Recursive}
\end{align}
Now, if $\lceil K/2 \rceil \leq m \leq K$, then similar to Appendix \ref{App:ClosedForm-ICFD-ICOF}, one can show that
\begin{align}
\DoFa_m(K)=\left(\frac{m}{K-m+1}\sum_{\ell=m}^K\frac{1}{\ell}\right)^{-1},\hspace{2cm} \lceil K/2 \rceil \leq m \leq K. \label{Eq:Case(i)-XFD1}
\end{align}
Otherwise, the recursive Eq. \eqref{Eq:XFD_Recursive} can be rewritten as
\begin{align}
\frac{1}{m^2\DoFa_m(M,K)}&=\frac{1}{m^2(m+1)}+\frac{1}{(m+1)^2\DoFa_{m+1}(M,K)} \nonumber \\
&=\sum_{\ell=m}^{\lceil \frac{K}{2}\rceil-1}\frac{1}{\ell^2(\ell+1)}+\frac{1}{\lceil \frac{K}{2}\rceil^2\DoFa_{\lceil \frac{K}{2}\rceil}(M,K)} \nonumber \\
&\stackrel{\textup{(a)}}{=}\frac{1}{\lceil \frac{K}{2}\rceil}-\frac{1}{m}+\sum_{\ell=m}^{\lceil \frac{K}{2}\rceil-1}\frac{1}{\ell^2}+\frac{1}{\lceil \frac{K}{2}\rceil(\lfloor \frac{K}{2}\rfloor+1)}\sum_{\ell=\lceil \frac{K}{2}\rceil}^K\frac{1}{\ell}, \hspace{1cm}1\leq m <\lceil K/2\rceil, \label{Eq:Case(i)-XFD2}
\end{align}
where (a) uses \cref{Eq:Case(i)-XFD1} with $m=\lceil \frac{K}{2}\rceil$, and the fact that $K-\lceil \frac{K}{2}\rceil=\lfloor \frac{K}{2} \rfloor$.

\Cref{Eq:Case(i)-XFD1,Eq:Case(i)-XFD2} yield
\begin{align}
\DoFa_m(M,K)=
\begin{cases}
\left(\frac{m^2}{\lceil \frac{K}{2}\rceil}-m+m^2\sum_{\ell=m}^{\lceil \frac{K}{2}\rceil-1}\frac{1}{\ell^2}+\frac{m^2}{\lceil \frac{K}{2}\rceil(\lfloor \frac{K}{2}\rfloor+1)}\sum_{\ell=\lceil \frac{K}{2}\rceil}^K\frac{1}{\ell}\right)^{-1}, & 1\leq m <\lceil \frac{K}{2}\rceil \\
\left(\frac{m}{K-m+1}\sum_{\ell=m}^K\frac{1}{\ell}\right)^{-1}, &\lceil \frac{K}{2} \rceil \leq m \leq K
\end{cases}.
\end{align}

(ii) $M-1<\lceil K/2 \rceil$: In this case, if $K-M+1\leq m \leq K$, then the same expression as \cref{Eq:Case(i)-XFD1} holds for $\DoFa_m(K)$. Otherwise, if $M-1\leq m < K-M+1$, then $Q_m(M,K)=M-1$, and we have
\begin{align}
\frac{1}{m\DoFa_m(M,K)}&=\frac{1}{mM}+\left(\frac{M-1}{M}\right)\frac{1}{(m+1)\DoFa_{m+1}(M,K)} \nonumber \\
&=\frac{1}{M}\sum_{\ell=m}^{K-M}\left(\frac{M-1}{M}\right)^{\ell-m}\frac{1}{\ell}+\frac{\left(\frac{M-1}{M}\right)^{K-M-m+1}}{(K-M+1)\DoFa_{K-M+1}(M,K)}\nonumber \\
&\stackrel{\textup{(a)}}{=}\frac{1}{M}\sum_{\ell_1=m}^{K-M}\left(\frac{M-1}{M}\right)^{\ell_1-m}\frac{1}{\ell_1}+\frac{1}{M}\left(\frac{M-1}{M}\right)^{K-M-m+1}\sum_{\ell_2=K-M+1}^K\frac{1}{\ell_2},
\end{align}
where (a) follows from \cref{Eq:Case(i)-XFD1} with $m=K-M+1$. Therefore, 
\begin{align}
\DoFa_m(M,K)=\left(\frac{m}{M}\sum_{\ell=m}^K\frac{1}{\ell}\left(\frac{M-1}{M}\right)^{\min(\ell,K-M+1)-m}\right)^{-1},\hspace{.5cm} M-1\leq m < K-M+1. \label{Eq:Case(ii)-XFD1}
\end{align}

Finally, if $1\leq m <M-1$, then
\begin{align}
\frac{1}{m^2\DoFa_m(M,K)}&=\frac{1}{m^2(m+1)}+\frac{1}{(m+1)^2\DoFa_{m+1}(M,K)} \nonumber \\
&=\frac{1}{M-1}-\frac{1}{m}+\sum_{\ell=m}^{M-2}\frac{1}{\ell^2}+\frac{1}{(M-1)^2\DoFa_{M-1}(M,K)} \nonumber \\
&\stackrel{\textup{(a)}}{=}\frac{1}{M-1}-\frac{1}{m}+\sum_{\ell_1=m}^{M-2}\frac{1}{\ell_1^2}+\frac{1}{M^2}\sum_{\ell_2=M-1}^K\frac{1}{\ell_2}\left(\frac{M-1}{M}\right)^{\min(\ell_2,K-M+1)-M},
\end{align}
where (a) uses \cref{Eq:Case(ii)-XFD1} with $m=M-1$. Thus,
\begin{align}
\DoFa_m(M,K)=\frac{1}{\frac{m^2}{M-1}-m+m^2\sum_{\ell_1=m}^{M-2}\frac{1}{\ell_1^2}+\left(\frac{m}{M}\right)^2\sum_{\ell_2=M-1}^{K}\frac{1}{\ell_2}\left(\frac{M-1}{M}\right)^{\min(\ell_2,K-M+1)-M}},\hspace{.3cm} 1\leq m <M-1. \label{Eq:Case(ii)-XFD2}
\end{align}

\section{Closed Form Expression for the Recursive Eq. \eqref{Eq:DoF_ICSF_phasem}}
\label{App:ClosedForm-ICSF}
Consider the recursive equation
\begin{align}
\DoFa_m(K)=\frac{(m+1)Q_m(K+1)}{m+1+\frac{m\times(Q_m(K+1)-1)}{\DoFa_{m+1}(K)}},\quad\quad 2\leq m \leq K-1,
\end{align}
with initial condition $\DoFa_K(K)=1$. For $\lfloor \frac{K}{2} \rfloor < m \leq K$, it is easily shown that $\DoFa_m(K)$ is given by \cref{Eq:Case(i)-XFD1}. For $2\leq m \leq \lfloor \frac{K}{2} \rfloor$, we have
\begin{align}
\frac{1}{\DoFa_m(K)}&=\frac{1}{m}+\left(\frac{m-1}{m+1}\right)\frac{1}{\DoFa_{m+1}(K)}\nonumber \\
&=\frac{1}{m}+m(m-1)\sum_{\ell=m}^{\lfloor \frac{K}{2}\rfloor-1}\frac{1}{\ell(\ell+1)^2}+\left(\frac{m(m-1)}{\lfloor \frac{K}{2}\rfloor(\lfloor \frac{K}{2}\rfloor+1)}\right)\frac{1}{\DoFa_{\lfloor \frac{K}{2}\rfloor+1}(K)} \nonumber \\
&\stackrel{\textup{(a)}}{=}\frac{1}{m}+m-1-\frac{m(m-1)}{\lfloor \frac{K}{2}\rfloor}-m(m-1)\sum_{\ell_1=m+1}^{\lfloor \frac{K}{2}\rfloor}\frac{1}{\ell_1^2}+\frac{m(m-1)}{\lfloor \frac{K}{2}\rfloor\lceil \frac{K}{2}\rceil}\sum_{\ell_2=\lfloor \frac{K}{2}\rfloor+1}^K\frac{1}{\ell_2},
\end{align}
where (a) uses \cref{Eq:Case(i)-XFD1} with $m=\lfloor \frac{K}{2}\rfloor+1$, and the fact that $K-\lfloor \frac{K}{2} \rfloor=\lceil \frac{K}{2}\rceil$. Therefore,
\begin{align}
\DoFa_m(K)=\left(\frac{1}{m}+m(m-1)\left[\frac{1}{m}-\frac{1}{\lfloor \frac{K}{2}\rfloor}-\!\!\!\sum_{\ell_1=m+1}^{\lfloor \frac{K}{2}\rfloor}\frac{1}{\ell_1^2}+\frac{1}{\lfloor \frac{K}{2}\rfloor\lceil \frac{K}{2}\rceil}\sum_{\ell_2=\lfloor \frac{K}{2}\rfloor+1}^K\frac{1}{\ell_2}\right]\right)^{-1}\!\!, \hspace{.5cm} 2\leq m \leq \lfloor K/2 \rfloor.
\end{align}

\bibliographystyle{IEEEtran}
\bibliography{FullDuplex_IC_X_Ref}

\begin{thebibliography}{10}
\providecommand{\url}[1]{#1}
\csname url@samestyle\endcsname
\providecommand{\newblock}{\relax}
\providecommand{\bibinfo}[2]{#2}
\providecommand{\BIBentrySTDinterwordspacing}{\spaceskip=0pt\relax}
\providecommand{\BIBentryALTinterwordstretchfactor}{4}
\providecommand{\BIBentryALTinterwordspacing}{\spaceskip=\fontdimen2\font plus
\BIBentryALTinterwordstretchfactor\fontdimen3\font minus
  \fontdimen4\font\relax}
\providecommand{\BIBforeignlanguage}[2]{{%
\expandafter\ifx\csname l@#1\endcsname\relax
\typeout{** WARNING: IEEEtran.bst: No hyphenation pattern has been}%
\typeout{** loaded for the language `#1'. Using the pattern for}%
\typeout{** the default language instead.}%
\else
\language=\csname l@#1\endcsname
\fi
#2}}
\providecommand{\BIBdecl}{\relax}
\BIBdecl

\bibitem{Abdoli2012Fullduplex}
M.~J. Abdoli, A.~Ghasemi, and A.~K. Khandani, ``Full-duplex transmitter
  cooperation, feedback, and the degrees of freedom of {SISO} {Gaussian}
  interference and {X} channels,'' in \emph{Information Theory, 2012. ISIT
  2012. IEEE International Symposium on}, 2012.

\bibitem{shannon1956zero}
C.~Shannon, ``The zero error capacity of a noisy channel,'' \emph{Information
  Theory, IRE Transactions on}, vol.~2, no.~3, pp. 8--19, 1956.

\bibitem{ozarow1984MACcapacity}
L.~Ozarow, ``The capacity of the white {Gaussian} multiple access channel with
  feedback,'' \emph{Information Theory, IEEE Transactions on}, vol.~30, no.~4,
  pp. 623--629, 1984.

\bibitem{ozarow1984BCCachievable}
------, ``An achievable region and outer bound for the {Gaussian} broadcast
  channel with feedback (corresp.),'' \emph{Information Theory, IEEE
  Transactions on}, vol.~30, no.~4, pp. 667--671, 1984.

\bibitem{suh2010feedback}
C.~Suh and D.~Tse, ``Feedback capacity of the {Gaussian} interference channel
  to within 2 bits,'' \emph{Arxiv preprint arXiv:1005.3338}, 2010.

\bibitem{Bhaskaran2008GaussianBCC}
S.~R. Bhaskaran, ``Gaussian broadcast channel with feedback,''
  \emph{Information Theory, IEEE Transactions on}, vol.~54, no.~11, pp.
  5252--5257, 2008.

\bibitem{tse2009fadingBC}
D.~Tse and R.~Yates, ``Fading broadcast channels with state information at the
  receivers,'' \emph{Arxiv preprint arXiv:0904.3165}, 2009.

\bibitem{Zhu2011ZIC}
Y.~Zhu and D.~Guo, ``Ergodic fading {Z}-interference channels without state
  information at transmitters,'' \emph{Information Theory, IEEE Transactions
  on}, vol.~57, no.~5, pp. 2627--2647, 2011.

\bibitem{vaze2009DoF_NoCSIT}
C.~S. Vaze and M.~K. Varanasi, ``The degrees of freedom regions of {MIMO}
  broadcast, interference, and cognitive radio channels with no {CSIT},''
  \emph{Arxiv preprint arXiv:0909.5424}, 2009.

\bibitem{cadambe2008interference}
V.~R. Cadambe and S.~A. Jafar, ``Interference alignment and degrees of freedom
  of the {$K$}-user interference channel,'' \emph{Information Theory, IEEE
  Transactions on}, vol.~54, no.~8, pp. 3425--3441, 2008.

\bibitem{cadambe2009X}
------, ``Interference alignment and the degrees of freedom of wireless {X}
  networks,'' \emph{Information Theory, IEEE Transactions on}, vol.~55, no.~9,
  pp. 3893--3908, 2009.

\bibitem{maddah2010DoF_BCC_Delayed_Arxiv}
M.~A. Maddah-Ali and D.~Tse, ``Completely stale transmitter channel state
  information is still very useful,'' \emph{Arxiv preprint arXiv:1010.1499v2},
  2010.

\bibitem{vaze2011DoF_BCC_Delayed}
C.~S. Vaze and M.~K. Varanasi, ``The degrees of freedom region of the two-user
  {MIMO} broadcast channel with delayed {CSIT},'' in \emph{Information Theory,
  2011. ISIT 2011. IEEE International Symposium on}, 2011, pp. 331--335.

\bibitem{abdoli2011BCC}
M.~J. Abdoli, A.~Ghasemi, and A.~K. Khandani, ``On the degrees of freedom of
  three-user {MIMO} broadcast channel with delayed {CSIT},'' in
  \emph{Information Theory, 2011. ISIT 2011. IEEE International Symposium on},
  2011, pp. 341--345.

\bibitem{Maleki_Jafar_Retro}
H.~Maleki, S.~A. Jafar, and S.~Shamai, ``Retrospective interference
  alignment,'' in \emph{Information Theory, 2011. ISIT 2011. IEEE International
  Symposium on}, 2011, pp. 2887--2891.

\bibitem{Ghasemi2011Xchannel}
A.~Ghasemi, A.~S. Motahari, and A.~K. Khandani, ``On the degrees of freedom of
  {X} channel with delayed {CSIT},'' in \emph{Information Theory, 2011. ISIT
  2011. IEEE International Symposium on}, 2011, pp. 909--912.

\bibitem{Ghasemi2012MIMOX}
A.~Ghasemi, M.~J. Abdoli, and A.~K. Khandani, ``On the degrees of freedom of
  {MIMO} {X} channel with delayed {CSIT},'' in \emph{Information Theory, 2012.
  ISIT 2012. IEEE International Symposium on}, 2012.

\bibitem{Abdoli2011Allerton}
M.~J. Abdoli, A.~Ghasemi, and A.~K. Khandani, ``On the degrees of freedom of
  {SISO} interference and {X} channels with delayed {CSIT},'' in
  \emph{Communication, Control, and Computing (Allerton), 2011 49th Annual
  Allerton Conference on}, 2011, pp. 625--632.

\bibitem{Abdoli2011IC-X-Arxiv}
------, ``On the degrees of freedom of {$K$}-user {SISO} interference and {X}
  channels with delayed {CSIT},'' \emph{Arxiv preprint arXiv:1109.4314}, 2011.

\bibitem{vaze2011DoF_IC_DelayedCSIT_Arxiv}
C.~S. Vaze and M.~K. Varanasi, ``The degrees of freedom region and interference
  alignment for the {MIMO} interference channel with delayed {CSI},''
  \emph{Arxiv preprint arXiv:1101.5809}, 2011.

\bibitem{ghasemi2011interference}
A.~Ghasemi, A.~S. Motahari, and A.~K. Khandani, ``Interference alignment for
  the {MIMO} interference channel with delayed local {CSIT},'' \emph{Arxiv
  preprint arXiv:1102.5673}, 2011.

\bibitem{Vaze2011ShannonFeedback}
C.~S. Vaze and M.~K. Varanasi, ``The degrees of freedom region of the {MIMO}
  interference channel with {Shannon} feedback,'' \emph{Arxiv preprint
  arXiv:1109.5779}, 2011.

\bibitem{Tandon2011ShannonFeedback}
R.~Tandon, S.~Mohajer, H.~V. Poor, and S.~Shamai, ``On interference networks
  with feedback and delayed {CSI},'' \emph{Arxiv preprint arXiv:1109.5373},
  2011.

\bibitem{Host2006CooperativeDiversity}
A.~Host-Madsen, ``Capacity bounds for cooperative diversity,''
  \emph{Information Theory, IEEE Transactions on}, vol.~52, no.~4, pp.
  1522--1544, 2006.

\bibitem{Prabhakaran2011FullDuplex}
V.~M. Prabhakaran and P.~Viswanath, ``Interference channels with source
  cooperation,'' \emph{Information Theory, IEEE Transactions on}, vol.~57,
  no.~1, pp. 156--186, 2011.

\bibitem{yang2011interference}
S.~Yang and D.~Tuninetti, ``Interference channel with generalized feedback
  (a.k.a. with source cooperation): Part {I}: Achievable region,''
  \emph{Information Theory, IEEE Transactions on}, vol.~57, no.~5, pp.
  2686--2710, 2011.

\bibitem{tandon2011dependence}
R.~Tandon and S.~Ulukus, ``Dependence balance based outer bounds for {Gaussian}
  networks with cooperation and feedback,'' \emph{Information Theory, IEEE
  Transactions on}, vol.~57, no.~7, pp. 4063--4086, 2011.

\bibitem{cao2007achievable}
Y.~Cao and B.~Chen, ``An achievable rate region for interference channels with
  conferencing,'' in \emph{Information Theory, 2007. ISIT 2007. IEEE
  International Symposium on}, 2007, pp. 1251--1255.

\bibitem{Host2005Multiplexing}
A.~Host-Madsen and A.~Nosratinia, ``The multiplexing gain of wireless
  networks,'' in \emph{Information Theory, 2005. ISIT 2005. IEEE International
  Symposium on}, 2005, pp. 2065--2069.

\bibitem{Cadambe2009FullDuplex}
V.~R. Cadambe and S.~A. Jafar, ``Degrees of freedom of wireless networks with
  relays, feedback, cooperation, and full duplex operation,'' \emph{Information
  Theory, IEEE Transactions on}, vol.~55, no.~5, pp. 2334--2344, 2009.

\bibitem{tandon2012x}
R.~Tandon, S.~Mohajer, H.~V. Poor, and S.~Shamai, ``On {X}-channels with
  feedback and delayed {CSI},'' \emph{Arxiv preprint arXiv:1201.6313}, 2012.

\end{thebibliography}
\end{document}